\documentclass[epj,nopacs]{svjour}

\usepackage[ngerman, english]{babel}   
\usepackage[utf8]{inputenc}
\usepackage{lmodern}
\usepackage{amssymb}
\usepackage{amsmath}
\usepackage{graphicx}
\usepackage{diagbox} 
\usepackage{hhline}  

\usepackage{xr} 
\usepackage{cite} 
  
\usepackage[usenames,dvipsnames]{xcolor} 
\usepackage[colorlinks, citecolor={blue!50!black}, urlcolor={blue!50!black}, linkcolor={red!50!black},hypertexnames=false]{hyperref}

\usepackage[figure]{hypcap}

\usepackage{subfigure}   
\usepackage{tikz}             

\usetikzlibrary{positioning}
\usetikzlibrary{arrows.meta}
\usetikzlibrary{decorations.markings}
\usetikzlibrary{decorations.pathmorphing}
\usetikzlibrary{decorations.pathreplacing}
\usetikzlibrary{calc}
   
\tikzset{
   linePlain/.style={draw=black, thick},
   lineBare/.style={draw=gray, thick},
   lineWithArrowEnd/.style={draw=black, thick, postaction={decorate},decoration={markings,mark=at position 1. with {\arrow[scale=1.2]{latex}}}},
   lineWithSmallArrowEnds/.style={draw=gray, postaction={decorate},decoration={markings,mark={at position 1. with {\arrow[scale=0.8]{latex}} }}},
   lineBareWithArrowEnd/.style={draw=gray, thick, postaction={decorate},decoration={markings,mark=at position 1. with {\arrow[scale=1.2]{latex}}}},
   lineWithArrowCenter/.style={draw=black, thick, postaction={decorate},decoration={markings,mark=at position .6 with {\arrow[scale=1.2]{latex}}}},
   lineWithArrowCenterCenter/.style={draw=black, thick, postaction={decorate},decoration={markings,mark=at position .5 with {\arrow[scale=1.2]{latex}}}},
   lineBareWithArrowCenter/.style={draw=gray, thick, postaction={decorate},decoration={markings,mark=at position .6 with {\arrow[scale=1.2]{latex}}}},
   lineWithArrowCenterEnd/.style={draw=black, thick, postaction={decorate},decoration={markings,mark=at position .85 with {\arrow[scale=1.2]{latex}}}},
   lineWithArrowCenterEnd/.style={draw=black, thick, postaction={decorate},decoration={markings,mark=at position .85 with {\arrow[scale=1.2]{latex}}}},
   lineWithArrowCenterStart/.style={draw=black, thick, postaction={decorate},decoration={markings,mark=at position .35 with {\arrow[scale=1.2]{latex}}}},
   lineWithArrowInline/.style={draw=black, semithick, postaction={decorate},decoration={markings,mark=at position .7 with {\arrow[scale=1.2]{latex}}}},
   vertex/.style={draw, shape=circle, fill=black, minimum size=1.1mm, inner sep=0mm, outer sep=0mm},
   bosonLine/.style={draw=black, thick, decorate, decoration={snake, segment length=2mm, amplitude=0.6mm}},
}

\newcommand{\tikzm}[2]{
   \tikz[baseline=-0.65ex]{#2}
}

\newcommand{\bosonfull}[4]{
   \draw[bosonLine, very thick] (#1,#2) -- (#3,#4);
}

\newcommand{\arrowslefthalf}[2]{
   \def\shift{0.3};
   \coordinate (center) at (#1,#2);
   \draw[lineWithArrowCenterEnd] (center)  -- ($(center) + (-\shift,-\shift)$);
   \draw[lineWithArrowCenterEnd] ($(center)     + (-\shift,+\shift)$) -- (center);
}
\newcommand{\arrowsrighthalf}[2]{
   \def\shift{0.3};
   \coordinate (center) at (#1,#2);
   \draw[lineWithArrowCenterEnd] ($(center) + (+\shift,-\shift)$) -- (center);
   \draw[lineWithArrowCenterEnd] (center)    -- ($(center)   + (+\shift,+\shift)$);
}
\newcommand{\arrowslowerhalf}[2]{
   \def\shift{0.3};
   \coordinate (center) at (#1,#2);
   \draw[lineWithArrowCenterEnd] (center)  -- ($(center) + (-\shift,-\shift)$);
   \draw[lineWithArrowCenterEnd] ($(center) + (+\shift,-\shift)$) -- (center);
}
\newcommand{\arrowsupperhalf}[2]{
   \def\shift{0.3};
   \coordinate (center) at (#1,#2);
   \draw[lineWithArrowCenterEnd] ($(center)     + (-\shift,+\shift)$) -- (center);
   \draw[lineWithArrowCenterEnd] (center)    -- ($(center)   + (+\shift,+\shift)$);
}

\newcommand{\arrowslefthalffull}[3]{
   \def\shift{0.3};
   \def\shiftbox{0.3*#3};
   \coordinate (center) at (#1,#2);
   \coordinate (bottomleft)  at ($(center) + (-\shiftbox,-\shiftbox)$);
   \coordinate (topleft)     at ($(center) + (-\shiftbox,+\shiftbox)$);
   \draw[lineWithArrowCenterEnd] (bottomleft)  -- ($(bottomleft) + (-\shift,-\shift)$);
   \draw[lineWithArrowCenterEnd] ($(topleft)     + (-\shift,+\shift)$) -- (topleft);
}
\newcommand{\arrowsrighthalffull}[3]{
   \def\shift{0.3};
   \def\shiftbox{0.3*#3};
   \coordinate (center) at (#1,#2);
   \coordinate (bottomright) at ($(center) + (+\shiftbox,-\shiftbox)$);
   \coordinate (topright)    at ($(center) + (+\shiftbox,+\shiftbox)$);
   \draw[lineWithArrowCenterEnd] ($(bottomright) + (+\shift,-\shift)$) -- (bottomright);
   \draw[lineWithArrowCenterEnd] (topright)    -- ($(topright)   + (+\shift,+\shift)$);
}
\newcommand{\arrowslowerhalffull}[3]{
   \def\shift{0.3};
   \def\shiftbox{0.3*#3};
   \coordinate (center) at (#1,#2);
   \coordinate (bottomleft)  at ($(center) + (-\shiftbox,-\shiftbox)$);
   \coordinate (bottomright) at ($(center) + (+\shiftbox,-\shiftbox)$);
   \draw[lineWithArrowCenterEnd] (bottomleft)  -- ($(bottomleft) + (-\shift,-\shift)$);
   \draw[lineWithArrowCenterEnd] ($(bottomright) + (+\shift,-\shift)$) -- (bottomright);
}
\newcommand{\arrowsupperhalffull}[3]{
   \def\shift{0.3};
   \def\shiftbox{0.3*#3};
   \coordinate (center) at (#1,#2);
   \coordinate (topleft)     at ($(center) + (-\shiftbox,+\shiftbox)$);
   \coordinate (topright)    at ($(center) + (+\shiftbox,+\shiftbox)$);
   \draw[lineWithArrowCenterEnd] ($(topleft)     + (-\shift,+\shift)$) -- (topleft);
   \draw[lineWithArrowCenterEnd] (topright)    -- ($(topright)   + (+\shift,+\shift)$);
}
\newcommand{\arrowsallfull}[3]{
   \arrowslefthalffull{#1}{#2}{#3}
   \arrowsrighthalffull{#1}{#2}{#3}
}

\newcommand{\arrowslefthalfp}[3]{
   \def\shift{0.3};
   \coordinate (center) at (#1,#2);
   \draw[lineWithArrowCenterEnd] (center) -- ($(center) + (-\shift,-\shift)$);
   \draw[lineWithArrowCenterEnd] (center) to [out=45, in=180] ($(center) + (1.2*#3+\shift,2*\shift)$);
}

\newcommand{\arrowsrighthalfp}[3]{
   \def\shift{0.3};
   \coordinate (center) at (#1,#2);
   \draw[lineWithArrowCenterEnd] ($(center) + (+\shift,-\shift)$) -- (center);
   \draw[lineWithArrowCenterStart] ($(center) + (-1.2*#3-\shift,2*\shift)$) to [out=0, in=135] (center);
}

\newcommand{\arrowslefthalffullp}[5]{
   \def\shift{0.3};
   \def\shiftboxl{0.3*#4};
   \def\shiftboxr{0.3*#5};
   \coordinate (center) at (#1,#2);
   \coordinate (bottomleft)  at ($(center) + (-\shiftboxl,-\shiftboxl)$);
   \coordinate (topright)    at ($(center) + (+\shiftboxl,+\shiftboxl)$);
   \draw[lineWithArrowCenterEnd] (bottomleft)  -- ($(bottomleft) + (-\shift,-\shift)$);
   \draw[lineWithArrowCenterEnd] (topright) to [out=30, in=180] ($(center) + (1.2*#3+\shiftboxl+2*\shiftboxr+\shift,\shiftboxr+1.85*\shift)$);
}
\newcommand{\arrowsrighthalffullp}[5]{
   \def\shift{0.3};
   \def\shiftboxl{0.3*#4};
   \def\shiftboxr{0.3*#5};
   \coordinate (center) at (#1,#2);
   \coordinate (topleft)     at ($(center) + (-\shiftboxr,+\shiftboxr)$);
   \coordinate (bottomright) at ($(center) + (+\shiftboxr,-\shiftboxr)$);
   \draw[lineWithArrowCenterEnd] ($(bottomright) + (+\shift,-\shift)$) -- (bottomright);
   \draw[lineWithArrowCenterStart] ($(center) + (-1.2*#3-2*\shiftboxl-\shiftboxr-\shift,\shiftboxl+1.85*\shift)$) to [out=0, in=150] (topleft);
}

\newcommand{\barevertex}[2]{
   \fill (#1,#2) circle (2pt);
}

\newcommand{\barevertexwithlegs}[2]{
   \fill (#1,#2) circle (2pt) coordinate (center);
   \arrowslefthalf{#1}{#2}
   \arrowsrighthalf{#1}{#2}
}

\newcommand{\fullvertex}[4]{
   \def\shift{0.3};   
   \def\shiftbox{0.3*#4};
   \coordinate (center) at (#2,#3);
   \coordinate (bottomleft)  at ($(center) + (-\shiftbox,-\shiftbox)$);
   \coordinate (topleft)     at ($(center) + (-\shiftbox,+\shiftbox)$);
   \coordinate (bottomright) at ($(center) + (+\shiftbox,-\shiftbox)$);
   \coordinate (topright)    at ($(center) + (+\shiftbox,+\shiftbox)$);
   \draw[linePlain, fill=verylightgray] (bottomleft) rectangle (topright);
   \node at (center) {#1};
}

\newcommand{\fullvertexwithlegs}[4]{
   \def\shift{0.3};   
   \def\shiftbox{0.3*#4};
   \coordinate (center) at (#2,#3);
   \coordinate (bottomleft)  at ($(center) + (-\shiftbox,-\shiftbox)$);
   \coordinate (topleft)     at ($(center) + (-\shiftbox,+\shiftbox)$);
   \coordinate (bottomright) at ($(center) + (+\shiftbox,-\shiftbox)$);
   \coordinate (topright)    at ($(center) + (+\shiftbox,+\shiftbox)$);
   \draw[linePlain, fill=verylightgray] (bottomleft) rectangle (topright);
   \node at (center) {#1};
   \draw[lineWithArrowCenterEnd] (bottomleft)  -- ($(bottomleft) + (-\shift,-\shift)$);
   \draw[lineWithArrowCenterEnd] ($(topleft)     + (-\shift,+\shift)$) -- (topleft);
   \draw[lineWithArrowCenterEnd] ($(bottomright) + (+\shift,-\shift)$) -- (bottomright);
   \draw[lineWithArrowCenterEnd] (topright)    -- ($(topright)   + (+\shift,+\shift)$);
}

\newcommand{\fullvertexwide}[5]{
   \def\shift{0.3};   
   \def\shiftbox{0.3*#4};
   \coordinate (center) at (#2,#3);
   \coordinate (bottomleft)  at ($(center) + (-\shiftbox,-\shiftbox)$);
   \coordinate (topleft)     at ($(center) + (-\shiftbox,+\shiftbox)$);
   \coordinate (bottomright) at ($(center) + (+\shiftbox+#5,-\shiftbox)$);
   \coordinate (topright)    at ($(center) + (+\shiftbox+#5,+\shiftbox)$);
   \draw[linePlain, fill=verylightgray] (bottomleft) rectangle (topright);
   \node at (#2+0.5*#5,#3) {#1};
}

\newcommand{\Konea}[4]{
   \draw[linePlain, fill=verylightgray] (#2,#3) to [out=45, in=135] (#2+1.4*#4,#3) to [out=225, in=315] (#2,#3);
	\node at (#2+0.7*#4,#3) {#1};
}

\newcommand{\Konet}[4]{
   \draw[linePlain, fill=verylightgray] (#2,#3) to [out=315, in=45] (#2,#3-1.4*#4) to [out=135, in=225] (#2,#3);
	\node at (#2,#3-0.7*#4) {#1};
}

\newcommand{\Ktwoa}[4]{
   \draw[linePlain, fill=verylightgray] (#2,#3+0.3*#4) to [out=0, in=135] (#2+1.2*#4,#3) to [out=225, in=0] (#2,#3-0.3*#4) -- (#2,#3+0.3*#4);
	\node at (#2+0.4*#4,#3) {#1};
}

\newcommand{\Ktwot}[4]{
   \draw[linePlain, fill=verylightgray] (#2+0.3*#4,#3) to [out=270, in=45] (#2,#3-1.2*#4) to [out=135, in=270] (#2-0.3*#4,#3) -- (#2+0.3*#4,#3);
	\node at (#2,#3-0.4*#4) {#1};
}

\newcommand{\Ktwoab}[4]{
   \draw[linePlain, fill=verylightgray] (#2,#3) to [out=45, in=180] (#2+1.2*#4,#3+0.3*#4) -- (#2+1.2*#4,#3-0.3*#4) to [out=180, in=315] (#2,#3);
	\node at (#2+0.8*#4,#3) {#1};
}

\newcommand{\Ktwotb}[4]{
   \draw[linePlain, fill=verylightgray] (#2,#3) to [out=315, in=90] (#2+0.3*#4,#3-1.2*#4) -- (#2-0.3*#4,#3-1.2*#4) to [out=90, in=225] (#2,#3);
	\node at (#2,#3-0.8*#4) {#1};
}

\newcommand{\abubblebarebarebare}[3]{
   \draw[lineBareWithArrowCenter] (#1,#2) to [out=45, in=135] (#1+1.2*#3,#2);
   \draw[lineBareWithArrowCenter] (#1+1.2*#3,#2) to [out=225, in=315] (#1,#2);
}

\newcommand{\abubblebarefull}[4]{
   \draw[lineWithArrowCenter] (#1,#2) to [out=45, in=135] (#1+1.2*#3,#2+0.3*#4);
   \draw[lineWithArrowCenter] (#1+1.2*#3,#2-0.3*#4) to [out=225, in=315] (#1,#2);
}

\newcommand{\abubblefullbare}[4]{
   \draw[lineWithArrowCenter] (#1,#2+0.3*#4) to [out=45, in=135] (#1+1.2*#3,#2);
   \draw[lineWithArrowCenter] (#1+1.2*#3,#2) to [out=225, in=315] (#1,#2-0.3*#4);
}

\newcommand{\abubblefullfull}[5]{
   \draw[lineWithArrowCenter] (#1,#2+0.3*#4) to [out=45, in=135] (#1+1.2*#3,#2+0.3*#5);
   \draw[lineWithArrowCenter] (#1+1.2*#3,#2-0.3*#5) to [out=225, in=315] (#1,#2-0.3*#4);
}
\newcommand{\abubblefullfullplain}[5]{
   \draw[linePlain] (#1,#2+0.3*#4) to [out=45, in=135] (#1+1.2*#3,#2+0.3*#5);
   \draw[linePlain] (#1+1.2*#3,#2-0.3*#5) to [out=225, in=315] (#1,#2-0.3*#4);
}

\newcommand{\abubblefullfulldiff}[5]{
   \draw[lineWithArrowCenterCenter] (#1,#2+0.3*#4) to [out=45, in=135] (#1+1.2*#3,#2+0.3*#5);
   \draw[lineWithArrowCenterCenter] (#1+1.2*#3,#2-0.3*#5) to [out=225, in=315] (#1,#2-0.3*#4);
   \draw[linePlain] (#1+0.6*#3-0.03,-0.7*#4) -- (#1+0.6*#3-0.03,+0.7*#4);
   \draw[linePlain] (#1+0.6*#3+0.03,-0.7*#4) -- (#1+0.6*#3+0.03,+0.7*#4);
}

\newcommand{\pbubblebarebarebare}[3]{
   \draw[lineBareWithArrowCenter] (#1+1.2*#3,#2) to [out=45, in=135] (#1,#2);
   \draw[lineBareWithArrowCenter] (#1+1.2*#3,#2) to [out=225, in=315] (#1,#2);
}

\newcommand{\pbubblefullfull}[5]{
   \draw[lineWithArrowCenter] (#1+1.2*#3+0.6*#5,#2+0.3*#5) to [out=45, in=135] (#1-0.6*#4,#2+0.3*#4);
   \draw[lineWithArrowCenter] (#1+1.2*#3,#2-0.3*#5) to [out=225, in=315] (#1,#2-0.3*#4);
}

\newcommand{\tbubblebarebarebare}[3]{
   \draw[lineBareWithArrowCenter] (#1,#2) to [out=225, in=135] (#1,#2-1.2*#3);
   \draw[lineBareWithArrowCenter] (#1,#2-1.2*#3) to [out=45, in=315] (#1,#2);
}

\newcommand{\tbubblefullfull}[5]{
   \draw[lineWithArrowCenter] (#1-0.3*#4,#2) to [out=225, in=135] (#1-0.3*#5,#2-0.9*#3);
   \draw[lineWithArrowCenter] (#1+0.3*#5,#2-0.9*#3) to [out=45, in=315] (#1+0.3*#4,#2);
}

\newcommand{\selfenergy}[4]{
   \def\shift{0.3*#4};
   \coordinate (center) at (#2,#3);
   \draw[linePlain, fill=verylightgray] (center) circle [radius=\shift];
   \node at (center) {#1};
}
\newcommand{\selfenergywithlegs}[4]{
   \selfenergy{#1}{#2}{#3}{#4}
   \draw[lineWithArrowCenterEnd] (#2-0.3*#4,#3) -- (#2-0.3*#4-0.45,#3);
   \draw[lineWithArrowCenterEnd] (#2+0.3*#4+0.45,#3) -- (#2+0.3*#4,#3);
}

\newcommand{\loopfullvertex}[4]{
   \def\shift{0.3*#4};
   \draw[#1] ($(#2,#3) + (\shift,\shift)$) .. controls ++(45:0.4) and ++(0:0.4) .. ($(#2,#3) + (0,0.6+0.3*#4)$) .. controls ++(180:0.4) and ++(135:0.4) .. ($(#2,#3) + (-\shift,\shift)$);
}

\newcommand{\tikzunderbrace}[5]{
   \draw[decoration={brace,mirror,amplitude=5},decorate,thick] (#2,#4) -- (#3,#4);
		\node at (0.5*#3+0.5*#2,#4+#5) {#1};
}

\newcommand{\threepointvertexleft}[4]{
   \draw[linePlain, fill=verylightgray] (#2,#3+0.4*#4) -- (#2+0.6*#4,#3) -- (#2,#3-0.4*#4) -- (#2,#3+0.4*#4);
   \node at (#2+0.2*#4,#3) {#1};
}

\newcommand{\threepointvertexleftarrows}[4]{
   \threepointvertexleft{#1}{#2}{#3}{#4}
   \arrowslefthalffull{#2+0.4*#4}{#3}{4./3.*#4}
}

\newcommand{\threepointvertexright}[4]{
   \draw[linePlain, fill=verylightgray] (#2,#3) -- (#2+0.6*#4,#3+0.4*#4) -- (#2+0.6*#4,#3-0.4*#4) -- (#2,#3);
   \node at (#2+0.4*#4,#3) {#1};
}

\newcommand{\threepointvertexrightarrows}[4]{
   \threepointvertexright{#1}{#2}{#3}{#4}
   \arrowsrighthalffull{#2+0.2*#4}{#3}{4./3.*#4}
}

\newcommand{\threepointvertexupper}[4]{
   \draw[linePlain, fill=verylightgray] (#2-0.4*#4,#3+0.3*#4) -- (#2+0.4*#4,#3+0.3*#4) -- (#2,#3-0.3*#4) -- (#2-0.4*#4,#3+0.3*#4);
   \node at (#2,#3+0.1*#4) {#1};
}

\newcommand{\threepointvertexupperarrows}[4]{
   \threepointvertexupper{#1}{#2}{#3}{#4}
   \arrowsupperhalffull{#2}{#3-0.1*#4}{4./3.*#4}
}

\newcommand{\threepointvertexlower}[4]{
   \draw[linePlain, fill=verylightgray] (#2-0.4*#4,#3-0.3*#4) -- (#2+0.4*#4,#3-0.3*#4) -- (#2,#3+0.3*#4) -- (#2-0.4*#4,#3-0.3*#4);
   \node at (#2,#3-0.1*#4) {#1};
}

\newcommand{\threepointvertexlowerarrows}[4]{
   \threepointvertexlower{#1}{#2}{#3}{#4}
   \arrowslowerhalffull{#2}{#3+0.1*#4}{4./3.*#4}
}

\usepackage{tabularx}                    
\usepackage{multirow, bigdelim}  
\usepackage{longtable}
\usepackage{makecell}                   
\usepackage{cellspace}

\usepackage{mdwlist}      

\usepackage{verbatim}       
\usepackage{bbm}              
\usepackage{dsfont}			
\usepackage{nicefrac}        
\usepackage[vcentermath]{youngtab} 
\usepackage{cancel}

\usepackage{etoolbox} 

\newcolumntype{L}[1]{>{\raggedright\arraybackslash}p{#1}} 
\newcolumntype{C}[1]{>{\centering\arraybackslash}p{#1}} 
\newcolumntype{R}[1]{>{\raggedleft\arraybackslash}p{#1}} 
\definecolor{schwarz}{RGB}{0,0,0}
\definecolor{braun}{RGB}{102,51,0}
\definecolor{blau}{RGB}{0,84,159}
\definecolor{tiefblau}{RGB}{0,0,255}
\definecolor{myblau}{RGB}{57,106,177}
\definecolor{maigruen}{RGB}{189,205,0}
\definecolor{mygruen}{RGB}{62,150,81}
\definecolor{rot}{RGB}{204,7,30}
\definecolor{tiefrot}{RGB}{255,0,0}
\definecolor{myrot}{RGB}{204,37,41}
\definecolor{bordeaux}{RGB}{161,16,53}
\definecolor{violett}{RGB}{97,33,88}
\definecolor{lila}{RGB}{122,111,172}
\definecolor{tieflila}{RGB}{204,0,204}
\definecolor{magenta}{RGB}{255,0,255}
\definecolor{orange}{RGB}{255,100,0}    
\definecolor{gelb}{RGB}{246,168,0}       
\definecolor{gruen}{RGB}{87,171,39}      
\definecolor{petrol}{RGB}{0,97,101}
\definecolor{rot2}{RGB}{205,2,37}        
\definecolor{blau2}{RGB}{0,86,153}      
\definecolor{darkpetrol}{RGB}{0,73,76}

\definecolor{verylightgray}{RGB}{240,240,240}

\definecolor{darkgreen}{rgb}{0,0.5,0}
\definecolor{purple}{rgb}{0.6,0,0.5}
\definecolor{orange}{rgb}{1,0.5,0}
\definecolor{darkred}{rgb}{.7,0,0}
\definecolor{darkblue}{rgb}{0,0,.3}
\definecolor{blue}{rgb}{0,0,1}
\definecolor{grey}{rgb}{.6,.6,.6}
\definecolor{dimgreen}{rgb}{0.2,0.6,0.1}
\hypersetup{colorlinks,linkcolor=blue,urlcolor=blue,citecolor=blue}

\newcommand{\bl}[1]{{\color{blau}{#1}}}

\newcommand{\rt}[1]{{\color{rot}{#1}}}

\usepackage[normalem]{ulem}

\newcommand{\al}[6]{
	^{
	\ifcase #5 
		\alpha_#1 \alpha_#2 
	\or 
		\alpha_#1^\prime \alpha_#2^\prime 
	\or 
		#1 #2
	\or 
		\bar{\alpha}_#1 \alpha_#2
	\or 
		\alpha_#1 \bar{\alpha}_#2
	\or 
		\bar{\alpha}_#1 \bar{\alpha}_#2
	\or 
		\bar{\alpha}_#1^\prime \alpha_#2^\prime
	\or 
		\alpha_#1^\prime \bar{\alpha}_#2^\prime
	\or 
		\bar{\alpha}_#1^\prime \bar{\alpha}_#2^\prime
	\else
		0
	\fi
	|
	\ifcase #6 
		\alpha_#3 \alpha_#4
	\or 
		\alpha_#3^\prime \alpha_#4^\prime
	\or 
		#3 #4
	\or 
		\bar{\alpha}_#3 \alpha_#4
	\or 
		\alpha_#3 \bar{\alpha}_#4
	\or 
		\bar{\alpha}_#3 \bar{\alpha}_#4
	\or 
		\bar{\alpha}_#3^\prime \alpha_#4^\prime
	\or 
		\alpha_#3^\prime \bar{\alpha}_#4^\prime
	\or 
		\bar{\alpha}_#3^\prime \bar{\alpha}_#4^\prime
	\else
		0
	\fi
	}
}
\newcommand{\sigm}[6]{
	_{
	\ifcase #5
		\sigma_#1 \sigma_#2 
	\or
		\sigma_#1^\prime \sigma_#2^\prime 
	\or
		#1 #2
	\else
		0
	\fi
	|
	\ifcase #6
		\sigma_#3 \sigma_#4
	\or
		\sigma_#3^\prime \sigma_#4^\prime
	\or
		#3 #4
	\else
		0
	\fi
	}
}
\newcommand{\sig}[2]{
	{
	\ifcase #1
		\sigma \sigma
	\or
		\sigma \bar{\sigma}
	\or
		\bar{\sigma} \sigma
	\else
		0
	\fi
	|
	\ifcase #2
		\sigma \sigma
	\or
		\sigma \bar{\sigma}
	\or
		\bar{\sigma} \sigma
	\else
		0
	\fi
	}
}
\newcommand{\q}[6]{
	\ifcase #5
		q_#1 q_#2 
	\or
		q_#1^\prime q_#2^\prime 
	\else
		0
	\fi
	|
	\ifcase #6
		q_#3 q_#4
	\or
		q_#3^\prime q_#4^\prime
	\else
		0
	\fi
}
\newcommand{\om}[6]{
	\ifcase #5
		\nu_#1 \nu_#2 
	\or
		\nu_#1^\prime \nu_#2^\prime 
	\else
		0
	\fi
	|
	\ifcase #6
		\nu_#3 \nu_#4
	\or
		\nu_#3^\prime \nu_#4^\prime
	\else
		0
	\fi
}

\newcommand{\sss}{{\sigma \sigma}}
\newcommand{\ssb}{{\sigma \bar{\sigma}}}

\newcommand{\Ktot}[1]{\mathcal{K}_{#1}}
\newcommand{\dKtot}[1]{\dot{\mathcal{K}}_{#1}}
\newcommand{\K}[2]{\mathcal{K}_{#1}^{#2}}
\newcommand{\dK}[2]{\dot{\mathcal{K}}_{#1}^{#2}}
\newcommand{\Kb}[2]{\mathcal{K}_{#1^\prime}^{#2}}
\newcommand{\dKb}[2]{\dot{\mathcal{K}}_{#1^\prime}^{#2}}
\newcommand{\Gammar}{\Gamma^r_2}
\newcommand{\Gammabr}{\Gamma^r_{2'}}
\newcommand{\bGammar}{\bar{\Gamma}^r_2}
\newcommand{\bGammabr}{\bar{\Gamma}^r_{2'}}

\newcommand{\fcirc}{\mathbin{\vcenter{\hbox{\scalebox{0.65}{$\bullet$}}}}} 
\newcommand{\doubleI}{\mathds{1}}     
\newcommand{\boldI}{\mathbf{1}}       
\newcommand{\pprime}{{\phantom{\prime}}}  

\newcommand{\Pir}{\mbox{$\Pi$-$r$}}
\newcommand{\Pia}{\mbox{$\Pi$-$a$}}

\newcommand{\Ur}{\mbox{$U$-$r$}}
\newcommand{\Ua}{\mbox{$U$-$a$}}
\newcommand{\Up}{\mbox{$U$-$p$}}
\newcommand{\Ut}{\mbox{$U$-$t$}}

\newcommand{\updown}{^{\uparrow\downarrow}}

\newcommand{\downup}{^{\downarrow\uparrow}}
\newcommand{\upup}{^{\uparrow\uparrow}}
\newcommand{\downdown}{^{\downarrow\downarrow}}

\renewcommand{\wr}{\omega_r}
\newcommand{\wa}{\omega_a}
\renewcommand{\wp}{\omega_p}
\newcommand{\wt}{\omega_t}

\newcommand{\vr}{\nu_r}
\newcommand{\va}{\nu_a}
\newcommand{\vp}{\nu_p}
\newcommand{\vt}{\nu_t}

\newcommand{\wrt}{w.r.t.\ }
\newcommand{\sutwo}{\text{SU}(2)} 
\newcommand{\Sec}[1]{Sec.~\ref{#1}}
\newcommand{\App}[1]{App.~\ref{#1}}
\newcommand{\Eq}[1]{Eq.~\eqref{#1}}
\newcommand{\Eqs}[1]{Eqs.~\eqref{#1}}
\newcommand{\Fig}[1]{Fig.~\ref{#1}}

\definecolor{darkgreen}{rgb}{0,0.5,0}
\definecolor{orange}{rgb}{1,0.5,.3}
\definecolor{darkred}{rgb}{.7,0,0}
\definecolor{purple}{rgb}{0.6,0,0.5}
\definecolor{darkpetrol}{RGB}{0,73,76}

\usepackage[normalem]{ulem} 
\definecolor{darkgreen}{rgb}{0,0.5,0}
\definecolor{purple}{rgb}{0.6,0,0.5}
\definecolor{orange}{rgb}{1,0.5,0}
\definecolor{darkred}{rgb}{.7,0,0}
\definecolor{darkblue}{rgb}{0,0,.6}
\definecolor{grey}{rgb}{.6,.6,.6}
\definecolor{dimgreen}{rgb}{0.2,0.6,0.1}

\newcommand{\expval}[1]{\langle #1 \rangle}

\newcommand\blfootnote[1]{%
  \begingroup
  \renewcommand\thefootnote{}\footnote{#1}%
  \addtocounter{footnote}{-1}%
  \endgroup
}

\begin{document}
	\sloppy 

\title{Multiloop flow equations for single-boson exchange fRG}

\author{
	Marcel Gievers\inst{1,2,*}
	\and
	Elias Walter\inst{1,*}
	\and
	Anxiang Ge\inst{1,*}
	\and
	Jan von Delft\inst{1}
	\and
	Fabian B. Kugler\inst{3}
}

\institute{
	Arnold Sommerfeld Center for Theoretical Physics, 
	Center for NanoScience,\looseness=-1\,  and Munich 
	Center for Quantum Science and Technology,\looseness=-2\, Ludwig-Maximilians-Universität München, 80333 Munich, Germany
	\and
	Max Planck Institute  of  Quantum  Optics,  Hans-Kopfermann-Straße  1,  85748  Garching,  Germany
	\and
	Department of Physics and Astronomy, Rutgers University, Piscataway, New Jersey 08854, USA
}

\date{May 11, 2022}

\abstract{
The recently introduced single-boson exchange (SBE) decomposition of the four-point vertex of interacting fermionic many-body systems is a conceptually and computationally appealing parametrization of the vertex. It relies on the notion of reducibility of vertex diagrams with respect to the bare interaction $U$, instead of a classification based on two-particle reducibility within the widely-used parquet decomposition. 
Here, we re-derive the SBE decomposition in a generalized framework (suitable for extensions to, e.g., inhomogeneous systems or real-frequency treatments) following from the parquet equations.
We then derive multiloop functional renormalization group (mfRG) flow equations for the ingredients of this SBE decomposition,
both in the parquet approximation, where the fully two-particle irreducible vertex is treated as an input,
and in the more restrictive SBE approximation, where this role is taken by the fully $U$-irreducible vertex.
Moreover, we give mfRG flow equations for the popular parametrization of the vertex in terms of asymptotic classes of the two-particle reducible vertices. Since the parquet and SBE decompositions are closely related, their mfRG flow equations are very similar in structure.
}

\maketitle

\section{Introduction}
\label{sec:Introduction}

\blfootnote{$^*$These authors contributed equally to this work.}

The understanding of strongly correlated many-body systems like the two-dimensional Hubbard model remains an important challenge of contemporary condensed-matter physics \cite{Schaefer2021}. For this, it is desirable to gain profound understanding of two-body interactions which are described by the full four-point vertex $\Gamma$.

A powerful technique for calculating the four-point vertex $\Gamma$ is the functional renormalization group (fRG) \cite{Metzner2012,Kopietz2010}. There, a scale parameter $\Lambda$ is introduced into the bare Green's function $G_0\to G_0^\Lambda$ in such a way that for an initial value $\Lambda\to\Lambda_i$ the theory (specifically, the calculation of the self-energy $\Sigma^\Lambda$ and the four-point vertex $\Gamma^\Lambda$) becomes solvable, and after successively integrating out higher-energy modes $\Lambda\to\Lambda_f$, the fully renormalized objects $\Sigma$ and $\Gamma$ are obtained.

Traditionally, fRG is formulated as an infinite hierarchy of exact flow equations for $n$-point vertex functions.
However, since already the six-point vertex is numerically intractable, truncations are needed. 
A frequently-used strategy employs a one-loop ($1\ell$) truncation of the exact hierarchy of flow equations by 
completely neglecting six-point and higher vertices. This can be justified, e.g., from a perturbative \cite{Metzner2012} or leading-log 
\cite{Diekmann2021} perspective.
Another truncation scheme is given by the multiloop fRG approach, mfRG, which includes all contributions of the six-point vertex to the flow
of the four-point vertex and self-energy that can be computed
with numerical costs proportional to the $1\ell$ flow \cite{Kugler2018a,Kugler2018b,Kugler2018e}. In doing so, it sums up all parquet diagrams, formally reconstructing the parquet approximation (PA) \cite{Roulet1969,Bickers2004} if loop convergence is achieved. 
Converged multiloop results thus inherit all the properties of the PA. These include self-consistency at the one- and two-particle level (in
that the PA is a solution of the self-consistent parquet equations \cite{Bickers2004}); the validity of one-particle conservation
laws (but not of two-particle ones); and the independence of the final results on the choice of regulator (since the parquet equations and PA do not involve specifying any regulator).
The mfRG approach was recently applied to the Hubbard model \cite{Tagliavini2019,Hille2020}, Heisenberg models \cite{Thoenniss2020,Kiese2020}, and the Anderson impurity model \cite{Chalupa2021}.

A full treatment of the frequency and momentum dependence of the four-point vertex generally requires tremendous numerical resources. 
Hence, it is important to parametrize these dependencies in an efficient  way, to reduce computational effort without losing  information on important physical properties.
One such scheme expresses the vertex as a sum of diagrammatic classes distinguished by their asymptotic frequency behavior \cite{Li2016,Wentzell2020}: Asymptotic classes which remain nonzero when one or two frequency arguments are sent to infinity do not depend on these arguments, while the class depending on all three frequency arguments decays in each direction.

A related strategy is to express parts of the vertex through fermion bilinears that interact via exchange bosons \cite{Husemann2009,Husemann2012}. Partial bosonization schemes, which approximate the vertex through one \cite{Stepanov2016b,Stepanov2018,Stepanov2019a} or several boson-exchange channels \cite{Stepanov2019,Harkov2021a,Stepanov2021}, have been employed within the dual boson formalism, used in diagrammatic extensions of dynamical mean field theory (DMFT) aiming to include nonlocal correlations.

A decomposition of the full vertex into single-boson exchange (SBE) parts, involving functions of at most two frequencies, and residual parts depending on three frequencies was developed in Refs.~\cite{Krien2019a,Krien2019,Krien2019b,Krien2020a,Krien2020b,Krien2021}. 
The guiding principle of the SBE decomposition is reducibility in the bare interaction $U$ \cite{Krien2019a}. This criterion distinguishes SBE contributions, that are $U$-reducible, from multi-boson exchange and other contributions, that are not. 
The SBE \textsl{approximation} retains only the $U$-reducible part while neglecting all $U$-irreducible terms \cite{Krien2019}.
The SBE terms are expressible through bosonic fluctuations and their (Yukawa) couplings to fermions---the Hedin vertices---and thus have a transparent physical interpretation. Numerically, two- and three-point objects can be computed and stored more easily than a genuine four-point vertex.

Studies of the two-dimensional Hubbard model have shown that the SBE decomposition is a promising technique for computing the frequency and momentum dependences of the vertex \cite{Krien2020a,Krien2020b,Krien2021}. 
In a $1\ell$ fRG calculation, it was found that some of its essential features are already captured by its $U$-reducible parts, which are much easier to compute numerically than the $U$-irreducible ones \cite{Bonetti2021}.
Reference~\cite{Bonetti2021} also obtained results at strong interaction using DMF$^2$RG, a method that makes use of a DMFT vertex as the starting point for the fRG flow \cite{Taranto2014,Vilardi2019,Katanin2019}. 
Here, a very interesting aspect of the SBE decomposition is that the SBE approximation 
(neglecting $U$-irreducible contributions)
remains a meaningful approximation also in the strong-coupling regime \cite{Harkov2021}, which is not the case for a similar approximation scheme based on the parametrization through asymptotic classes while using functions of at most two frequency arguments. 

Given these encouraging developments, it is of interest to have a strategy for computing the ingredients of the SBE approach---the bosonic propagators, the Hedin vertices, 
and the remaining $U$-irreducible terms%
---not only in $1\ell$ fRG \cite{Bonetti2021} but also in mfRG.
In this paper, we therefore derive multiloop flow equations for the SBE ingredients. To this end, we start from the parquet equations to derive a general form of the SBE decomposition where the structure of non-frequency arguments is not specified. We then derive multiloop flow equations for the SBE ingredients, and finally illustrate the relation of these objects to the parametrization of the vertex in terms of two-particle reducible asymptotic classes \cite{Wentzell2020,Bonetti2021}.
The numerical implementation of the resulting SBE multiloop flow equations goes beyond the scope of this purely analytical paper and is left for the future.

The paper is organized as follows: In \Sec{sec:mfRG}, we recapitulate the parquet equations, the corresponding mfRG flow equations, and the frequency parametrization of the four-point vertex adapted to each two-particle channel. 
In \Sec{sec:SBE-decomposition}, we deduce the SBE decomposition from the parquet equations and derive multiloop flow equations for the SBE ingredients in two different ways. We also discuss the SBE approximation and its associated mfRG flow.
In \Sec{sec:AsymptoticClasses}, we recall the definition of the asymptotic vertex classes and derive multiloop equations for these. We outline the relation between SBE ingredients and asymptotic classes and their respective mfRG equations. We conclude with a short outlook in \Sec{sec:Conclusion}. 
Appendices~\ref{Appendix:Diagrams_for_SBE} and \ref{Appendix:Diagrams_for_Ki} illustrate the SBE ingredients and asymptotic vertex classes diagrammatically, 
while App.~\ref{Appendix:physical_diagrammatic} describes the relation between our generalized notation of the SBE decomposition to that of the original papers.
Finally, Apps.~\ref{app:3-point-correlators} and \ref{Appendix:susceptibilities} give details on different definitions of correlators and susceptibilities and show their close relation to the SBE ingredients.

\section{Recap of parquet and mfRG equations}
\label{sec:mfRG}

The parquet equations and the associated multiloop fRG equations form the basis for the main outcomes of this paper.
For ease of reference and use in future sections, we recapitulate the notational conventions and compactly summarize the main ingredients and results of the mfRG approach \cite{Kugler2018a,Kugler2018b,Kugler2018e}.
To make the presentation self-contained, we also recall from the literature
the motivation for some of the definitions and conventions presented below.

\subsection{Parquet equations}
\label{sec:Parquet}

The action of a typical fermionic model reads
\begin{align}
S =&
 -  \bar c_{1'} [G_0^{-1}]_{1'|1} c_{1} -  \tfrac{1}{4} U_{ 1'2'|12} \, \bar c_{1'}\bar c_{2'} c_{2} c_{1},
\label{eq:Fermionic-action}
\end{align}
with the bare propagator $G_0$.
The Grassmann fields $c_i$ 
are labeled by a composite index $i$ describing
frequency and other quantum numbers, such as position or momentum, spin, etc. 
Throughout this paper, repeated $i$-indices are 
understood to be integrated over or summed over.
Furthermore, $U$ is the crossing symmetric bare interaction vertex, $U_{ 1'2'|12} = -U_{ 2'1'|12}$
(called $\Gamma_0$ in Refs.~\cite{Kugler2018b,Kugler2018e}).
We assume it to be energy-conserving without further frequency dependence, as in any action derived directly from a 
time-independent Hamiltonian. 
Our expression for the action \eqref{eq:Fermionic-action} and later definitions of correlation functions are given in the Matsubara formalism \cite{altland2010condensed} and for fermionic fields. 
However, our analysis can easily be transcribed to the Keldysh formalism \cite{kamenev2011field}, and/or to bosonic fields, by suitably adapting the content of the index $i$ on $c_i$ and adjusting some prefactors. 
Such changes do not modify the structure of the vertex decomposition and flow equations that are the focus of this paper.

The time-ordered one- and two-particle correlators, $G_{1|1'} = -\expval{c_1 \bar c_{1'}}$ and $G^{(4)}_{12|1'2'} = \expval{c_1 c_2 \bar c_{2'} \bar c_{1'}}$,
can be expressed in standard fashion \cite{Kopietz2010} through the self-energy and the four-point vertex,
\begin{align}
\label{eq:DefineSigmaGamma}
    \Sigma_{1'|1}
    =
    \tikzm{Definitions-Sigma}{
        \selfenergywithlegs{$\Sigma$}{0}{0}{1}
        \node at (-0.75,0.3) {$1'$};
		\node at (0.75,0.3) {$1\phantom{'}$};
    }
    ,\quad
    \Gamma_{1'2'|12}
    =
    \tikzm{Definitions-Gamma}{
        \fullvertexwithlegs{$\Gamma$}{0}{0}{1}
		\node[left] at (-0.6,0.6) {$2\phantom{'}$};
		\node[right] at (0.6,0.6) {$2'$};
		\node[left] at (-0.6,-0.6) {$1'$};
		\node[right] at (0.6,-0.6) {$1\phantom{'}$};
    }
    .
\end{align}
These contain all one-particle irreducible one- and two-particle vertex diagrams, respectively. Hence, these are (amputated connected) diagrams that cannot be split into two pieces by cutting a single bare propagator line.

The one-particle self-energy is related to the two-particle vertex via the Schwinger--Dyson equation (SDE) \cite{Bickers2004}. 
We do not discuss this equation much further because its treatment is similar for both vertex decompositions discussed below.
On the two-particle level, the starting point of parquet approaches \cite{Bickers2004} is the parquet decomposition,
\begin{align}
\Gamma = R + \gamma_a + \gamma_p + \gamma_t .
\label{eq:mfRG:mfRG:Parquet:PE}
\end{align}
It states that the set of all vertex diagrams can be divided into four disjoint classes: the diagrams in $\gamma_r$, $r = a,p,t$, are two-particle reducible in channel $r$, i.e.,
they can be split into two parts by cutting two antiparallel ($a$), parallel ($p$), or transverse antiparallel ($t$) propagator lines, respectively. The diagrams in $R$ do not fall apart by cutting two propagator lines and are thus fully two-particle irreducible.
This classification is exact and unambiguous \cite{Rohringer2012,Wentzell2020}. 
In the literature, the diagrammatic channels are also known as crossed particle-hole ($\overline{\text{ph}}\leftrightarrow a$), particle-particle (pp $\leftrightarrow p$), and particle-hole (ph $\leftrightarrow t$) channel. 

\begin{figure}
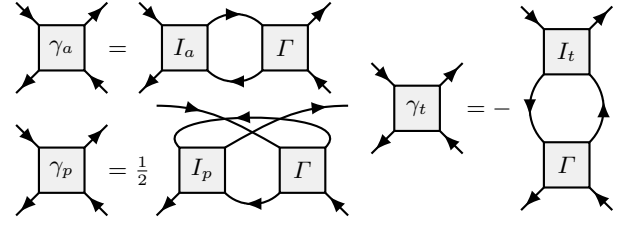

\begin{align*}
	\begin{array}{rl}
		\begin{array}{rl}
			\tikzm{mfRG-parquet-BSE_a}{ \fullvertexwithlegs{$\gamma_a$}{0}{0}{1} }
			\!\!\!
			&=
			\phantom{\frac{1}{2} \!\!\!}
			\tikzm{mfRG-parquet-BSE_a1}{
				\arrowslefthalffull{0}{0}{1}
				\fullvertex{$I_a$}{0}{0}{1}
				\draw[linePlain] (0.3,0.3) to [out=45, in=135] (1.02,0.3);
				\draw[lineWithArrowCenter] (0.759,0.445) -- (0.761,0.445);
				\draw[linePlain] (1.02,-0.3) to [out=225, in=315] (0.3,-0.3);
				\draw[lineWithArrowCenter] (0.561,-0.445) -- (0.559,-0.445);
				\fullvertex{$\Gamma$}{1.32}{0}{1}
				\arrowsrighthalffull{1.32}{0}{1}
			}
			\\
			\tikzm{mfRG-parquet-BSE_p}{ \fullvertexwithlegs{$\gamma_p$}{0}{0}{1} }
			\!\!\!
			&=
			\frac{1}{2} \!\!\!
			\tikzm{mfRG-parquet-BSE_p1}{
				\arrowslefthalffullp{0}{0}{0.6}{1}{1}
				\fullvertex{$I_p$}{0}{0}{1}
				\draw[lineWithArrowCenter] (1.62,0.3) to [out=45, in=135] (-0.3,0.3);
				\draw[linePlain] (1.02,-0.3) to [out=225, in=315] (0.3,-0.3);
				\draw[lineWithArrowCenter] (0.561,-0.445) -- (0.559,-0.445);
				\fullvertex{$\Gamma$}{1.32}{0}{1}
				\arrowsrighthalffullp{1.32}{0}{0.6}{1}{1}
				\node at (0.66,0.9) {};
			}
		\end{array}
		&
		\!\!
		\tikzm{mfRG-parquet-BSE_t}{ \fullvertexwithlegs{$\gamma_t$}{0}{0}{1} }
		\!
		=
		- \
		\tikzm{mfRG-parquet-BSE_t1}{
			\arrowsupperhalffull{0}{0.75}{1}
			\fullvertex{$I_t$}{0}{0.75}{1}
			\draw[linePlain] (-0.3,0.45) to [out=225, in=135] (-0.3,-0.45);
			\draw[lineWithArrowCenter] (-0.485,-0.099) -- (-0.485,-0.101);
			\draw[linePlain] (0.3,-0.45) to [out=45, in=315] (0.3,0.45);
			\draw[lineWithArrowCenter] (0.485,0.099) -- (0.485,0.101);
			\fullvertex{$\Gamma$}{0}{-0.75}{1}
			\arrowslowerhalffull{0}{-0.75}{1}
		}
	\end{array}
\end{align*}
\caption{Bethe--Salpeter equations in the antiparallel ($a$), 
parallel ($p$) and transverse ($t$) channels.
\label{fig:mfRG:mfRG:Parquet:BSE}
}
\end{figure}

Since the four classes in the parquet decomposition are disjoint, one can decompose $\Gamma$ \wrt its two-particle reducibility in one of the channels $r$, $\Gamma = I_r + \gamma_r$. Here, $I_r$ comprises the sum of all diagrams irreducible in channel $r$ and fulfills $I_r = R + \gamma_{\bar{r}}$ with $\gamma_{\bar{r}} = \sum_{r^\prime \neq r}\gamma_{r^\prime}$. The Bethe--Salpeter equations (BSEs) relate the reducible diagrams to the irreducible ones and can be summarized by 
\begin{align}
\gamma_r = I_r \circ \Pi_r \circ \Gamma = \Gamma \circ \Pi_r \circ I_r.
\label{eq:Bethe-Salpeter-equations}
\end{align}
The $\Pi_r $ \textsl{bubble}, defined as 
\begin{subequations}
\begin{align}
	\Pi_{a;34|3'4'} &= \phantom{-} G_{3|3'}G_{4|4'} ,
	\\
	\Pi_{p;34|3'4'} &= \tfrac{1}{2} G_{3|3'}G_{4|4'} ,
	\label{eq:Definition_Pi_p}
	\\
	\Pi_{t;43|3'4'} &=  - G_{3|3'}G_{4|4'} ,
\end{align}
\end{subequations}
represents the corresponding propagator pair in channel $r$, see \Fig{fig:mfRG:mfRG:Parquet:BSE}.
(Note that $\Pi_{a;34|3'4'} = -\Pi_{t;43|3'4'}$ is consistent with crossing symmetry.) The connector symbol $\circ$ denotes summation over internal frequencies and quantum numbers ($5,6$ in \Eq{eq:Bubble_summation} below) and its definition depends on the channel $r\in\{a,p,t\}$: When connecting $\Pi_r$ (or other four-leg objects labeled by $r$) to some vertex, it gives 
\begin{subequations}
	\label{eq:Bubble_summation}
	\begin{align}
	a:
	\quad
	[A \circ B]_{12|34}
	&=
	A_{ 1\textcolor{black}6|\textcolor{black}54} B_{\textcolor{black}52|3\textcolor{black}6},
	\\
	p:
	\quad
	[A \circ B]_{12|34}
	&=
	A_{ 12|\textcolor{black}5\textcolor{black}6} B_{\textcolor{black}5\textcolor{black}6|34},
	\\
	t:
	\quad
	[A \circ B]_{12|34}
	&=
	A_{ \textcolor{black}62 | \textcolor{black}54} B_{1\textcolor{black}5|3\textcolor{black}6}.
	\end{align}
\end{subequations}
By combining $\Gamma = I_r + \gamma_r$ with the BSEs \eqref{eq:Bethe-Salpeter-equations}, one can eliminate $\gamma_r$ to get the 
``extended BSEs'' \cite{Kugler2018e} needed later:
\begin{subequations}
	\label{eq:extended-BSE}	
	\begin{align}
		\doubleI_r + \Pi_r\circ\Gamma &= (\doubleI_r-\Pi_r\circ I_r)^{-1}, 
	\\
		\doubleI_r + \Gamma\circ\Pi_r &= (\doubleI_r-I_r\circ\Pi_r)^{-1}. 
	\end{align}
\end{subequations}
Here, the channel-specific unit vertices $\doubleI_r$, defined by the requirement $\Gamma = \doubleI_r\circ \Gamma = \Gamma \circ\doubleI_r$, are given by 
\begin{subequations}
	\begin{align}
		\doubleI_{a;12|34} &= \delta_{13}\delta_{24},\\
		\doubleI_{p;12|34} &= \tfrac{1}{2}(\delta_{13}\delta_{24}-\delta_{14}\delta_{23}),\\ 
		\doubleI_{t;12|34} &= \delta_{14}\delta_{23}.
		\label{eq:1_r}
	\end{align}
\end{subequations}
(For the $p$ channel,  the internal sum in
$\doubleI_p\circ \Gamma = \Gamma \circ\doubleI_p$ runs over both outgoing (or ingoing) legs
of $\Gamma$. Therefore, the crossing symmetry of the vertex, i.e., $\Gamma_{12|34}=-\Gamma_{21|34}=-\Gamma_{12|43}$, is transferred to $\doubleI_p$,  resulting in an expression more involved than for the other two channels.)

The combination of the Dyson equation 
$G = G_0(1 + \Sigma G)$, the 
SDE, the parquet decomposition \eqref{eq:mfRG:mfRG:Parquet:PE}, the three BSEs \eqref{eq:Bethe-Salpeter-equations}, and the definitions $I_r = \Gamma-\gamma_r$ constitutes the self-consistent \textsl{parquet equations}. 
The only truly independent object is the fully irreducible vertex $R$. 
If $R$ is specified, everything else can be computed self-consistently via the parquet equations. However, $R$ is the most complicated object: 
its diagrams contain several nested integrals/sums over internal arguments, whereas the integrals in reducible diagrams partially factorize. 
A common simplification, the \textsl{parquet approximation} (PA), replaces $R$ by $U$, closing the set of parquet equations.

\subsection{Parquet mfRG}
\label{sec:mfRG-equations}

The conventional mfRG flow equations can be derived from the parquet equations by introducing a regulator $\Lambda$ into the bare propagator $G_0$, thus making all objects in the parquet equations $\Lambda$-dependent \cite{Kugler2018e}.
The fully irreducible vertex $R$ is treated as an input and is thus assumed to be $\Lambda$-independent, $R^\Lambda \approx R$. 
For instance, this assumption arises both in the PA where $R \approx U$ or in the dynamical vertex approximation D$\Gamma$A \cite{Toschi2007,Held2008} where $R \approx R^{\mathrm{DMFT}}$ is taken from DMFT%
---here, we will not distinguish these cases explicitly.
Taking the derivative of the SDE
and the BSEs \wrt $\Lambda$ then yields flow equations for $\Sigma$ and $\Gamma$. Within the context of this paper, we will call this mfRG approach \textsl{parquet mfRG}, in order to distinguish it from an \textsl{SBE mfRG} approach to be discussed in Sec.~\ref{sec:mfRG_equations_SBE}. 

\begin{figure}
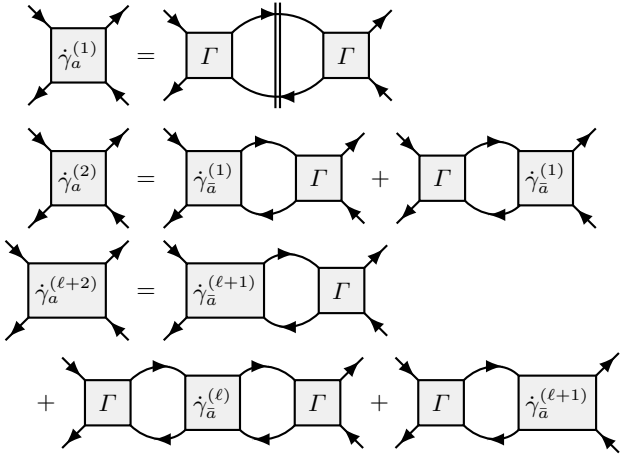

\begin{align*} 
	\tikzm{mfRG-ml-dgamma1}{
		\fullvertexwithlegs{$\dot{\gamma}_a^{(1)}$}{0}{0}{1.2}
	}
	&=
	\tikzm{mfRG-ml-dgamma1_rhs}{
		\arrowslefthalffull{0}{0}{1}
		\fullvertex{$\Gamma$}{0}{0}{1}
		\abubblefullfulldiff{0.3}{0}{1}{1}{1}
		\fullvertex{$\Gamma$}{1.8}{0}{1}
		\arrowsrighthalffull{1.8}{0}{1}
	}
	\\[1mm] 
	\tikzm{mfRG-ml-dgamma2}{
		\fullvertexwithlegs{$\dot{\gamma}_a^{(2)}$}{0}{0}{1.2}
	}
	&= 
	\tikzm{mfRG-ml-dgamma2_rhs_L}{
		\arrowslefthalffull{0}{0}{1.2}
		\fullvertex{$\dot{\gamma}_{\bar{a}}^{(1)}$}{0}{0}{1.2}
		\abubblefullfullplain{0.36}{0}{0.6}{1.2}{1}
		\draw[lineWithArrowCenter] (0.759,0.485) -- (0.761,0.485);
		\draw[lineWithArrowCenter] (0.561,-0.485) -- (0.559,-0.485);
		\fullvertex{$\Gamma$}{1.38}{0}{1}
		\arrowsrighthalffull{1.38}{0}{1}
	}
	+
	\tikzm{mfRG-ml-dgamma2_rhs_R}{
		\arrowslefthalffull{0}{0}{1}
		\fullvertex{$\Gamma$}{0}{0}{1}
		\abubblefullfullplain{0.3}{0}{0.6}{1}{1.2}
		\draw[lineWithArrowCenter] (0.809,0.485) -- (0.811,0.485);
		\draw[lineWithArrowCenter] (0.611,-0.485) -- (0.609,-0.485);
		\fullvertex{$\dot{\gamma}_{\bar{a}}^{(1)}$}{1.36}{0}{1.2}
		\arrowsrighthalffull{1.36}{0}{1.2}
	}
	\\
	\tikzm{mfRG-ml-dgamma3}{
		\arrowslefthalffull{0}{0}{1.2}
		\fullvertexwide{$\dot{\gamma}_a^{(\ell+2)}$}{0}{0}{1.2}{0.3}
		\arrowsrighthalffull{0.3}{0}{1.2}
	}
	&=
	\tikzm{mfRG-ml-dgamma3_rhs_L}{
		\arrowslefthalffull{0}{0}{1.2}
		\fullvertexwide{$\dot{\gamma}_{\bar{a}}^{(\ell+1)}$}{0}{0}{1.2}{0.3}
		\abubblefullfullplain{0.66}{0}{0.6}{1.2}{1}
		\draw[lineWithArrowCenter] (1.059,0.485) -- (1.061,0.485);
		\draw[lineWithArrowCenter] (0.861,-0.485) -- (0.859,-0.485);
		\fullvertex{$\Gamma$}{1.68}{0}{1}
		\arrowsrighthalffull{1.68}{0}{1}
	}
	\\ \nonumber
	& \hspace{-1.3cm} +
	\tikzm{mfRG-ml-dgamma3_rhs_C}{
		\arrowslefthalffull{0}{0}{1}
		\fullvertex{$\Gamma$}{0}{0}{1}
		\abubblefullfullplain{0.3}{0}{0.6}{1}{1.2}
		\draw[lineWithArrowCenter] (0.809,0.485) -- (0.811,0.485);
		\draw[lineWithArrowCenter] (0.611,-0.485) -- (0.609,-0.485);
		\fullvertex{$\dot{\gamma}_{\bar{a}}^{(\ell)}$}{1.38}{0}{1.2}
		\abubblefullfullplain{1.74}{0}{0.6}{1.2}{1}
		\draw[lineWithArrowCenter] (2.109,0.485) -- (2.111,0.485);
		\draw[lineWithArrowCenter] (1.911,-0.485) -- (1.909,-0.485);;
		\fullvertex{$\Gamma$}{2.76}{0}{1}
		\arrowsrighthalffull{2.76}{0}{1}
	}
	+
	\tikzm{mfRG-ml-dgamma3_rhs_R}{
		\arrowslefthalffull{0}{0}{1}
		\fullvertex{$\Gamma$}{0}{0}{1}
		\abubblefullfullplain{0.3}{0}{0.6}{1}{1.2}
		\draw[lineWithArrowCenter] (0.809,0.485) -- (0.811,0.485);
		\draw[lineWithArrowCenter] (0.611,-0.485) -- (0.609,-0.485);
		\fullvertexwide{$\dot{\gamma}_{\bar{a}}^{(\ell+1)}$}{1.38}{0}{1.2}{0.3}
		\arrowsrighthalffull{1.68}{0}{1.2}
	}
\end{align*}
\caption{Diagrammatic depiction of the mfRG flow equations \eqref{eq:mfRG-equations_gamma_r} in the $a$ channel. The double-dashed bubble $\dot\Pi_a$ represents a sum of two terms, $G \dot{G} + \dot{G} G$, where double-dashed propagators $\dot{G}$ are fully differentiated ones (cf.~\Eq{eq:mfRG:mfRG:dLambda_G}).}
\label{fig:mfRG_equations_a_channel}
\end{figure}

When computing $\dot\gamma_r = \partial_\Lambda \gamma_r$ 
via the BSEs, one obtains terms including $\dot{I}_r = \sum_{r^\prime \neq r} \dot\gamma_{r^\prime}$. Thus, one has to iteratively insert the flow equation for $\gamma_r$ into the equations of the other channels $r'\neq r$, yielding an infinite set of contributions of increasing \textsl{loop order}:
\begin{align}
\label{eq:loop-expansions}
    \dot{\Gamma} = \dot{\gamma}_a + \dot{\gamma}_p + \dot{\gamma}_t ,
    \qquad
    \dot{\gamma}_r = \sum_{\ell=1}^\infty \dot{\gamma}_r^{(\ell)} .
\end{align}
The individual $\ell$-loop contributions read \cite{Kugler2018a,Kugler2018e}
\begin{subequations}
	\label{eq:mfRG-equations_gamma_r}
	\begin{align}
		\label{eq:mfRG-equations_gamma_r_1l}
		\dot\gamma_r^{(1)} &= \Gamma \circ \dot\Pi_r \circ \Gamma 		
		,
		\\
		\label{eq:mfRG-equations_gamma_r_2l}
		\dot\gamma_r^{(2)} &= \dot\gamma_{\bar r}^{(1)} \circ \Pi_r \circ \Gamma + \Gamma \circ \Pi_r \circ \dot\gamma_{\bar r}^{(1)} 		
		\\
		\nonumber\dot\gamma_r^{(\ell+2)} &= \dot\gamma_{\bar r}^{(\ell+1)} \circ \Pi_r \circ \Gamma + \Gamma \circ \Pi_r \circ \dot\gamma_{\bar r}^{(\ell)} \circ \Pi_r \circ \Gamma
		\\
		&
		\label{eq:mfRG-equations_gamma_r_l}
		\quad + \Gamma \circ \Pi_r \circ \dot\gamma_{\bar r}^{(\ell+1)}.
	\end{align}
\end{subequations}
where $\dot\gamma_{\bar{r}}^{(\ell)} = \sum_{r'\neq r} \! \dot\gamma_{r'}^{(\ell)}$ 
and \Eq{eq:mfRG-equations_gamma_r_l} applies for $\ell+2 \geq 3$.
In general, all terms at loop order $\ell$ contain 
$\ell-1$ factors of $\Pi$ and one $\dot \Pi$ (i.e., $\ell$ loops,
one of which is differentiated), connecting $\ell$ renormalized vertices $\Gamma$. We have
$\dot\Pi_r \sim G \dot{G} + \dot{G} G$, where 
\begin{align}
	\dot{G} = S + G \,\dot{\Sigma}\, G ,
	\label{eq:mfRG:mfRG:dLambda_G}
\end{align}
with the single-scale propagator $S = \dot{G} |_{\Sigma=\text{const}}$.
Figure~\ref{fig:mfRG_equations_a_channel} illustrates \Eqs{eq:mfRG-equations_gamma_r} diagrammatically in the $a$ channel.

The flow equation for the self-energy, derived in Ref.~\cite{Kugler2018e}
by requiring $\Sigma$ to satisfy the
SDE throughout the flow, reads 
\begin{align}
	\tikzm{mfRG-ml-dSigma}{
		\selfenergywithlegs{$\dot\Sigma$}{0}{0}{1.2}
	}
	\
	&= \
	- \
	\tikzm{mfRG-ml-dSigma_std}{
		\fullvertex{$\Gamma$}{0}{0}{1.2}
		\arrowslowerhalffull{0}{0}{1.2}
		\loopfullvertex{lineWithArrowCenterCenter}{0}{0}{1.2}
		\draw[linePlain] (0,0.76) -- (0,1.16);
	}
	\
	\underbrace{
		- \
		\tikzm{mfRG-ml-dSigma_tbar}{
			\fullvertex{$\dot\gamma_{\bar{t},C}$}{0}{0}{1.2}
			\arrowslowerhalffull{0}{0}{1.2}
			\loopfullvertex{lineWithArrowCenterCenter}{0}{0}{1.2}
			\node at (0,1.01) {};
		}
	}_{\dot\Sigma_{\bar{t}}}
	\
	\underbrace{
		- \
		\tikzm{mfRG-ml-dSigma_t}{
			\fullvertex{$\Gamma$}{0}{0}{1.2}
			\arrowslowerhalffull{0}{0}{1.2}
			\selfenergy{$\dot\Sigma_{\bar{t}}$}{0}{1}{1.2}
			\draw[lineWithArrowCenter] (0.36,0.36) to [out=45, in=270] (0.55,0.65) to [out=90, in=0] (0.36,1);
			\draw[lineWithArrowCenter] (-0.36,1) to [out=180, in=90] (-0.55,0.65) to [out=270, in=135] (-0.36,0.36);
		}
	}_{\dot\Sigma_t}
	\,.
	\label{eq:mfRG:mfRG:Selfenergy_flow}
\end{align}
It has $\Gamma$ and $\dot\gamma_{\bar{t},C}= \sum_\ell
\Gamma \circ \Pi_r \circ \dot\gamma_{\bar t}^{(\ell)} \circ \Pi_r \circ \Gamma$ as input and holds irrespective of the choice of
vertex parametrization. For this reason, we do not discuss the self-energy flow further in this paper, but it should of course be implemented for numerical work.

\begin{figure*}
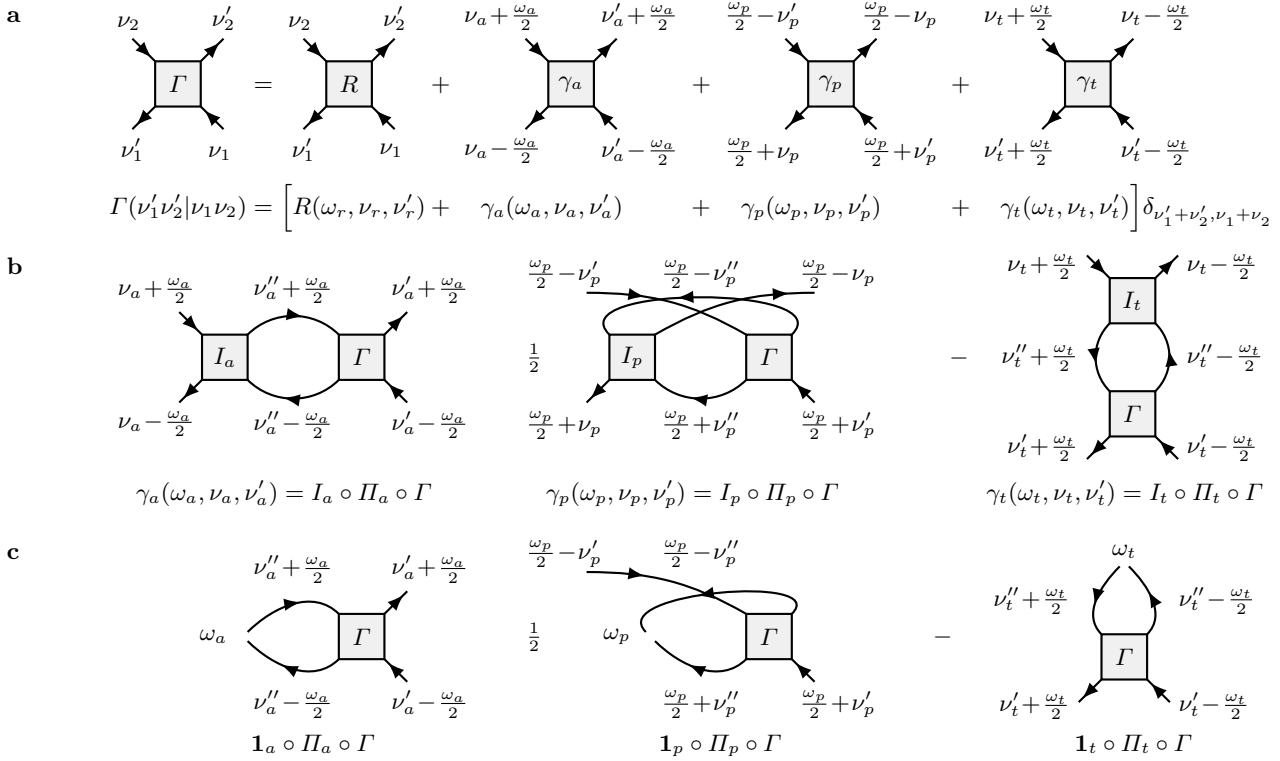

	\textbf{a}
	\vspace{-0.7cm}
	\begin{alignat}{5}
		\tikzm{Vertex_parametrization-channeldec-PE}{
			\fullvertexwithlegs{$\Gamma$}{0}{0}{1}
			\node[above] at (-0.6,0.6) {\small$\nu_2\phantom{'}$};
			\node[above] at (0.6,0.6) {\small$\nu_2'$};
			\node[below] at (-0.6,-0.6) {\small$\nu_1'$};
			\node[below] at (0.6,-0.6) {\small$\nu_1\phantom{'}$};
		}
		&=
		\tikzm{Vertex_parametrization-channeldec-PE_R}{
			\fullvertexwithlegs{$R$}{0}{0}{1}
			\node[above] at (-0.6,0.6) {\small$\nu_2\phantom{'}$};
			\node[above] at (0.6,0.6) {\small$\nu_2'$};
			\node[below] at (-0.6,-0.6) {\small$\nu_1'$};
			\node[below] at (0.6,-0.6) {\small$\nu_1$\phantom{'}};
		}
		&&+
		\tikzm{Vertex_parametrization-channeldec-PE_a}{
			\fullvertexwithlegs{$\gamma_a$}{0}{0}{1}
			\node[above] at (-0.9,0.6) {\small$\va \! + \! \tfrac{\wa}{2}$};
			\node[above] at (0.9,0.6) {\small$\va' \! + \! \tfrac{\wa}{2}$};
			\node[below] at (-0.9,-0.6) {\small$\va \! - \! \tfrac{\wa}{2}$};
			\node[below] at (0.9,-0.6) {\small$\va' \! - \! \tfrac{\wa}{2}$};
		}
		&&+
		\tikzm{Vertex_parametrization-channeldec-PE_p}{
			\fullvertexwithlegs{$\gamma_p$}{0}{0}{1}
			\node[above] at (-0.9,0.6) {\small$\tfrac{\wp}{2} \! - \! \vp'$};
			\node[above] at (0.9,0.6) {\small$\tfrac{\wp}{2} \! - \! \vp$};
			\node[below] at (-0.9,-0.6) {\small$\tfrac{\wp}{2} \! + \! \vp$};
			\node[below] at (0.9,-0.6) {\small$\tfrac{\wp}{2} \! + \! \vp'$};
		}
		&&+
		\tikzm{Vertex_parametrization-channeldec-PE_t}{
			\fullvertexwithlegs{$\gamma_t$}{0}{0}{1}
			\node[above] at (-0.9,0.6) {\small$\vt \! + \! \tfrac{\wt}{2}$};
			\node[above] at (0.9,0.6) {\small$\vt \! - \! \tfrac{\wt}{2}$};
			\node[below] at (-0.9,-0.6) {\small$\vt' \! + \! \tfrac{\wt}{2}$};
			\node[below] at (0.9,-0.6) {\small$\vt' \! - \! \tfrac{\wt}{2}$};
		}
		\nonumber \\
		\Gamma(\nu_1'\nu_2'|\nu_1\nu_2)
		&= \Bigl[R(\wr,\vr,\vr')
		&&+ \quad \gamma_a(\wa,\va,\va') &&+ \quad \gamma_p(\wp,\vp,\vp') &&+ \quad \gamma_t(\wt,\vt,\vt') \Bigr] \delta_{\nu_1'+\nu_2',\nu^\pprime_1+\nu^\pprime_2} \nonumber 
	\end{alignat}
	\textbf{b}
	\vspace{-0.7cm}
	\begin{alignat*}{3}
		\tikzm{Vertex_parametrization-Ba_rhs}{
			\arrowslefthalffull{0}{0}{1}
			\fullvertex{$I_a$}{0}{0}{1}
			\abubblefullfull{0.3}{0}{1}{1}{1}
			\fullvertex{$\Gamma$}{1.8}{0}{1}
			\arrowsrighthalffull{1.8}{0}{1}
			\node[above] at (-0.9,0.6) {\small$\va \! + \! \tfrac{\wa}{2}$};
			\node[above] at (0.9,0.6) {\small$\va'' \! + \! \tfrac{\wa}{2}$};
			\node[above] at (2.7,0.6) {\small$\va' \! + \! \tfrac{\wa}{2}$};
			\node[below] at (-0.9,-0.6) {\small$\va \! - \! \tfrac{\wa}{2}$};
			\node[below] at (0.9,-0.6) {\small$\va'' \! - \! \tfrac{\wa}{2}$};
			\node[below] at (2.7,-0.6) {\small$\va' \! - \! \tfrac{\wa}{2}$};
		}
		&&
		\qquad
		\tfrac{1}{2}
		\!\!\!\!\!\!
		\tikzm{Vertex_parametrization-Bp_rhs}{
			\arrowslefthalffullp{0}{0}{1}{1}{1}
			\fullvertex{$I_p$}{0}{0}{1}
			\pbubblefullfull{0.3}{0}{1}{1}{1}
			\fullvertex{$\Gamma$}{1.8}{0}{1}
			\arrowsrighthalffullp{1.8}{0}{1}{1}{1}
			\node[above] at (-0.9,0.8) {\small$\tfrac{\wp}{2} \! - \! \vp'$};
			\node[above] at (0.9,0.8) {\small$\tfrac{\wp}{2} \! - \! \vp''$};
			\node[above] at (2.7,0.8) {\small$\tfrac{\wp}{2} \! - \! \vp$};
			\node[below] at (-0.9,-0.6) {\small$\tfrac{\wp}{2} \! + \! \vp$};
			\node[below] at (0.9,-0.6) {\small$\tfrac{\wp}{2} \! + \! \vp''$};
			\node[below] at (2.7,-0.6) {\small$\tfrac{\wp}{2} \! + \! \vp'$};
		}
		&&
		\qquad
		-\quad
		\tikzm{Vertex_parametrization-Bt_rhs}{
			\arrowsupperhalffull{0}{0.75}{1}
			\fullvertex{$I_t$}{0}{0.75}{1}
			\tbubblefullfull{0}{0.45}{1}{1}{1}
			\fullvertex{$\Gamma$}{0}{-0.75}{1}
			\arrowslowerhalffull{0}{-0.75}{1}
			\node[left] at (-0.6,1.2) {\small$\vt \! + \! \tfrac{\wt}{2}$};
			\node[right] at (0.6,1.2) {\small$\vt \! - \! \tfrac{\wt}{2}$};
			\node[left] at (-0.6,0) {\small$\vt'' \! + \! \tfrac{\wt}{2}$};
			\node[right] at (0.6,0) {\small$\vt'' \! - \! \tfrac{\wt}{2}$};
			\node[left] at (-0.6,-1.2) {\small$\vt' \! + \! \tfrac{\wt}{2}$};
			\node[right] at (0.6,-1.2) {\small$\vt' \! - \! \tfrac{\wt}{2}$};
		}
		&
		\\
		\gamma_a(\wa,\va,\va') = I_a \circ \Pi_a \circ \Gamma \phantom{\Pi\Pi} &&
		\gamma_p(\wp,\vp,\vp') = I_p \circ \Pi_p \circ \Gamma \phantom{\Pi\Pi} &&
		\gamma_t(\wt,\vt,\vt')= I_t \circ \Pi_t \circ \Gamma \phantom{\ } &
		\intertext{\textbf{c}}
		\\[-1.4cm]
		\tikzm{Vertex_parametrization-Ba_rhs_boldI}{
			\abubblefullfull{0.3}{0}{1}{0.2}{1}
			\fullvertex{$\Gamma$}{1.8}{0}{1}
			\arrowsrighthalffull{1.8}{0}{1}
			\node at (-0.15,0) {\small $\wa$};
			\node[above] at (0.9,0.6) {\small$\va'' \! + \! \tfrac{\wa}{2}$};
			\node[above] at (2.7,0.6) {\small$\va' \! + \! \tfrac{\wa}{2}$};
			\node[below] at (0.9,-0.6) {\small$\va'' \! - \! \tfrac{\wa}{2}$};
			\node[below] at (2.7,-0.6) {\small$\va' \! - \! \tfrac{\wa}{2}$};
		}
		&&
		\qquad
		\tfrac{1}{2}
		\!\!\!\!\!\!
		\tikzm{Vertex_parametrization-Bp_rhs_boldI}{
			\pbubblefullfull{0.3}{0}{1}{0.2}{1}
			\fullvertex{$\Gamma$}{1.8}{0}{1}
			\arrowsrighthalffullp{1.8}{0}{1}{1}{1}
			\node at (-0.2,0) {\small $\wp$};
			\node[above] at (-0.9,0.8) {\small$\tfrac{\wp}{2} \! - \! \vp'$};
			\node[above] at (0.9,0.8) {\small$\tfrac{\wp}{2} \! - \! \vp''$};
			\node[below] at (0.9,-0.6) {\small$\tfrac{\wp}{2} \! + \! \vp''$};
			\node[below] at (2.7,-0.6) {\small$\tfrac{\wp}{2} \! + \! \vp'$};
		}
		&&
		\qquad 
		-\quad
		\raisebox{0.48cm}{\tikzm{Vertex_parametrization-Bt_rhs_boldI}{
			\tbubblefullfull{0}{0.45}{1}{0.2}{1}
			\fullvertex{$\Gamma$}{0}{-0.75}{1}
			\arrowslowerhalffull{0}{-0.75}{1}
			\node at (0,0.65) {\small $\wt$};
			\node[left] at (-0.6,0) {\small$\vt'' \! + \! \tfrac{\wt}{2}$};
			\node[right] at (0.6,0) {\small$\vt'' \! - \! \tfrac{\wt}{2}$};
			\node[left] at (-0.6,-1.4) {\small$\vt' \! + \! \tfrac{\wt}{2}$};
			\node[right] at (0.6,-1.4) {\small$\vt' \! - \! \tfrac{\wt}{2}$};
		}}
		&
		\\[-1mm]
		\boldI_a \circ \Pi_a \circ \Gamma \phantom{\Pi\Pi\Pi\Pi\circ} &&
		\boldI_p \circ \Pi_p \circ \Gamma \phantom{\Pi\Pi\Pi\Pi\circ} &&
		\boldI_t \circ \Pi_t \circ \Gamma \phantom{\Pi\Pi\Pi\circ} &
	\end{alignat*}
\caption{
Definition of the three channel-specific frequency parametrizations of the four-point vertex. \textbf{a} The vertex is nonzero only if the four fermionic frequencies satisfy ${\nu_1'+\nu_2' =\nu^\pprime_1+\nu^\pprime_2}$. In that case, they can be expressed in three different ways through one bosonic transfer frequency, $\wr$, and two fermionic frequencies, $\vr^\pprime$, $\vr'$.
Of course, each term can also be expressed through the frequencies $(\wr^\pprime,\vr^\pprime,\vr')$ of any of the three channels, as indicated here for $R$.
\textbf{b} The choice of frequency arguments in each channel $\gamma_a$, $\gamma_p$, and $\gamma_t$ is motivated by the structure of their BSEs \eqref{eq:Bethe-Salpeter-equations}.
\textbf{c} Diagrammatic depiction of $\boldI_r\circ \Pi_r \circ 
\Gamma = \sum_{\nu''_r} \Pi_r \protect\fcirc  \Gamma$ (Eqs.~\eqref{subeq:define-Pi-Gamma}, third line), a four-leg object obtained
by inserting $\boldI_r$ between $U$ and $\Pi_r$ (Eq.~\eqref{eq:shorthand-U-Pi-Gamma-b}). The multiplication of $\boldI_r\circ$ 
onto $\Pi_r \circ \Gamma$ carries two instructions: draw $\Pi_r$ such that the endpoints of the lines connected to $\boldI_r$ lie close together (awaiting being connected to $U$), and perform the sum over the fermionic frequency $\nu_r''$ of $\Pi_r$.
}
\label{fig:Vertex_frequency_parametrization}
\end{figure*}

The $1\ell$ contribution \eqref{eq:mfRG-equations_gamma_r_1l} of the vertex flow, 
with the fully-differentiated $\dot G$ replaced by the single-scale propagator $S$ in $\dot \Pi_r$ 
is equivalent to the usual $1\ell$ flow equation. 
Using $\dot G$ instead of $S$, as done in \Eq{eq:mfRG-equations_gamma_r_1l}, corresponds to the so-called Katanin substitution \cite{Katanin2004}: it contains the feedback of the differentiated self-energy into the vertex flow and already goes beyond the standard $1\ell$ approximation. By adding higher-loop contributions until convergence is reached, one effectively solves the self-consistent parquet equations through an fRG flow. 
On the one hand, this ensures two-particle self-consistency and related properties mentioned in the introduction.
On the other hand, it also provides a way of reaching a solution of the parquet equations by integrating differential equations. This may be numerically favorable compared to an iteration of the self-consistent equations.
Particularly, when computing diagrammatic extensions of DMFT via DMF$^2$RG, one then needs only the \textsl{full} DMFT vertex as an input, and not the $r$-(ir)reducible ones entering the parquet equations. This is helpful in the Matsubara formalism, where the $r$-(ir)reducible vertices sometimes exhibit divergences \cite{Schaefer2013,Schaefer2016,Gunnarsson2017,Chalupa2018,Thunstroem2018},
and even more so when aiming for real-frequency approaches \cite{Kugler2021,Lee2021}.

\subsection{Frequency parametrization}
\label{sec:Frequency-parametrization}
The four-point vertex $\Gamma$ is a highly complicated object and must be parametrized efficiently. 
In this section, we summarize the frequency parametrization of the vertex adapted to the three diagrammatic channels.
This parametrization is the building block for the SBE decomposition discussed in \Sec{sec:SBE-decomposition}.

Focusing on the frequency dependence, we
switch from the compact notation $\Gamma_{1'2'|12}$ to the more elaborate $\Gamma_{1'2'|12} (\nu_1'\nu_2'|\nu^\pprime_1\nu^\pprime_2)$, with frequency arguments written in brackets, and the subscripts now referring  to non-frequency quantum numbers (position or momentum, spin, etc.). As mentioned earlier, we assume the bare vertex $U$ to have the form 
\begin{align}
	\label{eq:bareU-frequency-independent}
	U_{1'2'|12} (\nu_1'\nu_2'|\nu^\pprime_1\nu^\pprime_2) =\delta_{\nu_1'+\nu_2',\nu^\pprime_1+\nu^\pprime_2}
U_{1'2'|12} , 
\end{align}
with $U_{1'2'|12}$ independent of frequency. If $U$ is momentum-conserving without further momentum dependence, our
treatment of frequency sums below may be extended to
include momentum sums. 
To keep the discussion general, we refrain from elaborating this in detail.
Note that, e.g., in the repulsive Hubbard model, our sign convention in Eq.~\eqref{eq:Fermionic-action} is such that $U^{\sigma\bar{\sigma}|\sigma\bar{\sigma}}=-U^{\bar{\sigma}\sigma|\sigma\bar{\sigma}}<0$
(where, as usual, $\sigma \in \{ \uparrow, \downarrow \}$, $\bar{\uparrow} = \downarrow$, $ \bar{\downarrow} = \uparrow$).

Due to frequency conservation, one-particle correlators depend on
only one frequency, 
\begin{equation}
\label{eq:G-frequency-relation}
 	G_{1'1}(\nu'_1,\nu^\pprime_1) = \delta_{\nu'_1,\nu^\pprime_1}
	G_{1'1}(\nu^\pprime_1) . 
 \end{equation} 
Likewise, three frequencies are sufficient to parametrize the vertex.
For each channel $\gamma_r$, we express the four fermionic frequencies $\nu'_1, \nu'_2, \nu^\pprime_1, \nu^\pprime_2$ at the vertex legs through a choice of three frequencies, a bosonic transfer frequency, $\omega^\pprime_r$, and two fermionic frequencies, $\nu^\pprime_r$ and $\nu'_r$. These are chosen differently for each channel (see Fig.~\ref{fig:Vertex_frequency_parametrization}a) and reflect its asymptotic behavior \cite{Wentzell2020} as discussed in Sec.~\ref{subsec:Asymptotic-classes}. We have
\begin{align}
& \gamma_{r; 1'2'|12}(\nu_1'\nu_2'|\nu^\pprime_1\nu^\pprime_2) = \delta_{\nu_1'+\nu_2',\nu^\pprime_1 + \nu^\pprime_2} \gamma_{r;1'2'|12}(\wr^\pprime,\vr^\pprime,\vr') , 
\end{align}\vspace{-1.3\baselineskip}

\noindent
with $\wr^\pprime$, $\vr^\pprime$, $\vr'$ related to
$\nu'_1$, $\nu^\pprime_1$, $\nu^\pprime_2$ through 
\begin{align}
\nonumber
 	\nu'_1 & = \va^\pprime - \tfrac{\wa}{2} = \phantom{+} \vp^\pprime + \tfrac{\wp}{2} = \vt' + \tfrac{\wt}{2} ,
\\ 
\label{eq:define-omegar-nur-nur'}
 	\nu^\pprime_1 & = \va' - \tfrac{\wa}{2} = \phantom{+} \vp' + \tfrac{\wp}{2} = \vt' - \tfrac{\wt}{2} , 
\\
\nonumber 
 	\nu^\pprime_2 & = \va^\pprime + \tfrac{\wa}{2} = - \vp' + \tfrac{\wp}{2} = \vt^\pprime + 
 	\tfrac{\wt}{2} .  
\end{align}%
This parametrization symmetrically assigns $\pm\frac{\omega_r}{2}$ shifts to all external legs. (In the Matsubara formalism, the bosonic Matsubara frequency closest to $\pm\frac{\omega_r}{2}$ is chosen for the shift.)
With these shifts, crossing symmetries ensure that prominent vertex peaks are centered around $\omega_r=0$, which is convenient for numerical work.
However, other conventions are of course possible, too.

Though the frequencies $\omega^\pprime_r, \nu^\pprime_r, \nu_r'$ are tailored to a specific channel $\gamma_r$, one may also use them to
define the $r$ parametrization of the full vertex, writing
\begin{flalign}
\label{eq:r-representation-Gamma}
& \Gamma_{1'2'|12}(\nu_1'\nu_2'|\nu^\pprime_1\nu^\pprime_2) = \delta_{\nu_1'+\nu_2',\nu^\pprime_1 + \nu^\pprime_2} \Gamma_{1'2'|12}(\wr^\pprime,\vr^\pprime,\vr') . \hspace{-1cm} & 
\end{flalign}
Likewise, $R, \gamma_a, \gamma_p, \gamma_t$ can each be expressed as a $\delta$ symbol times a
function of \textsl{any} of the variable sets
$ (\wr,\vr,\vr')$.
The $r$ parametrization of $\Gamma \circ \Pi_r$ or 
$\Pi_r \circ \Gamma$ is obtained by inserting
\Eqs{eq:G-frequency-relation} and \eqref{eq:r-representation-Gamma}
into \Eq{eq:Bubble_summation}. The summations  
$\sum_{\nu_5 \nu_6}$ over internal frequencies can be collapsed using frequency-conserving $\delta$ symbols, leading to 
\begin{subequations}
	\label{eq:Gamma-circ-Pi}
\begin{align}
[\Gamma \circ \Pi_r] (\omega^\pprime_r, \nu^\pprime_r, \nu''_r) 
	& = \Gamma(\omega^\pprime_r, \nu^\pprime_r, \nu''_r) \fcirc
\Pi (\omega^\pprime_r, \nu''_r) , \\
[\Pi_r \circ \Gamma] (\omega^\pprime_r, \nu''_r, \nu'_r) 
	 & = \Pi (\omega^\pprime_r, \nu''_r) \fcirc
	\Gamma(\omega^\pprime_r, \nu''_r, \nu'_r) 
 ,
\end{align}
\end{subequations}
where the bubble factors $\Pi_r (\omega_r, \nu''_r)$ are given by  
\begin{subequations}
\begin{flalign}
\label{eq:frequency-bubbles}
	\Pi_{a;34|3'4'} (\omega^\pprime_a, \nu''_a) &= 
	\phantom{-} 
	G_{3|3'} \bigl(\nu''_a \!-\! \tfrac{\omega_a}{2}\bigr) 
	G_{4|4'} \bigl(\nu''_a \!+\! \tfrac{\omega_a}{2}\bigr)  ,
	\hspace{-1cm} & \\
	\Pi_{p;34|3'4'} (\omega^\pprime_p, \nu''_p) &= \tfrac{1}{2} 
	G_{3|3'} \bigl(\tfrac{\omega_p}{2} \!+\! \nu''_p \bigr) 
	G_{4|4'} \bigl(\tfrac{\omega_p}{2} \!-\! \nu''_p \bigr) ,
	\hspace{-1cm} & 
	\\
	\Pi_{t;43|3'4'} (\omega^\pprime_t, \nu''_t) &=  - 
	G_{3|3'} \bigl(\nu''_t \!-\! \tfrac{\omega_t}{2}\bigr) 
	G_{4|4'} \bigl(\nu''_t \!+\! \tfrac{\omega_t}{2}\bigr)   . 
	\hspace{-1cm} & 
\end{flalign}
\end{subequations}
In \Eqs{eq:Gamma-circ-Pi}, the connector  $\fcirc$  
by definition denotes an internal summation analogous to $\circ$,
except that only non-frequency quantum numbers (position, spin, etc.) are summed over.  Correspondingly,
the bubble $\tilde \Gamma \circ \Pi_r \circ \Gamma$, involving
two $\circ$ connectors, has the $r$ parametrization
	\begin{align}
		\nonumber&[\tilde \Gamma \circ\Pi_r\circ \Gamma]
		(\omega^\pprime_r,\nu^\pprime_r,\nu'_r)\\
		&=\sum_{\nu_r''}\tilde \Gamma(\omega^\pprime_r,\nu^\pprime_r,\nu_r'')
		\fcirc \Pi_r(\omega^\pprime_r,\nu''_r)\fcirc 
		\Gamma(\omega^\pprime_r,\nu_r'',\nu_r'),
		\label{eq:Bubble_summation_frequencies}
	\end{align}
see Fig.~\ref{fig:Vertex_frequency_parametrization}b. Here, one frequency sum survives, running over the fermionic frequency $\nu''_r$ associated with $\Pi_r$.

\begin{figure*}[h]
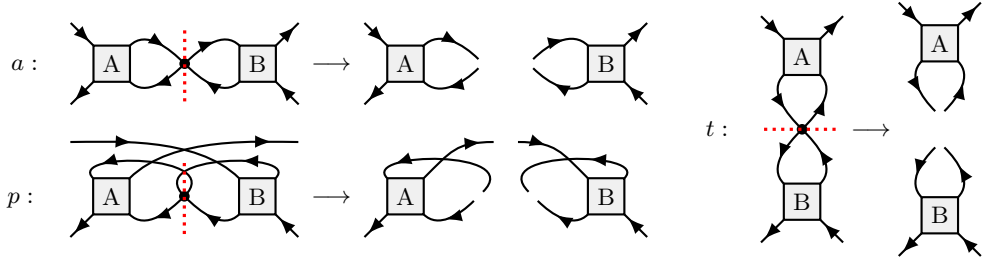

	\begin{align*}
		\begin{array}{ccc}
			a: \quad
			\tikzm{U-a-reducibility}{
				\arrowslefthalffull{0}{0}{0.8}
				\fullvertex{A}{0}{0}{0.8}
				\abubblefullbare{0.24}{0}{0.6}{0.8}
				\barevertex{0.96}{0}
				\abubblebarefull{0.96}{0}{0.6}{0.8}
				\fullvertex{B}{1.92}{0}{0.8}
				\arrowsrighthalffull{1.92}{0}{0.8}
				\draw[red,very thick,dotted] (0.96,-0.5) -- (0.96,0.5);
			}
			&\longrightarrow&
			\tikzm{U-a-reducibility-leftright}{
				\arrowslefthalffull{0}{0}{0.8}
				\fullvertex{A}{0}{0}{0.8}
				\abubblefullfull{0.24}{0}{0.6}{0.8}{0.2}
				\abubblefullfull{1.68}{0}{0.6}{0.2}{0.8}
				\fullvertex{B}{2.64}{0}{0.8}
				\arrowsrighthalffull{2.64}{0}{0.8}
			}
			\\
			\\
			p: \quad
			\tikzm{U-p-reducibility}{
				\draw[lineWithArrowCenterEnd] (-0.24,-0.24) -- (-0.54,-0.54);
				\draw[lineWithArrowCenterEnd] (0.24,0.24) .. controls ++(45:0.8) and ++(180:0.5) .. (2.46,0.72);
				\fullvertex{A}{0}{0}{0.8}
				\draw[lineWithArrowCenterEnd] (0.96,0) .. controls ++(45:0.8) and ++(135:0.5) .. (-0.24,0.24);
				\draw[lineWithArrowCenter] (0.96,0) to [out=225, in=315] (0.24,-0.24);
				\barevertex{0.96}{0}
				\draw[lineWithArrowCenterStart] (2.16,0.24) .. controls ++(45:0.5) and ++(135:0.8) .. (0.96,0);
				\draw[lineWithArrowCenter] (1.68,-0.24) to [out=225, in=315] (0.96,0);
				\fullvertex{B}{1.92}{0}{0.8}
				\draw[lineWithArrowCenterEnd] (2.46,-0.54) -- (2.16,-0.24);
				\draw[lineWithArrowCenterStart] (-0.54,0.72) .. controls ++(0:0.5) and ++(135:0.8) .. (1.68,0.24);
				\draw[red,very thick,dotted] (0.96,-0.5) -- (0.96,0.5);
			}
			&\longrightarrow&
			\tikzm{U-p-reducibility-leftright}{
				\draw[lineWithArrowCenterEnd] (-0.24,-0.24) -- (-0.54,-0.54);
				\draw[lineWithArrowCenterEnd] (0.24,0.24) .. controls ++(45:0.5) and ++(180:0.5) .. (1.16,0.72);
				\node at (1.16,0.72) {};
				\fullvertex{A}{0}{0}{0.8}
				\draw[lineWithArrowCenterEnd] (1.02,0.06) .. controls ++(45:0.8) and ++(135:0.5) .. (-0.24,0.24);
				\draw[lineWithArrowCenter] (0.9,-0.06) to [out=225, in=315] (0.24,-0.24);
				\draw[lineWithArrowCenterStart] (2.88,0.24) .. controls ++(45:0.5) and ++(135:0.8) .. (1.62,0.06);
				\draw[lineWithArrowCenter] (2.4,-0.24) to [out=225, in=315] (1.74,-0.06);
				\fullvertex{B}{2.64}{0}{0.8}
				\draw[lineWithArrowCenterEnd] (3.18,-0.54) -- (2.88,-0.24);
				\draw[lineWithArrowCenterStart] (1.48,0.72) .. controls ++(0:0.5) and ++(135:0.5) .. (2.4,0.24);
				\node at (1.48,0.72) {};
			}
		\end{array}
		\qquad
		t: \quad
		\tikzm{U-t-reducibility}{
			\arrowsupperhalffull{0}{0.96}{0.8}
			\fullvertex{A}{0}{0.96}{0.8}
			\draw[lineWithArrowCenter] (-0.24,0.72) to [out=225, in=135] (0,0);
			\draw[lineWithArrowCenter] (0,0) to [out=45, in=315] (0.24,0.72);
			\barevertex{0}{0}
			\draw[lineWithArrowCenter] (0,0) to [out=225, in=135] (-0.24,-0.72);
			\draw[lineWithArrowCenter] (0.24,-0.72) to [out=45, in=315] (0,0);
			\fullvertex{B}{0}{-0.96}{0.8}
			\arrowslowerhalffull{0}{-0.96}{0.8}
			\draw[red,very thick,dotted] (-0.5,0) -- (0.5,0);
		}
		\longrightarrow
		\tikzm{U-t-reducibility-leftright}{
			\arrowsupperhalffull{0}{1.14}{0.8}
			\fullvertex{A}{0}{1.14}{0.8}
			\draw[lineWithArrowCenter] (-0.24,0.9) to [out=225, in=135] (-0.06,0.24);
			\draw[lineWithArrowCenter] (0.06,0.24) to [out=45, in=315] (0.24,0.9);
			\draw[lineWithArrowCenter] (-0.06,-0.24) to [out=225, in=135] (-0.24,-0.9);
			\draw[lineWithArrowCenter] (0.24,-0.9) to [out=45, in=315] (0.06,-0.24);
			\fullvertex{B}{0}{-1.14}{0.8}
			\arrowslowerhalffull{0}{-1.14}{0.8} 
		} 
	\end{align*} \vspace{-3mm}
	\caption{Illustration of \Ur-reducibility, analogous to Fig.~4 of \cite{Krien2019}. $A$ and $B$ can be any vertex diagram or simply $\doubleI_r$.
		\label{fig:U-r-reducibility}
	} 
\end{figure*}

For future reference, we define unit vertices for non-frequency quantum numbers, $\boldI_r$, by $\Gamma = \boldI_r \fcirc \Gamma = \Gamma \fcirc \boldI_r$.
(For a bare vertex with momentum conservation and no further
momentum dependence, one could include a momentum 
sum, $\sum_{k''_r}$, in \Eq{eq:Bubble_summation_frequencies} 
and exclude momentum indices from the $\fcirc$ summation
and $\boldI_r$.) The distinction between $\circ$, $\doubleI$ and $\fcirc$, $\boldI$,
indicating if connectors and unit vertices
include summations and $\delta$ symbols for frequency variables or not, will be needed for the SBE decomposition of \Sec{sec:SBE-decomposition}. There,
we will encounter bubbles involving one or two bare vertices, $U \circ \Pi_r \circ U$, $\tilde \Gamma \circ \Pi_r \circ U$,
or $U \circ \Pi_r \circ \Gamma$. Expressing these in the
form \eqref{eq:Bubble_summation_frequencies}, the bare vertex $U$,
since it is frequency independent, can be pulled out of the sum over $\nu_r''$. To make this explicit, we insert unit operators 
$\boldI_r$ next to $U$:
\begin{subequations}
	\label{subeq:shorthand-U-Pi-Gamma}
		\begin{align}
		\label{eq:UPiU}
		U \circ \Pi_r \circ U &= U \fcirc \boldI_r \circ \Pi_r \circ \boldI_r \fcirc U, \\ 
		\tilde \Gamma \circ \Pi_r \circ U
		& = \tilde \Gamma \circ \Pi_r \circ \boldI_r \fcirc U , 
		\\ 
		U \circ \Pi_r \circ \Gamma 
		 & =  U \fcirc \boldI_r \circ \Pi_r \circ \Gamma  .
	\label{eq:shorthand-U-Pi-Gamma-b}
	\end{align}
\end{subequations}
We suppressed frequency arguments for brevity, it being understood that
equations linking $\Pi_r$ and $\boldI_r$ use the $r$ parametrization. 
Making the frequency sum involved in $\circ \, \Pi_r \circ$ explicit, we
obtain four-leg objects, 
\begin{flalign}
	\label{subeq:define-Pi-Gamma} 
		[\boldI_r \circ \Pi_r \circ \boldI_r](\omega^\pprime_r) &= \sum_{\nu_r''}\Pi_r(\omega^\pprime_r,\nu''_r), & 
		\nonumber
		\\
		[\tilde \Gamma \circ \Pi_r \circ \boldI_r] (\omega^\pprime_r,\nu^\pprime_r)
		&=\sum_{\nu_r''}\tilde \Gamma(\omega^\pprime_r,\nu^\pprime_r,\nu_r'')
		\fcirc \Pi_r(\omega^\pprime_r,\nu''_r) , & 
		\nonumber
		\\
		[\boldI_r \circ \Pi_r \circ \Gamma](\omega^\pprime_r,\nu'_r)
		&=\sum_{\nu_r''}\Pi_r(\omega^\pprime_r,\nu''_r)\fcirc 
		\Gamma(\omega^\pprime_r,\nu_r'',\nu_r'), & \hspace{-1cm}
\end{flalign}
that depend on only one or two frequency arguments (cf. Fig.~\ref{fig:Vertex_frequency_parametrization}c) and are thus numerically cheaper than $\Gamma$. Note that, in general, $\boldI_r$ is not the unit operator \wrt the $\circ$ connector, i.e., $\boldI_r \circ \Gamma \neq \Gamma \neq \Gamma \circ \boldI_r$, since $\circ$ involves a frequency summation which does not affect $\boldI_r$.

\section{SBE decomposition}
\label{sec:SBE-decomposition}

We now turn to the SBE decomposition. It also yields 
an exact, unambiguous classification of vertex diagrams, now according to their \textsl{$U$-reducibility} in each channel.
This notion of reducibility, introduced in Ref. \cite{Krien2019}, is very analogous to $\Pi$-reducibility, i.e., two-particle reducibility.
A diagram is called $U$-reducible if it can be split into two parts by splitting apart a bare vertex $U$ (in ways specified below) in either of the three channels. 
Otherwise, it is fully $U$-irreducible.

The SBE decomposition was originally formulated in terms of physical (charge, spin, and singlet pairing) channels which involve linear combinations of spin components. 
For our purposes, it is more convenient not to use such linear combinations (the relation between both formulations is given in \App{Appendix:physical_diagrammatic}). Moreover, the original SBE papers considered models with translational invariance, with vertices labeled by three momentum variables. We here present a generalization of the SBE decomposition applicable to models without translational invariance, requiring four position or momentum labels. Starting from the BSEs, we use arguments inspired by Ref.~\cite{Krien2019} to arrive at a set of self-consistent equations for SBE ingredients which will also enable us to derive multiloop flow equations directly within this framework.
In terms of notation, we follow Ref.~\cite{Krien2019}
for the objects $\nabla_r$, $w_r$, $\bar \lambda_r$, $\lambda_r$---with $\varphi^{\text{firr}}$ there denoted $\varphi^{U\text{irr}}$ here---while we 
follow Ref.~\cite{Krien2021} for $M_r$ and $T_r$ (the latter instead of $\varphi_r$ from Ref.~\cite{Krien2019}).

\subsection{Derivation of SBE decomposition from BSEs}

\begin{figure*}
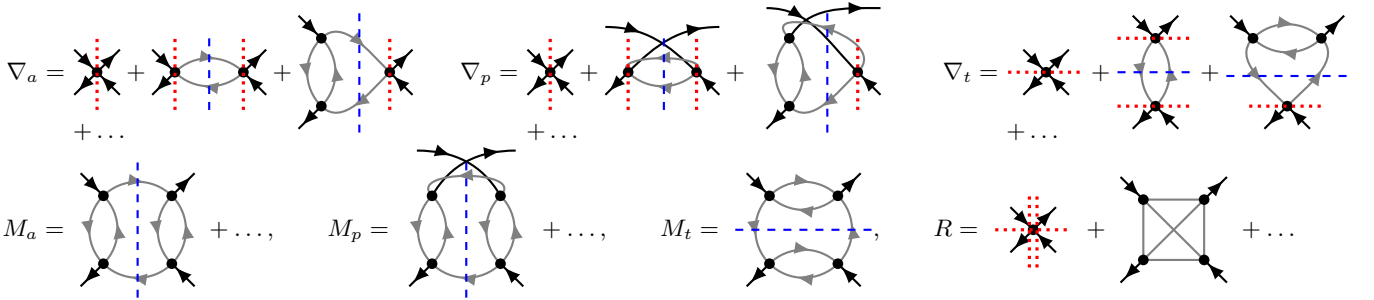

	\begin{align*}
		&
		\begin{array}{rlrlrl}
			\nabla_a =&
			\tikzm{nabla_a_PT-barevertex}{
				\barevertexwithlegs{0}{0}
				\draw[red,very thick,dotted] (0,-0.5) -- (0,0.5);
			}
			+
			\tikzm{nabla_a_PT-K1a}{
				\arrowslefthalf{0}{0}
				\abubblebarebarebare{0}{0}{0.75}
				\arrowsrighthalf{0.9}{0}
				\barevertex{0}{0}
				\barevertex{0.9}{0}
				\draw[red,very thick,dotted] (0,-0.5) -- (0,0.5);
				\draw[red,very thick,dotted] (0.9,-0.5) -- (0.9,0.5);
				\draw[blue,thick,dashed] (0.45,-0.5) -- (0.45,0.5);
			}
			+
			\tikzm{nabla_a_PT-K2a}{
				\arrowslefthalffull{0.45}{0}{1.5}
				\tbubblebarebarebare{0}{0.45}{0.75}
				\draw[lineBareWithArrowCenter] (0,0.45) to [out=45, in=135] (0.9,0);
				\draw[lineBareWithArrowCenter] (0.9,0) to [out=225, in=315] (0,-0.45);
				\arrowsrighthalf{0.9}{0};
				\barevertex{0}{0.45}
				\barevertex{0}{-0.45}
				\barevertex{0.9}{0}
				\draw[red,very thick,dotted] (0.9,-0.5) -- (0.9,0.5);
				\draw[blue,thick,dashed] (0.5,-0.8) -- (0.5,0.8);
			}
			\phantom{--}
			&
			\nabla_p =&
			\tikzm{nabla_p_PT-barevertex}{
				\barevertexwithlegs{0}{0}
				\draw[red,very thick,dotted] (0,-0.5) -- (0,0.5);
			}
			+
			\tikzm{nabla_p_PT-K1p}{
				\arrowslefthalfp{0}{0}{0.75}
				\arrowsrighthalfp{0.9}{0}{0.75}
				\node at (0,0.6) {};
				\pbubblebarebarebare{0}{0}{0.75}
				\node at (0.9,0.6) {};
				\barevertex{0}{0}
				\barevertex{0.9}{0}
				\draw[red,very thick,dotted] (0,-0.5) -- (0,0.5);
				\draw[red,very thick,dotted] (0.9,-0.5) -- (0.9,0.5);
				\draw[blue,thick,dashed] (0.47,-0.5) -- (0.47,0.5);
			}
			+
			\tikzm{nabla_p_PT-K2p}{
				\draw[lineWithArrowCenterEnd] (1.2,-0.3) -- (0.9,0);
				\draw[lineWithArrowCenterEnd] (0,-0.45) -- (-0.3,-0.75);
				\draw[lineWithArrowCenterEnd] (0,0.45) to [out=60, in=180] (1.2,0.85);
				\draw[lineWithArrowCenterStart] (-0.3,0.85) to [out=0, in=135] (0.9,0);
				\tbubblebarebarebare{0}{0.45}{0.75}
				\draw[lineBareWithArrowCenter] (0.9,0) .. controls ++(45:0.7) and ++(135:0.7) .. (0,0.45);
				\draw[lineBareWithArrowCenter] (0.9,0) to [out=225, in=315] (0,-0.45);
				\barevertex{0}{0.45}
				\barevertex{0}{-0.45}
				\barevertex{0.9}{0}
				\draw[red,very thick,dotted] (0.9,-0.5) -- (0.9,0.5);
				\draw[blue,thick,dashed] (0.5,-0.8) -- (0.5,0.8);
			}
			\phantom{--}
			&
			\nabla_t =&
			\tikzm{nabla_t_PT-barevertex}{
				\barevertexwithlegs{0}{0}
				\draw[red,very thick,dotted] (-0.5,0) -- (0.5,0);
			}
			+
			\tikzm{nabla_t_PT-K1t}{
				\arrowsupperhalf{0}{0.45}
				\arrowslowerhalf{0}{-0.45}
				\tbubblebarebarebare{0}{0.45}{0.75}
				\barevertex{0}{0.45}
				\barevertex{0}{-0.45}
				\draw[red,very thick,dotted] (-0.5,0.45) -- (0.5,0.45);
				\draw[red,very thick,dotted] (-0.5,-0.45) -- (0.5,-0.45);
				\draw[blue,thick,dashed] (-0.5,0.0) -- (0.5,0.0);
			}
			+
			\tikzm{nabla_t_PT-K2t}{
				\arrowsupperhalffull{0}{0}{1.5}
				\abubblebarebarebare{-0.45}{0.45}{0.75}
				\arrowslowerhalf{0}{-0.45}
				\draw[lineBareWithArrowCenter] (-0.45,0.45) to [out=225, in=135] (0,-0.45);
				\draw[lineBareWithArrowCenter] (0,-0.45) to [out=45, in=315] (0.45,0.45);
				\barevertex{-0.45}{0.45}
				\barevertex{0.45}{0.45}
				\barevertex{0}{-0.45}
				\draw[red,very thick,dotted] (-0.5,-0.45) -- (0.5,-0.45);
				\draw[blue,thick,dashed] (-0.8,-0.05) -- (0.8,-0.05);
			}
			\\[-2mm]
			&+ \dots &
			&+ \dots &
			&+ \dots
		\end{array}
		\\[-2mm]
		&
		M_a = \
		\tikzm{M_a_PT}{
			\tbubblebarebarebare{-0.45}{0.45}{0.75}
			\tbubblebarebarebare{0.45}{0.45}{0.75}
			\draw[lineBareWithArrowCenter] (-0.45,0.45) to [out=45, in=135] (0.45,0.45);
			\draw[lineBareWithArrowCenter] (0.45,-0.45) to [out=225, in=315] (-0.45,-0.45);
			\arrowsallfull{0}{0}{1.5}
			\barevertex{0.45}{0.45}
			\barevertex{0.45}{-0.45}
			\barevertex{-0.45}{0.45}
			\barevertex{-0.45}{-0.45}
			\draw[blue,thick,dashed] (-0.,-0.9) -- (0.,0.9);
		}
		\ + \dots
		, \quad \quad
		M_p = \
		\tikzm{M_p_PT}{
			\arrowslowerhalffull{0}{0}{1.5}
			\draw[lineWithArrowCenterEnd] (-0.45,0.45) to [out=60, in=180] (0.65,1.05);
			\node at (0.65,1.05) {};
			\draw[lineWithArrowCenterStart] (-0.65,1.05) to [out=0, in=120] (0.45,0.45);
			\node at (-0.65,1.05) {};
			\tbubblebarebarebare{-0.45}{0.45}{0.75}
			\tbubblebarebarebare{0.45}{0.45}{0.75}
			\draw[lineBareWithArrowCenter] (0.45,0.45) .. controls ++(45:0.5) and ++(135:0.5) .. (-0.45,0.45);
			\draw[lineBareWithArrowCenter] (0.45,-0.45) to [out=225, in=315] (-0.45,-0.45);
			\barevertex{0.45}{0.45}
			\barevertex{0.45}{-0.45}
			\barevertex{-0.45}{0.45}
			\barevertex{-0.45}{-0.45}
			\draw[blue,thick,dashed] (-0.,-0.9) -- (-0.,0.9);
		}
		\ + \dots
		, \quad \quad
		M_t = \
		\tikzm{M_t_PT}{
			\arrowsupperhalffull{0}{0}{1.5}
			\arrowslowerhalffull{0}{0}{1.5}
			\abubblebarebarebare{-0.45}{0.45}{0.75}
			\abubblebarebarebare{-0.45}{-0.45}{0.75}
			\draw[lineBareWithArrowCenter] (-0.45,0.45) to [out=225, in=135] (-0.45,-0.45);
			\draw[lineBareWithArrowCenter] (0.45,-0.45) to [out=45, in=315] (0.45,0.45);
			\barevertex{0.45}{0.45}
			\barevertex{0.45}{-0.45}
			\barevertex{-0.45}{0.45}
			\barevertex{-0.45}{-0.45}
			\draw[blue,thick,dashed] (-0.9,-0.) -- (0.9,0.);
		}
		, \quad \quad
		R = \
		\tikzm{R_a_PT-barevertex}{
			\barevertexwithlegs{0}{0}
			\draw[red,very thick,dotted] (-0.05,-0.5) -- (-0.05,0.5);
			\draw[red,very thick,dotted] (0.05,-0.5) -- (0.05,0.5);
			\draw[red,very thick,dotted] (-0.5,0) -- (0.5,0);
		}
		\ + \
		\tikzm{R_a_PT-fourthorder}{
			\draw[lineBare] (-0.4,0.4) -- (0.4,-0.4);
			\draw[lineBare] (0.4,0.4) -- (-0.4,-0.4);
			\draw[lineBare] (-0.4,0.4) -- (0.4,0.4);
			\draw[lineBare] (-0.4,-0.4) -- (0.4,-0.4);
			\draw[lineBare] (0.4,-0.4) -- (0.4,0.4);
			\draw[lineBare] (-0.4,-0.4) -- (-0.4,0.4);
			\barevertex{-0.4}{0.4}
			\barevertex{-0.4}{-0.4}
			\barevertex{0.4}{0.4}
			\barevertex{0.4}{-0.4}
			\arrowsallfull{0}{0}{4./3.}
		}
		\ + \dots
	\end{align*} \vspace{-2mm}
\caption{Low-order diagrams for $\nabla_r$, $M_r$, and $R$, illustrating $\Pir$-reducibility (blue dashed  lines)  and \Ur-reducibility (red dotted lines; their meaning is made explicit in \Fig{fig:U-r-reducibility}). $\nabla_r$ contains all $\Ur$-reducible diagrams;
except for the bare vertex, they all are $\Pir$-reducible, too. $M_a$ contains all diagrams that are $\Pia$- but not $\Ua$-reducible. All diagrams in $R$ are neither  $\Pir$- nor $\Ur$-reducible, except for the bare vertex, which is $\Ua$-, $\Up$- and $\Ut$-reducible (as indicated by three red dotted lines).}
\label{fig:SBE_constituents_PT}
\end{figure*}

As mentioned earlier, a vertex diagram is called two-particle reducible in a specified channel $r \in \{a,p,t\}$, or $\Pir$-\textsl{reducible} for short, if it 
can be split into two parts by cutting the two lines of a $\Pi_r$ bubble (to be called \textsl{linking bubble}); 
if such a split is not possible, the diagram is $\Pir$-irreducible. 
The two-particle reducible vertex $\gamma_r$ is the sum of all $\Pir$-reducible
diagrams. Following Ref.~\cite{Krien2019}, we now introduce  a further channel-specific classification criterion. A $\Pir$-reducible diagram is called $\Ur$-\textsl{reducible} if a linking  bubble $\Pi_r$ has two of its legs attached to the same bare vertex in the combination $U \circ \Pi_r$ or $\Pi_r \circ U$.  Then, that bare vertex $U$, too, constitutes a link that, when ``cut out’’, splits the diagram into two parts. (To visualize the meaning of ``cutting out $U$'' diagrammatically, one may replace $U$ by $\boldI_r 
\fcirc U \fcirc \boldI_r$ and then remove $U$. This results
in two pairs of legs ending close together, ready to be connected through reinsertion of $U$,
see \Fig{fig:Vertex_frequency_parametrization}c and \Fig{fig:U-r-reducibility}.) The lowest-order $\Ur$-reducible contribution to 
$\gamma_r $ is $U \circ \Pi_r \circ U$. The lowest-order term of $\Gamma$, the bare vertex $U$ (which is $\Pir$-irreducible),
is viewed as $\Ur$-reducible in all three channels, corresponding to the three possible ways of splitting its four legs into two pairs of two.
All $\Ur$-reducible diagrams describe ``single-boson exchange’’ processes, in the sense that each link $U$ connecting two otherwise separate parts of the diagram mediates a single bosonic transfer frequency, $\omega_r$ 
(as defined in Fig.~\ref{fig:Vertex_frequency_parametrization}), across that link, as will become explicit below. 

All vertex diagrams that are not $\Ur$-reducible are called $\Ur$-irreducible.
These comprise all multi-boson exchange (i.e., \textsl{not} single-boson exchange) diagrams from $\gamma_r$, and all $\Pir$-irreducible diagrams except the bare vertex (which is trivially $\Ur$-reducible),  i.e., all diagrams from $I_r-U = R-U + \sum_{r^\prime \neq r} \gamma_{r^\prime}$. 

Next, we rewrite the parquet equations in terms of $\Ur$-reducible and $\Ur$-irreducible parts. 
We define $\nabla_r$ as the sum of all $\Ur$-reducible diagrams, including 
(importantly) the bare vertex $U$, and $M_r$ as the sum of all diagrams that are $\Pir$-reducible but $\Ur$-irreducible, thus describing multi-boson exchange processes. Then, the $\Pir$-reducible vertex $\gamma_r$, which does not include $U$, fulfills
\begin{align}
	\label{eq:splitgammar}
	\gamma_r = \nabla_r - U  + M_r   .
\end{align}

Inserting \Eq{eq:splitgammar} for 
$\gamma_r$ into the parquet decomposition
 \eqref{eq:mfRG:mfRG:Parquet:PE} yields
 \begin{subequations}
 	\label{eq:SBE_diagrammatic_channels}
  \begin{align}
\Gamma & = \varphi^{U\text{irr}}+{\textstyle \sum_r} \nabla_r-2U, 
\label{eq:SBE_diagrammatic_channels-Gamma}
\\
\label{eq:SBE_diagrammatic_channels-Phi}
	\varphi^{U\text{irr}} & = R - U + {\textstyle \sum_r} M_r \ , 
\end{align}
\end{subequations}
where $\varphi^{U\text{irr}}$ is the fully $U$-irreducible part of $\Gamma$.
The $U$ subtractions ensure that the bare vertex $U$, which is contained once in each $\nabla_r$ but not in $\varphi^{U\text{irr}}$, is not over-counted. 
Some low-order diagrams of $\nabla_r$, $M_r$, and $R$ are shown in \Fig{fig:SBE_constituents_PT}.

\begin{figure}
\includegraphics[width=\linewidth]{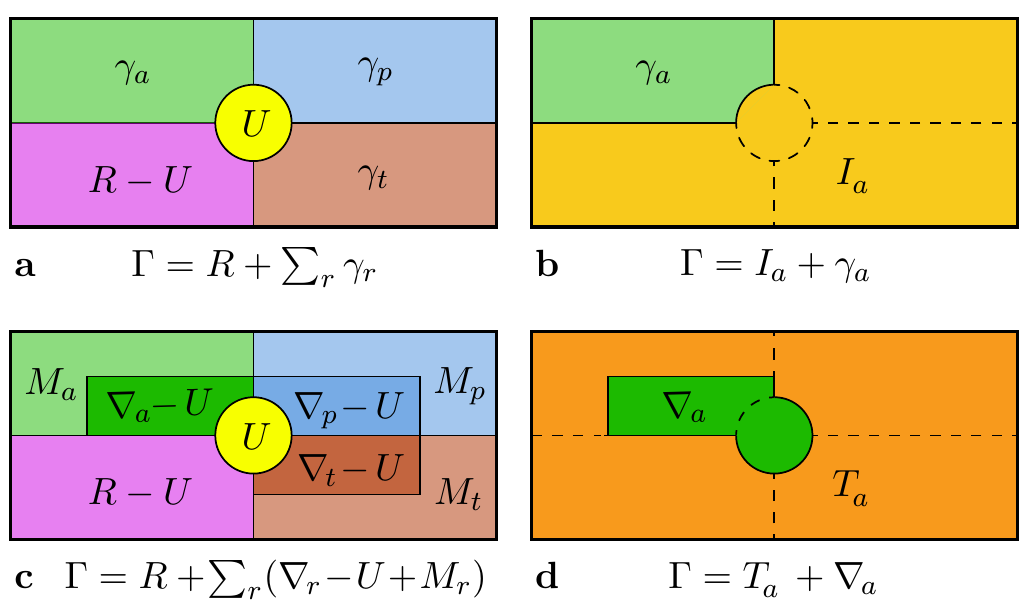}
\caption{Venn diagrams illustrating various ways of splitting the full vertex into 
distinct contributions. Panel \textbf{a} depicts the parquet decomposition \eqref{eq:mfRG:mfRG:Parquet:PE},  \textbf{b} the $\Pia$-reducible part $\gamma_a$ and its complement $I_a$, \textbf{c} the
SBE decomposition \eqref{eq:SBE_diagrammatic_channels} (mimicking Fig.~6 of \cite{Krien2019}), and \textbf{d} the $\Ua$-reducible part $\nabla_a$ and its complement $T_a$. For $r= p,t$, the $\Pir$- and $\Ur$-reducible parts and their complements can be depicted analogously.}
\label{fig:VennDiagr} 
\end{figure}

Just as $\gamma_r$, its parts $\nabla_r$ and $M_r$ satisfy 
Bethe--Salpeter-type equations, which we derive next.
Inserting \Eq{eq:splitgammar} into the full vertex $\Gamma = I_r + \gamma_r$, we split it into a $\Ur$-reducible part, $\nabla_r$,
and a $\Ur$-irreducible remainder, $T_r$:
\begin{subequations}
	\label{subeq:splitGamma-nablar-Iur}
\begin{align}
	\label{eq:splitGamma-nablar-Iur}
	\Gamma & = \nabla_r + T_r , \\
	\label{eq:IrU}
	T_r & = I_r - U + M_r . 
\end{align}
\end{subequations}
The relation between the different decompositions of the full vertex implied by Eqs.~\eqref{eq:splitgammar}--\eqref{subeq:splitGamma-nablar-Iur} is illustrated in \Fig{fig:VennDiagr}.
Inserting Eqs.~\eqref{eq:splitgammar} and  \eqref{eq:splitGamma-nablar-Iur} into either of the two forms of the BSEs \eqref{eq:Bethe-Salpeter-equations}
for $\gamma_r$, we obtain
\begin{align}
	\nonumber
	\nabla_r - U + M_r & = 
I_r \circ \Pi_r \circ \nabla_r  +  I_r \circ \Pi_r \circ  T_r 
	\\  \label{eq:BSE-Ur-version-1}
	& = 
	\nabla_r   \circ \Pi_r \circ I_r  + 
	  T_r \circ \Pi_r \circ I_r . 
\end{align}
This single set of equations can be split into two separate ones,
one for $\nabla_r -U$, the other for $M_r$, 
containing only $\Ur$-reducible or only $\Ur$-irreducible terms,
respectively. The first terms on the right are 
clearly $\Ur$-reducible, since they contain $\nabla_r$.
For the second terms on the right, we write $I_r$ as the sum 
of $U$ and $I_r - U$, yielding $\Ur$-reducible and $\Ur$-irreducible contributions, respectively. We thus obtain two separate sets of equations,
\begin{align}
\nonumber
\nabla_r - U 
	&= 	 I_r\circ\Pi_r \circ \nabla_r + U \circ\Pi_r \circ  T_r 
\label{eq:nabla_r-BSE}	 
\\
	& = \nabla_r\circ\Pi_r \circ I_r + T_r \circ\Pi_r \circ U,\\
\nonumber 
 M_r &= (I_r - U) \circ\Pi_r\circ  T_r  
 \\ \label{eq:M_r-BSE}
 & = 	T_r \circ \Pi_r \circ (I_r-U),
\end{align}
the latter of which corresponds to Eq.~(17) in Ref.~\cite{Krien2021}.
In \Eqs{eq:nabla_r-BSE},
we now bring all $\nabla_r$ contributions to the left, 
\begin{align}
\nonumber 
(\doubleI_r - I_r \circ \Pi_r) \circ \nabla_r  
	&= 	 U \circ (\doubleI_r + \Pi_r  \circ T_r) , 
	\\
\nabla_r \circ (\doubleI_r -  \Pi_r \circ I_r ) 
	&= 	 (\doubleI_r + T_r \circ \Pi_r ) \circ U, 
\end{align}
and solve for $\nabla_r$ by 
evoking the extended BSEs~\eqref{eq:extended-BSE}:
\begin{align}
	\nonumber\nabla_r &= (\doubleI_r+\Gamma\circ\Pi_r)\circ U\circ(\doubleI_r+\Pi_r\circ T_r)\,\\
	&= (\doubleI_r+T_r\circ\Pi_r)\circ U\circ(\doubleI_r+\Pi_r\circ\Gamma).
	\label{eq:nabla_r-long}	
\end{align}
This directly exhibits the $\Ur$-reducibility of $\nabla_r$.

\begin{figure}
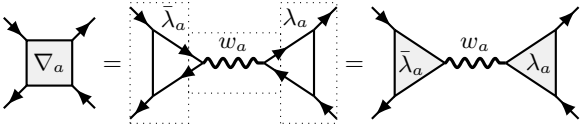

	\begin{align*}
		\tikzm{SBE_decomposition_indices_frequencies_nabla}{
			\fullvertexwithlegs{$\nabla_a$}{0}{0}{1}
		}
		=
		\tikzm{SBE_decomposition_indices_frequencies_boxes}{
			\arrowslefthalffull{0.44}{0}{4./3.*1.1}
			\draw[lineWithArrowCenterEnd] (0,0.44) -- (0.66,0);
			\draw[lineWithArrowCenterCenter] (0.66,0) -- (0,-0.44);
			\draw[linePlain] (0,0.44) -- (0,-0.44);
			\node at (0.28,0.6) {$\bar\lambda_a$};
			\bosonfull{0.66}{0}{1.46}{0}
			\node at (1.06,0.25) {$w_a$};
			\draw[lineWithArrowCenterCenter] (1.46,0) -- (1.46+0.66,0.44);
			\draw[lineWithArrowCenterEnd] (1.46+0.66,-0.44) -- (1.46,0);
			\draw[linePlain] (1.46+0.66,0.44) -- (1.46+0.66,-0.44);
			\arrowsrighthalffull{1.46+0.22}{0}{4./3.*1.1}
			\node at (1.88,0.6) {$\lambda_a$};
			\draw[dotted] (-0.3,-0.8) rectangle (0.47,0.8);
			\draw[dotted] (0.47,-0.4) rectangle (1.68,0.4);
			\draw[dotted] (1.68,-0.8) rectangle (2.42,0.8);
		}
		=
		\tikzm{SBE_decomposition_indices_frequencies}{
			\threepointvertexleftarrows{$\bar\lambda_a$}{0}{0}{1.1}
			\bosonfull{0.66}{0}{1.46}{0}
			\node at (1.06,0.25) {$w_a$};
			\threepointvertexrightarrows{$\lambda_a$}{1.46}{0}{1.1}
		}
	\end{align*}
\caption{Diagrammatic depiction of Eq.~\eqref{eq:nabla_r} (exemplified for the $a$ channel), expressing the $\Ur$-reducible vertex $\nabla_r = \bar{\lambda}_r\protect\fcirc w_r\protect\fcirc\lambda_r$ through two Hedin vertices, $\bar \lambda_r$, $\lambda_r$, and a screened interaction, $w_r$. 
The dashed boxes emphasize that $\bar \lambda_r$, $w_r$, $\lambda_r$ all have four fermionic legs; those of $w_r$ and the outer legs of $\bar \lambda_r$ and $\lambda_r$ are amputated. 
Still, $w_r$ depends on just a single, bosonic frequency and can hence be interpreted as an effective bosonic interaction. 
Its four legs lie pairwise close together, since each pair stems from a bare vertex (see Eq.~\eqref{eq:definition_screened_interaction} and Fig.~\ref{fig:Vertex_frequency_parametrization}c). The two inward-facing legs of both $\bar \lambda_r$ and $\lambda_r$, connecting to 
$w_r$, are therefore also drawn close together, whereas the outward-facing legs are not. To depict this asymmetry in a compact manner, triangles are used on the right. For explicit index summations for all three channels, see \Fig{fig:lambda_w_lambda} in App.~\ref{Appendix:Diagrams_for_SBE}. 
 }
\label{fig:compact-nabla=lambda-w-lambda}
\end{figure}

\begin{figure*}
	\begin{alignat*}{6}
		\tikzm{SBE_decomposition}{
			\fullvertexwithlegs{$\Gamma$}{0}{0}{1}
			\node[above] at (-0.6,0.6) {\small$\nu_2\phantom{'}$};
			\node[above] at (0.6,0.6) {\small$\nu_2'$};
			\node[below] at (-0.6,-0.6) {\small$\nu_1'$};
			\node[below] at (0.6,-0.6) {\small$\nu_1\phantom{'}$};
		}
		&=
		\tikzm{SBE_decomposition_irr}{
			\fullvertexwithlegs{$\varphi^{U\text{irr}}$}{0}{0}{1.3}
			\node[above] at (-0.7,0.7) {\small$\nu_2\phantom{'}$};
			\node[above] at (0.7,0.7) {\small$\nu_2'$};
			\node[below] at (-0.7,-0.7) {\small$\nu_1'$};
			\node[below] at (0.7,-0.7) {\small$\nu_1\phantom{'}$};
		}
		&&+ \!
			\tikzm{SBE_decomposition_a}{
				\threepointvertexleftarrows{$\bar\lambda_a$}{0}{0}{1.1}
				\bosonfull{0.66}{0}{1.46}{0}
				\node at (1.06,0.3) {$w_a$};
				\threepointvertexrightarrows{$\lambda_a$}{1.46}{0}{1.1}
				\node[above] at (-0.2,0.7) {\small$\va \! + \! \tfrac{\wa}{2}$};
				\node[above] at (2.42,0.7) {\small$\va' \! + \! \tfrac{\wa}{2}$};
				\node[below] at (-0.3,-0.7) {\small$\va \! - \! \tfrac{\wa}{2}$};
				\node[below] at (2.42,-0.7) {\small$\va' \! - \! \tfrac{\wa}{2}$};
			}
		\! &&+ \!
			\tikzm{SBE_decomposition_p}{
				\threepointvertexleft{$\bar\lambda_p$}{0}{0}{1.1}
				\bosonfull{0.66}{0}{1.46}{0}
				\node at (1.06,0.3) {$w_p$};
				\node at (1.06,0.74) {};
				\threepointvertexright{$\lambda_p$}{1.46}{0}{1.1}
				\draw[lineWithArrowCenterEnd] (0,-0.44) -- (-0.3,-0.74);
				\draw[lineWithArrowCenterEnd] (2.42,-0.74) -- (2.12,-0.44);
				\draw[lineWithArrowCenterEnd] (0,0.44) to [out=20, in=180] (2.42,0.74);
				\draw[lineWithArrowCenterStart] (-0.3,0.74) to [out=0, in=160] (2.12,0.44);
				\node[above] at (-0.3,0.8) {\small$\tfrac{\wp}{2} \! - \! \vp'$};
				\node[above] at (2.3,0.8) {\small$\tfrac{\wp}{2} \! - \! \vp$};
				\node[below] at (-0.3,-0.7) {\small$\tfrac{\wp}{2} \! + \! \vp$};
				\node[below] at (2.3,-0.7) {\small$\tfrac{\wp}{2} \! + \! \vp'$};
			}
		\! &&+ \!
			\tikzm{SBE_decomposition_t}{
				\threepointvertexupperarrows{$\bar\lambda_t$}{0}{0.73}{1.1}
				\bosonfull{0}{0.4}{0}{-0.4}
				\node at (0.4,0) {$w_t$};
				\threepointvertexlowerarrows{$\lambda_t$}{0}{-0.73}{1.1}
				\node[left] at (-0.7,1.2) {\small$\vt \! + \! \tfrac{\wt}{2}$};
				\node[right] at (0.7,1.2) {\small$\vt \! - \! \tfrac{\wt}{2}$};
				\node[left] at (-0.7,-1.2) {\small$\vt' \! + \! \tfrac{\wt}{2}$};
				\node[right] at (0.7,-1.2) {\small$\vt' \! - \! \tfrac{\wt}{2}$};
			}
		\! &&-2
		\tikzm{SBE_decomposition_Gamma0}{
			\barevertexwithlegs{0}{0}
		}
		\\ 
			\Gamma(\wr,\vr,\vr')
		&=  \varphi^{U\text{irr}}(\wr,\vr,\vr')
		&&+ \qquad \quad \nabla_a(\wa,\va,\va') &&+ 
		\qquad \quad \nabla_p(\wp,\vp,\vp') &&+ 
		\qquad \quad \nabla_t(\wt,\vt,\vt') && -2 U 
	\end{alignat*}
\caption{SBE decomposition of the vertex $\Gamma$ into $\Ur$-irreducible and $\Ur$-reducible contributions, with $r=a,p,t$.
When connecting Hedin vertices to other objects, the two fermionic legs require a $\circ$ connector, the bosonic leg a $\protect\fcirc$ connector.
}
\label{fig:SBE_diagrammatic_channels}
\end{figure*}

We now adopt the $r$ parametrization and note a key structural feature of \Eq{eq:nabla_r-long} for $\nabla_r$: it  contains a central bare vertex $U$,  connected via
$\circ \, \Pi_r \, \circ$ to either 
$\Gamma$ or $T_r$ or both.  We may thus  
pull the frequency-independent $U$ out of the frequency summations,
so that $\circ \, \Pi_r \, \circ$ leads to 
$\fcirc \, \boldI_r \circ \, \Pi_r \, \circ$ or $\circ \,  \Pi_r  \, \circ \boldI_r \fcirc$, where the multiplication with $\boldI_r$ includes a sum over an internal fermionic frequency (recall \Eqs{subeq:shorthand-U-Pi-Gamma}, \eqref{subeq:define-Pi-Gamma}
and Fig.~\ref{fig:Vertex_frequency_parametrization}).
Thus, \Eq{eq:nabla_r-long} leads to
\begin{align}
	\nonumber
	\nabla_r &= 
	(\boldI_r+ \Gamma \circ\Pi_r \circ \boldI_r)\fcirc U\fcirc 
	(\boldI_r + \boldI_r \circ \Pi_r \circ T_r) \, \\
	&= ( \boldI_r + T_r \circ \Pi_r \circ \boldI_r) \fcirc U\fcirc(\boldI_r+
	\boldI_r\circ\Pi_r \circ \Gamma ) .
	\label{eq:nabla_r-Hedin}	
\end{align}
In the first or second line, the expressions on the right or left of $\fcirc \, U \, \fcirc$, respectively, are $\Ur$-irreducible. These factors 
are the so-called Hedin vertices \cite{Hedin1965}
(cf.\ Ref.~\cite{Krien2021}, Eq.~(5)),
\begin{subequations}
	\label{eq:Hedin_definitions}
\begin{flalign}
		\bar{\lambda}_r(\wr^\pprime,\vr^\pprime) 
		&\equiv \boldI_r + [T_r \circ \Pi_r \circ \boldI_r] (\wr^\pprime,\vr^\pprime) ,\\
		\lambda_r(\wr^\pprime,\vr') 
		&\equiv \boldI_r + [ \boldI_r\circ\Pi_r \circ T_r ] (\wr^\pprime,\vr') .
\end{flalign}
\end{subequations}
In our notation, the Hedin vertices have four fermionic legs, but (importantly) depend on only two frequencies. 
Indeed, regarding their frequency dependence, they
can
be viewed as the $U$-irreducible, amputated parts of three-point response functions (see App.~\ref{app:3-point-correlators} and Ref.~\cite{Krien2019}).
Then, \Eqs{eq:Hedin_definitions} have the structure of SDEs for a three-point vertex with a bare three-point vertex $\boldI_r$ (cf.\ Refs.~\cite{Kopietz2010, Kugler2018e}). 
Via the Hedin vertices, $\nabla_r$ factorizes into functions of at most two frequency arguments and is thus computationally cheaper than, e.g., $\gamma_r$. Following Refs.~\cite{Krien2019,Krien2021}, we write
\begin{align}
	\nabla_r & = \bar{\lambda}_r\fcirc w_r\fcirc\lambda_r, 
	\label{eq:nabla_r}
\end{align}
where two $\Ur$-irreducible Hedin vertices sandwich a 
$\Ur$-reducible object, $w_r(\omega_r)$ (see Fig.~\ref{fig:compact-nabla=lambda-w-lambda}).
The object $w_r$ depends only on the bosonic frequency $\omega_r$  and can be interpreted 
as a screened interaction.
To find $w_r$ explicitly, we first express \Eq{eq:nabla_r-Hedin} through Hedin vertices,
\begin{align}
	\nonumber
	\nabla_r &= 
	(\boldI_r + \Gamma \circ\Pi_r \circ\boldI_r)\fcirc  U\fcirc \lambda_r \,\\
	&= \bar \lambda_r \fcirc U \fcirc(\boldI_r +
	\boldI_r \circ \Pi_r\circ \Gamma ).
	\label{eq:nabla_r-Hedin-0}	
\end{align}
Then, $\Gamma =  T_r + \nabla_r$ leads to implicit relations for $\nabla_r$: 
\begin{align}
	\nonumber
	\nabla_r &= 
	(\bar \lambda_r + \nabla_r \circ\Pi_r \circ \boldI_r )\fcirc U\fcirc \lambda_r \,\\
	&= \bar \lambda_r \fcirc U\fcirc(\lambda_r +
	\boldI_r\circ\Pi_r\circ \nabla_r ) .
	\label{eq:nabla_r-Hedin-a}	
\end{align}
Next, we insert \Eq{eq:nabla_r} for $\nabla_r$ on both sides to obtain
\begin{align}
	\nonumber
	\bar{\lambda}_r\fcirc w_r\fcirc\lambda_r &= 
	\bar \lambda_r \fcirc 
	( U + w_r \fcirc \lambda_r \circ \Pi_r \circ U)\fcirc \lambda_r \,\\
	&= \bar \lambda_r \fcirc (U + 
	U\circ\Pi_r\circ\bar\lambda_r\fcirc w_r ) \fcirc \lambda_r .
	\label{eq:nabla_r-Hedin-b}	
\end{align}
This implies that $w_r$ satisfies a pair of Dyson equations,
\begin{align}
\label{eq:w_r-Dyson}
	\nonumber w_r &= U + w_r \fcirc  \lambda_r\circ\Pi_r  \circ U
	\\
	&= U + U \circ  \Pi_r\circ\bar\lambda_r  \fcirc w_r , 
\end{align}
which can be formally solved as
\begin{align}
		\nonumber  w_r &= U\fcirc (\boldI_r -  \lambda_r\circ\Pi_r \circ U)^{-1} 
		\\
		&= (\boldI_r - U\circ  \Pi_r \circ\bar\lambda_r )^{-1}\fcirc U
		.
		\label{eq:w_r-Dyson-inverted}
\end{align}

As desired, the screened interaction $w_r$ is manifestly $\Ur$-reducible, 
and depends on only a single, bosonic frequency, $\omega_r$.
To emphasize this fact, \Eq{eq:w_r-Dyson-inverted} can be written as 
\begin{align}
		\nonumber  w_r & = U\fcirc (\boldI_r -  P_r \fcirc U)^{-1}
		\\
		&= (\boldI_r - U\fcirc  P_r )^{-1}\fcirc U,
		\label{eq:w_r-polarization}
\end{align}%
where $P_r (\omega_r)$ is the polarization  \cite{Krien2021},
	\begin{align}
	\label{eq:polarization}
		P_r  & = \lambda_r \circ \Pi_r \circ \boldI_r 
		= \boldI_r \circ \Pi_r \circ \bar \lambda_r  . 
	\end{align}
Regarding frequency dependencies, $w_r$ can be viewed as a bosonic propagator and $P_r$
as a corresponding self-energy; 
\Eq{eq:polarization} then has the structure of a SDE for $P_r$ involving
the bare three-point vertex $\boldI_r$ \cite{Kopietz2010, Kugler2018e}.

Inserting \Eq{eq:nabla_r} for $\nabla_r$
into \Eq{eq:SBE_diagrammatic_channels-Gamma} for $\Gamma$, we arrive at 
the SBE decomposition of the full vertex of Ref.~\cite{Krien2019} in our generalized notation,
\begin{subequations}
\label{eq:SBE-equations}
\begin{align}
\label{eq:FinalSBE-decomposition}
\Gamma & = \varphi^{U\text{irr}}+
{\textstyle \sum_r} \bar{\lambda}_r\fcirc w_r\fcirc\lambda_r -2U, 
\end{align}
depicted diagrammatically in Fig.~\ref{fig:SBE_diagrammatic_channels}.
For ease of reference, we gather all necessary relations for its ingredients:
\begin{flalign}
\label{eq:w_r-Dyson-in-list}
w_r &= U + U \fcirc P_r \fcirc w_r = U + w_r \fcirc P_r \fcirc U,
&
\\
\label{eq:P_r-in-list}
P_r  & = \lambda_r \circ \Pi_r \circ \boldI_r = \boldI_r \circ \Pi_r \circ \bar \lambda_r  ,
&
\\
\label{eq:summary-barlambda}
\bar{\lambda}_r &= \boldI_r + T_r \circ \Pi_r \circ \boldI_r,
&
\\
\label{eq:lambda_r-list}
\lambda_r &= \boldI_r + \boldI_r \circ \Pi_r \circ T_r,
&
\\
\label{eq:summary-Tr}
T_r &= \Gamma - \bar{\lambda}_r \fcirc w_r \fcirc \lambda_r,
&
\\
\label{eq:summary-varphiUirr}
\varphi^{U\text{irr}} & = R - U + {\textstyle \sum_r} M_r ,
&
\\
M_r &= (T_r \!-\! M_r ) 
\circ\Pi_r\circ  T_r 
= T_r \circ \Pi_r \circ 
(T_r \!-\! M_r) 
\label{eq:M_r-BSE-lambda-w-lambda-version} .
\hspace{-0.5cm} & 
\end{flalign}
\end{subequations}
We collectively call \Eqs{eq:SBE-equations} the \textsl{SBE equations}. Together with the SDE for the self-energy and an input for the two-particle irreducible vertex $R$, the SBE equations are a self-consistent set of equations and thus fully define the four-point vertex $\Gamma$. They can either be solved 
self-consistently (as by Krien et al.\ in Refs. \cite{Krien2019b,Krien2020a,Krien2020b,Krien2021}, where an analogous 
set of equations was set up), or via multiloop flow equations, derived in Sec.~\ref{sec:mfRG_equations_SBE}.

To conclude this section, let us point out the physical meaning of $\bar\lambda_r$, $w_r$, $\lambda_r$ by showing their relation to three-point vertices and susceptibilities. For this, a symmetric expression for $w_r$ is needed, which
can be obtained by comparing \Eqs{eq:nabla_r} and \eqref{eq:nabla_r-Hedin-0}
to deduce
\begin{subequations}
	\label{eq:definition_Hedin-vertex}
	\begin{align}
		\bar \lambda_r \fcirc w_r & = U + \Gamma \circ \Pi_r \circ U,
		\\ 
		w_r \fcirc \lambda_r  & = U + U \circ \Pi_r \circ \Gamma, 
	\end{align}
\end{subequations}
and inserting these into the Dyson equations~\eqref{eq:w_r-Dyson}:
\begin{align}
	w_r &= U + U\circ \Pi_r \circ U + 
	U \circ \Pi_r\circ\Gamma\circ\Pi_r \circ U.
	\label{eq:definition_screened_interaction}
\end{align}
Equations \eqref{eq:definition_Hedin-vertex} and \eqref{eq:definition_screened_interaction} can be expressed as
\begin{subequations}
	\label{eq:lambda_r-w_r-susceptibility}
	\begin{align}
		[\bar\lambda_r\fcirc w_r](\omega_r,\nu_r) &= \bar\Gamma^{(3)}_r(\omega_r,\nu_r)\fcirc U,\\
		[w_r\fcirc\lambda_r](\omega_r,\nu'_r) &= U\fcirc\Gamma^{(3)}_r(\omega_r,\nu'_r),\\
		w_r(\wr) & =  U + U\fcirc\chi_r(\omega_r)\fcirc U,
		\label{eq:w_r_chi}
	\end{align}
\end{subequations}
where $\bar\Gamma^{(3)}_r$, $\Gamma^{(3)}_r$ represent full three-point vertices and $\chi_r$ susceptibilities, defined by
\begin{subequations}
	\label{eq:3point-susceptibility}
	\begin{align}
		\label{eq:barGamma(3)}
		\bar\Gamma^{(3)}_r(\omega_r,\nu_r) &= \boldI_r + [\Gamma\circ\Pi_r\circ\boldI_r](\omega,\nu_r),\\
		\label{eq:Gamma(3)}
		\Gamma^{(3)}_r(\omega_r,\nu'_r) &= \boldI_r + [\boldI_r\circ\Pi_r\circ\Gamma](\omega_r,\nu'_r), 
		\\ 
		\label{eq:susceptibility}
		\chi_r(\omega_r) &= [\boldI_r\circ\Pi_r\circ\boldI_r](\omega_r)
		\\
		\nonumber&\quad + [\boldI_r\circ\Pi_r\circ\Gamma\circ\Pi_r\circ\boldI_r](\omega_r).
	\end{align}
\end{subequations}
(The bare vertices were pulled out in front of the frequency sums,
exploiting their frequency independence.) The relation of $\bar\Gamma^{(3)}_r$ and $\Gamma^{(3)}_r$ to three-point correlators and response functions is described in App.~\ref{app:3-point-correlators}; the relation of $\chi_r$ to physical susceptibilities
for a local bare interaction $U$
is discussed in App.~\ref{Appendix:susceptibilities}.

\subsection{SBE mfRG from parquet mfRG}
\label{sec:mfRG_equations_SBE}

\begin{figure*}
	\begin{align*}
		\tikzm{SBE-flow_eq-dgamma1_l}{
			\fullvertexwithlegs{$\dot{\gamma}_a$}{0}{0}{1.1}
		}
		&= 
		\partial_\Lambda
		\tikzm{SBE-flow_eq-dgamma1_l1}{
			\threepointvertexleftarrows{}{0}{0}{3/4}
			\bosonfull{0.45}{0}{1.25}{0}
			\threepointvertexrightarrows{}{1.25}{0}{3/4}
		}
		+
		\tikzm{SBE-flow_eq-dgamma1_l2}{
			\fullvertexwithlegs{$\dot{M}_a$}{0}{0}{1.1}
		}
		=
		\tikzm{SBE-flow_eq-dgamma1_l3}{
			\threepointvertexleftarrows{}{0}{0}{3/4}
			\node at (0.225,0.4) {\tiny $\bullet$};
			\bosonfull{0.45}{0}{1.25}{0}
			\threepointvertexrightarrows{}{1.25}{0}{3/4}
		}
		+
		\tikzm{SBE-flow_eq-dgamma1_l4}{
			\threepointvertexleftarrows{}{0}{0}{3/4}
			\bosonfull{0.45}{0}{1.25}{0}
			\node at (0.85,0.4) {\tiny $\bullet$};
			\threepointvertexrightarrows{}{1.25}{0}{3/4}
		}
		+
		\tikzm{SBE-flow_eq-dgamma1_l5}{
			\threepointvertexleftarrows{}{0}{0}{3/4}
			\bosonfull{0.45}{0}{1.25}{0}
			\threepointvertexrightarrows{}{1.25}{0}{3/4}
			\node at (1.475,0.4) {\tiny $\bullet$};
		}
		+
		\tikzm{SBE-flow_eq-dgamma1_l6}{
			\fullvertexwithlegs{$\dot{M}_a$}{0}{0}{1.1}.
		}
		\\
		\tikzm{SBE-flow_eq-dgamma1_r}{
			\fullvertexwithlegs{$\dot{\gamma}_a^{(1)}$}{0}{0}{1.1}
		}
		&=
		\tikzm{SBE-flow_eq-dgamma1_r2}{
			\arrowslefthalffull{0}{0}{1.1}
			\fullvertex{$T_a$}{0}{0}{1.1}
			\abubblefullfulldiff{0.33}{0}{0.8/1.2}{1.1}{1}
			\threepointvertexleft{}{1.13}{0}{3/4}
			\bosonfull{1.58}{0}{2.18}{0}
			\threepointvertexrightarrows{}{2.18}{0}{3/4}
		}
		+
		\tikzm{SBE-flow_eq-dgamma1_r1}{
			\threepointvertexleftarrows{}{0}{0}{3/4}
			\bosonfull{0.45}{0}{1.05}{0}
			\threepointvertexright{}{1.05}{0}{3/4}
			\abubblefullfulldiff{1.5}{0}{0.8/1.2}{1}{1}
			\threepointvertexleft{}{2.3}{0}{3/4}
			\bosonfull{2.75}{0}{3.55}{0}
			\threepointvertexrightarrows{}{3.35}{0}{3/4}
		}
		+
		\tikzm{SBE-flow_eq-dgamma1_r3}{
			\threepointvertexleftarrows{}{0}{0}{3/4}
			\bosonfull{0.45}{0}{1.05}{0}
			\threepointvertexright{}{1.05}{0}{3/4}
			\abubblefullfulldiff{1.5}{0}{0.8/1.2}{1}{1.1}
			\fullvertex{$T_a$}{2.63}{0}{1.1}
			\arrowsrighthalffull{2.63}{0}{1.1}
		}
		+
		\tikzm{SBE-flow_eq-dgamma1_r4}{
			\arrowslefthalffull{0}{0}{1.1}
			\fullvertex{$T_a$}{0}{0}{1.1}
			\abubblefullfulldiff{0.33}{0}{0.8/1.2}{1.1}{1.1}
			\fullvertex{$T_a$}{1.46}{0}{1.1}
			\arrowsrighthalffull{1.46}{0}{1.1}
		}
	\end{align*}
	\caption{SBE decomposition of the left and right sides of the $1\ell$ flow equation \eqref{eq:mfRG-equations_gamma_r_1l} (\Fig{fig:mfRG_equations_a_channel}) in the $a$ channel. The first line depicts \Eq{eq:derivative_gamma_r}, the second \Eq{eq:derivation_mfRG_SBE_1l}. Equating terms with matching structure yields \Eq{eq:SBE-mfRG_1l}, depicted in Fig.~\ref{fig:SBE_multiloop_equations}, first line.}
	\label{fig:Flow_equations_SBE:1l_lhs_rhs}
\end{figure*}

Having defined all the SBE ingredients, we are now ready to derive mfRG flow equations for them---the main goal of this work. Our strategy is to insert the SBE decomposition
of \Eqs{eq:splitgammar} and \eqref{eq:SBE_diagrammatic_channels} into the parquet mfRG flow equations \eqref{eq:mfRG-equations_gamma_r}  for the $\Pir$-reducible vertices $\gamma_r$.
An alternative derivation, starting directly from the SBE equations 
\eqref{eq:SBE-equations}, is given in Sec.~\ref{Appendix:alt_deriv2}.

We begin by differentiating the decomposition 
of the $\Pi_r$-reducible vertex $\gamma_r = \bar{\lambda}_r\fcirc w_r \fcirc\lambda_r-U + M_r$ (\Eq{eq:splitgammar}) \wrt the flow
parameter. Since $\dot{U}=0$ (the bare vertex does not depend on the regulator), we obtain 
\begin{flalign}
	\dot{\gamma}_r &= \dot{\bar{\lambda}}_r\fcirc w_r \fcirc \lambda_r + \bar{\lambda}_r\fcirc\dot{w}_r\fcirc\lambda_r + \bar{\lambda}_r\fcirc w_r\fcirc\dot{\lambda}_r + \dot{M}_r .  \hspace{-1cm} & 
	\label{eq:derivative_gamma_r}
\end{flalign}
The loop expansion $\dot \gamma_r = \sum_{\ell}\dot \gamma_{r}^{(\ell)}$ implies similar expansions for $\dot w_r$, $\dot{\bar \lambda}_r$, $\dot{\lambda}_r$, and $\dot M_r$. 
Each term at a given loop order $\ell$
can be found from the mfRG flow \eqref{eq:mfRG-equations_gamma_r} for $\dot \gamma_r^{(\ell)}$, by inserting the decomposition of the full vertex, $\Gamma=\bar{\lambda}_r\fcirc w_r\fcirc\lambda_r+ T_r$ (\Eq{eq:splitGamma-nablar-Iur}) on the right of \Eqs{eq:mfRG-equations_gamma_r}.

The $1\ell$ flow equation \eqref{eq:mfRG-equations_gamma_r_1l} for $\dot \gamma_r^{(1)}$ has four contributions (shown
diagrammatically for $\gamma_a^{(1)}$ in Fig.~\ref{fig:Flow_equations_SBE:1l_lhs_rhs}): 
\begin{align}
	\nonumber
	\dot{\gamma}_r^{(1)} &= \left(\bar{\lambda}_r\fcirc w_r\fcirc\lambda_r
	+ T_r\right)\circ\dot{\Pi}_r\circ\left(\bar{\lambda}_r\fcirc w_r\fcirc\lambda_r + T_r \right)\\
	\nonumber &= 
	 T_r\circ\dot{\Pi}_r\circ\bar{\lambda}_r\fcirc w_r\fcirc\lambda_r \\ 
	 & \quad \; + \bar{\lambda}_r\fcirc w_r\fcirc\lambda_r\circ\dot{\Pi}_r\circ\bar{\lambda}_r\fcirc w_r \fcirc\lambda_r \nonumber \\
	&\quad \;  
	+ 	\bar{\lambda}_r\fcirc w_r\fcirc\lambda_r\circ\dot{\Pi}_r\circ T_r
    + T_r\circ\dot{\Pi}_r\circ T_r .
	\label{eq:derivation_mfRG_SBE_1l}
\end{align}
By matching terms in \Eqs{eq:derivative_gamma_r} and \eqref{eq:derivation_mfRG_SBE_1l} containing 
factors of $ \bar \lambda_r$ and $\lambda_r$ or not,
we obtain the $1\ell$ SBE flow:
\begin{subequations}
\label{eq:SBE-mfRG_details}
	\begin{align}
	\nonumber \dot{w}_r^{(1)} & =w_r\fcirc\lambda_r\circ\dot{\Pi}_r\circ\bar{\lambda}_r\fcirc w_r,
		\\
		\nonumber
		\dot{\bar{\lambda}}_r^{(1)}&= T_r\circ\dot{\Pi}_r\circ\bar{\lambda}_r,
		\\ \nonumber 
	\dot{\lambda}_r^{(1)} &= \lambda_r\circ\dot{\Pi}_r\circ T_r,
	\\
	\dot{M}_r^{(1)}&=T_r\circ\dot{\Pi}_r\circ T_r.
		\label{eq:SBE-mfRG_1l}
	\end{align}
This reproduces the $1\ell$ SBE flow derived in Ref.~\cite{Bonetti2021} (their Eq.~(18)).
The higher-loop terms can be found similarly from 
$\dot \gamma_{r}^{(2)}$ and $\dot \gamma_{r}^{(\ell +2)}$
of \Eqs{eq:mfRG-equations_gamma_r_2l} and 
\eqref{eq:mfRG-equations_gamma_r_l}.
For each loop order $\ell$, the 
$\dot{\gamma}_{\bar{r}}^{(\ell)}$ factors on the right side
of these equations can be expressed through 
the already known flow of $\dot{w}_{r'}^{(\ell)}$, $\dot{\bar \lambda}_{r'}^{(\ell)}$ $\dot\lambda_{r'}^{(\ell)}$ and $\dot{M}^{(\ell)}_{r'}$. 
We obtain the flow equations ($\ell+2 \geq 3$)
\begin{align}
\nonumber
\dot{w}_r^{(2)}&=0,\\
\nonumber
\dot{\bar{\lambda}}_r^{(2)}&=\dot{\gamma}_{\bar{r}}^{(1)}\circ\Pi_r\circ\bar{\lambda}_r,\\
\nonumber
\dot{\lambda}_r^{(2)}&=\lambda_r\circ\Pi_r\circ\dot{\gamma}_{\bar{r}}^{(1)},\\		
\dot{M}_r^{(2)}&=\dot{\gamma}_{\bar{r}}^{(1)}\circ\Pi_r\circ T_r+ T_r\circ\Pi_r\circ\dot{\gamma}_{\bar{r}}^{(1)},
\label{eq:SBE-mfRG_2l}
\\
\nonumber
\dot{w}_r^{(\ell+2)}&= w_r\fcirc\lambda_r\circ\Pi_r\circ\dot{\gamma}_{\bar{r}}^{(\ell)}\circ\Pi_r\circ\bar{\lambda}_r\fcirc w_r,
\\
\nonumber
\dot{\bar{\lambda}}_r^{(\ell+2)}&=\dot{\gamma}_{\bar{r}}^{(\ell+1)}\circ\Pi_r\circ\bar{\lambda}_r
+ T_r\circ\Pi_r\circ\dot{\gamma}_{\bar{r}}^{(\ell)}\circ\Pi_r\circ\bar{\lambda}_r,\\
\nonumber
\dot{{\lambda}}_r^{(\ell+2)}&=\lambda_r\circ\Pi_r\circ\dot{\gamma}_{\bar{r}}^{(\ell)}\circ\Pi_r\circ T_r
+\lambda_r\circ\Pi_r\circ\dot{\gamma}_{\bar{r}}^{(\ell+1)},\\
\nonumber\dot{M}_r^{(\ell+2)}&=\dot{\gamma}_{\bar{r}}^{(\ell+1)}\circ\Pi_r\circ T_r
+ T_r\circ\Pi_r\circ\dot{\gamma}_{\bar{r}}^{(\ell)}\circ\Pi_r\circ T_r\\
&\quad +  T_r\circ\Pi_r\circ\dot{\gamma}_{\bar{r}}^{(\ell+1)}.
\label{eq:SBE-mfRG_l}
\end{align}
\end{subequations}
Here, $\dot\gamma_{\bar{r}}^{(\ell)}$, required for the flow at loop orders $\ell+1$ and $\ell+2$, can directly be constructed from the SBE ingredients using \Eq{eq:derivative_gamma_r}. 
Similarly as in \Eq{eq:mfRG-equations_gamma_r}, all terms at loop order $\ell$ contain $\ell-1$ factors of $\Pi$ and one $\dot \Pi$, now connecting the renormalized objects $w_r$, $\bar \lambda_r$,  $\lambda_r$, $T_r$.

\begin{figure*}
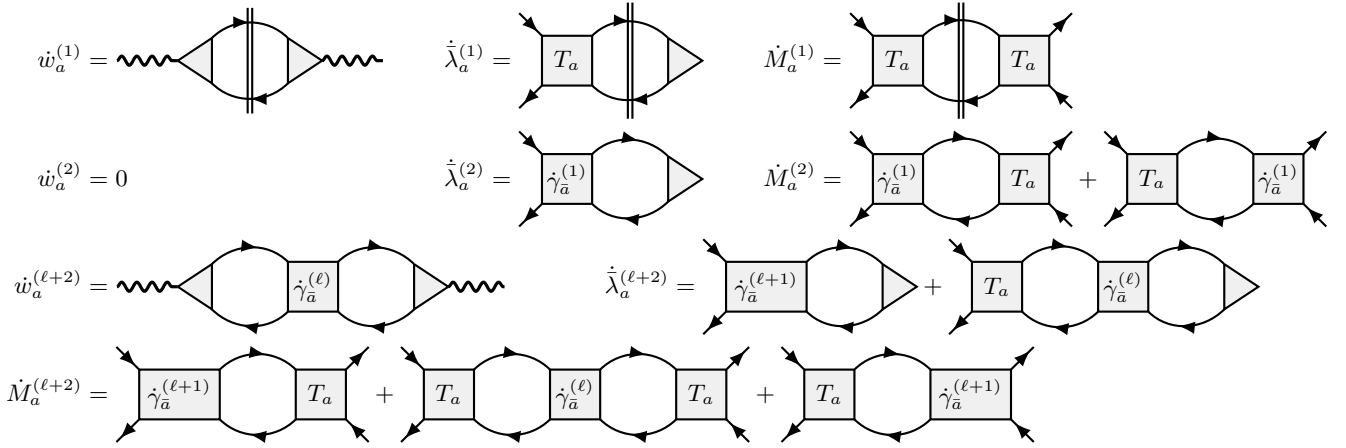

	\begin{align*}
		\dot{w}_a^{(1)}
		&=
		\tikzm{SBE-flow_eq-dw1}{
			\bosonfull{0}{0}{0.8}{0}
			\threepointvertexright{}{0.8}{0}{3/4}
			\abubblefullfulldiff{1.25}{0}{1/1.2}{1}{1}
			\threepointvertexleft{}{2.25}{0}{3/4}
			\bosonfull{2.7}{0}{3.5}{0}
		}
		\ & 
		\quad
		\dot{\bar\lambda}_a^{(1)}
		&=
		\tikzm{SBE-flow_eq-dlambda1}{
			\arrowslefthalffull{0}{0}{1.1}
			\fullvertex{$T_a$}{0}{0}{1.1}
			\abubblefullfulldiff{0.33}{0}{1/1.2}{1.1}{1}
			\threepointvertexleft{}{1.33}{0}{3/4}
		}
		\ & 
		\quad
		\dot{M}^{(1)}_a
		&=
		\tikzm{SBE-flow_eq-dphi1}{
			\arrowslefthalffull{0}{0}{1.1}
			\fullvertex{$T_a$}{0}{0}{1.1}
			\abubblefullfulldiff{0.33}{0}{1/1.2}{1.1}{1.1}
			\fullvertex{$T_a$}{1.66}{0}{1.1}
			\arrowsrighthalffull{1.66}{0}{1.1}
		}
		\\
		\dot{w}_{a}^{(2)}
		&=
		0 \ &
		\quad
		\dot{\bar\lambda}_{a}^{(2)}
		&=
		\tikzm{SBE-flow_eq-dlambda2}{
			\arrowslefthalffull{0}{0}{1.1}
			\fullvertex{$\dot{\gamma}_{\bar{a}}^{(1)}$}{0}{0}{1.1}
			\abubblefullfull{0.33}{0}{1/1.2}{1.1}{1}
			\threepointvertexleft{}{1.33}{0}{3/4}
		}
		\ & 
		\quad
		\dot{M}^{(2)}_a
		&=
		\tikzm{SBE-flow_eq-dphi2_L}{
			\arrowslefthalffull{0}{0}{1.1}
			\fullvertex{$\dot{\gamma}_{\bar{a}}^{(1)}$}{0}{0}{1.1}
			\abubblefullfull{0.33}{0}{1/1.2}{1.1}{1.1}
			\fullvertex{$T_a$}{1.66}{0}{1.1}
			\arrowsrighthalffull{1.66}{0}{1.1}
		}
		+
		\tikzm{SBE-flow_eq-dphi2_R}{
			\arrowslefthalffull{0}{0}{1.1}
			\fullvertex{$T_a$}{0}{0}{1.1}
			\abubblefullfull{0.33}{0}{1/1.2}{1.1}{1.1}
			\fullvertex{$\dot{\gamma}_{\bar{a}}^{(1)}$}{1.66}{0}{1.1}
			\arrowsrighthalffull{1.66}{0}{1.1}
		}
		\\
		\dot{w}_{a}^{(\ell+2)}
		&=
		\tikzm{SBE-flow_eq-dw3}{
			\bosonfull{0}{0}{0.8}{0}
			\threepointvertexright{}{0.8}{0}{3/4}
			\abubblefullfull{1.25}{0}{1/1.2}{1}{1.1}
			\fullvertex{$\dot{\gamma}_{\bar{a}}^{(\ell)}$}{2.58}{0}{1.1}
			\abubblefullfull{2.91}{0}{1/1.2}{1.1}{1}
			\threepointvertexleft{}{3.91}{0}{3/4}
			\bosonfull{4.36}{0}{5.1}{0}
		}
		\qquad \qquad
		\dot{\bar\lambda}_{a}^{(\ell+2)}
		=
		\tikzm{SBE-flow_eq-dlambda3_L}{
			\arrowslefthalffull{0}{0}{1.1}
			\fullvertexwide{$\dot{\gamma}_{\bar{a}}^{(\ell+1)}$}{0}{0}{1.1}{0.4}
			\abubblefullfull{0.73}{0}{1/1.2}{1.1}{1}
			\threepointvertexleft{}{1.73}{0}{3/4}
		}
		+
		\tikzm{SBE-flow_eq-dlambda3_C}{
			\arrowslefthalffull{0}{0}{1.1}
			\fullvertex{$T_a$}{0}{0}{1.1}
			\abubblefullfull{0.33}{0}{1/1.2}{1.1}{1.1}
			\fullvertex{$\dot{\gamma}_{\bar{a}}^{(\ell)}$}{1.66}{0}{1.1}
			\abubblefullfull{1.99}{0}{1/1.2}{1.1}{1}
			\threepointvertexleft{}{2.99}{0}{3/4}
		}
		\hspace{-20cm}
		\\
		\dot{M}^{(\ell+2)}_a
		&=
		\tikzm{SBE-flow_eq-dphi3_L}{
			\arrowslefthalffull{0}{0}{1.1}
			\fullvertexwide{$\dot{\gamma}_{\bar{a}}^{(\ell+1)}$}{0}{0}{1.1}{0.4}
			\abubblefullfull{0.73}{0}{1/1.2}{1.1}{1.1}
			\fullvertex{$T_a$}{2.06}{0}{1.1}
			\arrowsrighthalffull{2.06}{0}{1.1}
		}
		+
		\tikzm{SBE-flow_eq-dphi3_C}{
			\arrowslefthalffull{0}{0}{1.1}
			\fullvertex{$T_a$}{0}{0}{1.1}
			\abubblefullfull{0.33}{0}{1/1.2}{1.1}{1.1}
			\fullvertex{$\dot{\gamma}_{\bar{a}}^{(\ell)}$}{1.66}{0}{1.1}
			\abubblefullfull{1.99}{0}{1/1.2}{1.1}{1.1}
			\fullvertex{$T_a$}{3.32}{0}{1.1}
			\arrowsrighthalffull{3.32}{0}{1.1}
		}
		+
		\tikzm{SBE-flow_eq-dphi3_R}{
			\arrowslefthalffull{0}{0}{1.1}
			\fullvertex{$T_a$}{0}{0}{1.1}
			\abubblefullfull{0.33}{0}{1/1.2}{1.1}{1.1}
			\fullvertexwide{$\dot{\gamma}_{\bar{a}}^{(\ell+1)}$}{1.66}{0}{1.1}{0.4}
			\arrowsrighthalffull{2.06}{0}{1.1}
		}
		\hspace{-20cm}
	\end{align*}
	\caption{Multiloop flow equations \eqref{eq:SBE-mfRG_details} for the ingredients of the SBE decomposition in the $a$ channel.
		\label{fig:SBE_multiloop_equations}
	}
\end{figure*}

The SBE mfRG flow equations \eqref{eq:SBE-mfRG_details} are the most important result of this work. For the $a$ channel, they are depicted diagrammatically in \Fig{fig:SBE_multiloop_equations}. Equations~\eqref{eq:SBE-mfRG_details} can be condensed into more compact ones, giving the full flow (summed over all loop orders, $\dot w_r = \sum_{\ell \ge 1} \dot w_r^{(\ell)}$, etc.) of the SBE ingredients; see the next section.
The multiloop flow equation for the self-energy \cite{Kugler2018a,Kugler2018e} is given in \Eq{eq:mfRG:mfRG:Selfenergy_flow}.

\subsection{SBE mfRG from SBE equations}
\label{Appendix:alt_deriv2}

In the previous section, we derived the SBE mfRG flow equations by inserting the SBE decomposition into the known parquet mfRG flow equations of the two-particle reducible vertices $\gamma_r$.
They can also be derived without prior knowledge on the flow of $\gamma_r$,
by using the techniques of Ref.~\cite{Kugler2018e}.

In the parquet setting of Ref.~\cite{Kugler2018e},
one can view the $\Pi$-$r$-irreducible vertex $I_r$ as the key ingredient for all equations related to channel $r$.
In step (i), one uses $I_r$ to generate $\gamma_r$ and thus $\Gamma$ through a BSE.
Then, a post-processing of attaching and closing external legs yields (ii) (full) three-point vertices $\bar\Gamma_r^{(3)}$, $\Gamma^{(3)}_r$ and (iii) a susceptibility $\chi_r$.
The SBE setting can be understood in close analogy, with the only exception that one purposefully avoids generating $U$-$r$-reducible contributions, because these can (more efficiently) be constructed via $\nabla_r = \bar \lambda_r \fcirc w_r \fcirc \lambda_r$.
To exclude $U$-$r$-reducible contributions, one uses in step (i)
$I_r-U$ to generate $M_r$ and thus $T_r$ through a BSE.
The same post-processing as before yields (ii) $\bar\lambda_r$, $\lambda_r$ and then (iii) $w_r$ or $P_r$.

Because of this structural analogy, the SBE mfRG flow equations can be derived in the exact same fashion as the parquet mfRG flow equation of Ref.~\cite{Kugler2018e}. One merely has to replace the variables according to the dictionary
\begin{alignat}{3}
	\label{eq:correspondences-3-point}
I_r &\to I_r - U 
, & \quad\quad
\gamma_r &\to M_r
, & \quad\quad
\Gamma &\to T_r
,
\nonumber
\\
\bar\Gamma^{(3)}_r &\to \bar\lambda_r
, & \quad\quad
\Gamma^{(3)}_r &\to \lambda_r
, & \quad\quad
\chi_r &\to P_r
.
\end{alignat} 
For clarity, we now spell out the structural analogies between the original parquet formalism and its SBE version,
presenting similarly-structured expressions in pairs of equations, a) and b).
For both approaches, the full vertex can be decomposed in several ways: 
\begin{subequations}
	\label{subeq:define-starting-point2}
	\begin{align}
		\Gamma  & = R + \sum_r \gamma_r = I_r + \gamma_r , \\ 
		\label{eq:define-starting-point-Gamma-1}	
		\Gamma  & = R + \sum_r M_r + \sum_r (\nabla_r - U) = T_r + \nabla_r . 
	\end{align}
\end{subequations}
Here, $\gamma_r$ and $M_r$ satisfy analogous BSEs,
\begin{subequations}
	\label{eq:BetheSalpeter-Tr-Mr2}
	\begin{align}
		\gamma_r & = I_r \circ \Pi_r \circ \Gamma
		,
		\label{eq:BetheSalpeter-Ir-gammar}
		\\
		\label{eq:BetheSalpeter-Tr-Mr-b2}
		M_r & = (I_r-U) \circ \Pi_r \circ T_r ,
	\end{align}
\end{subequations}
where the objects on the left reappear on the right through
\begin{subequations}
	\label{eq:BetheSalpeter-recursion2}
	\begin{align}
		\Gamma & = I_r + \gamma_r
		,
		\\
		\label{eq:T_r(M_r)2}
		T_r & = (I_r-U) + M_r
		. 
	\end{align}
\end{subequations}
Relations \eqref{eq:BetheSalpeter-Tr-Mr2} and \eqref{eq:BetheSalpeter-recursion2} are used for step (i). Differentiation of \Eq{eq:BetheSalpeter-Ir-gammar} yields the mfRG flow of $\dot{\gamma}_r$ as in Eq.~(10) and Fig.~2(a) of Ref.~\cite{Kugler2018e}.
Here, we replace the variables as above and start by differentiating \Eq{eq:BetheSalpeter-Tr-Mr-b2}:
	\begin{align}
		\label{eq:alt_der_dot_Mr_start}
		\dot{M}_r &= \dot{I}_r \circ \Pi_r \circ T_r + (I_r-U)\circ\dot{\Pi}_r\circ T_r \nonumber \\
		\nonumber&\phantom{=} + (I_r-U)\circ\Pi_r\circ\dot{I}_r + (I_r-U)\circ\Pi_r\circ\dot{M}_r\\
		\nonumber\Rightarrow \dot{M}_r &= (\doubleI_r-(I_r-U)\circ\Pi_r)^{-1}\circ\left[\dot{I}_r \circ \Pi_r \circ T_r\right.\\
		&\phantom{=}\left.\quad + (I_r-U)\circ\dot{\Pi}_r\circ T_r + (I_r-U)\circ\Pi_r\circ\dot{I}_r\right].
\end{align}%
For the first argument of \Eq{eq:alt_der_dot_Mr_start}, 
we used $\partial_\Lambda (I_r - U) = \dot{I}_r$,
as $\dot{U}=0$. Next, we use the extended BSE $\doubleI_r + T_r\circ\Pi_r = \left(\doubleI_r-(I_r-U)\circ\Pi_r\right)^{-1}$ for $M_r$, cf.\ \Eqs{eq:extended-BSE} and \eqref{eq:BetheSalpeter-Tr-Mr2}. Recollecting the terms, we obtain
the flow of $\dot{M}_r$ as
\begin{align}
		\dot{M}_r &= T_r \circ \dot\Pi_r \circ T_r 
		+ \dot{I}_r \circ \Pi_r \circ T_r
		\nonumber \\
		& \qquad +
		T_r \circ \Pi_r \circ \dot{I}_r \circ \Pi_r \circ T_r
		+ T_r \circ \Pi_r \circ \dot{I}_r.
		\label{eq:alt_der_dot_Mr2}
\end{align}
A loop expansion with $\dot I_r = \dot \gamma_{\bar r} = \sum_{\ell} \dot \gamma_{\bar{r}}^{(\ell)}$
then yields our \Eqs{eq:SBE-mfRG_details} and \Fig{fig:SBE_multiloop_equations}. 

For step (ii), we have the analogous relations
\begin{subequations}
\begin{alignat}{2}
\hspace{-0.5cm}
\bar{\Gamma}^{(3)}_r &= \boldI_r + \Gamma \circ \Pi_r \circ \boldI_r
, & 
\Gamma^{(3)}_r & = \boldI_r + \boldI_r \circ \Pi_r \circ \Gamma
,
\label{eq:SDE_Gamma3}
\\
\bar{\lambda}_r &= \boldI_r + T_r \circ \Pi_r \circ \boldI_r
, & 
\lambda_r & = \boldI_r + \boldI_r \circ \Pi_r \circ T_r
\label{eq:SDE_lambda}
.
\end{alignat}
\end{subequations}
Differentiation of \Eq{eq:SDE_Gamma3} yields the mfRG flow of $\Gamma^{(3)}_r$ as in Eq.~(42) and Fig.~7 of Ref.~\cite{Kugler2018e}.
Here, we again replace the variables as above and differentiate \Eq{eq:SDE_lambda}:
	\begin{align}
		\nonumber		
		\dot{\bar{\lambda}}_r &= \dot{T}_r \circ \Pi_r \circ\boldI_r + T_r \circ \dot{\Pi}_r\circ\boldI_r,
		\\
		\dot{\lambda}_r &= \boldI_r\circ\dot{\Pi}_r \circ T_r + \boldI_r\circ\Pi_r \circ \dot{T}_r.
 		\label{eq:BetheSalpeter-Hedin2}
		\end{align}
As $\dot{T}_r = \dot{I}_r + \dot{M}_r$ (cf.~\Eq{eq:T_r(M_r)2}), we insert the flow equation~\eqref{eq:alt_der_dot_Mr2} for $\dot{M}_r$ into \Eq{eq:BetheSalpeter-Hedin2} and use again \Eq{eq:SDE_lambda}
This yields the  flow equations
	\begin{align}
		\nonumber 
		\dot{\bar{\lambda}}_r &= T_r \circ \dot{\Pi}_r \circ \bar\lambda_r + \dot{I}_r \circ \Pi_r \circ \bar\lambda_r + T_r \circ {\Pi}_r \circ \dot{I}_r \circ \Pi_r \circ \bar\lambda_r,
		\\
		\dot{\lambda}_r &= \lambda_r \circ \dot{\Pi}_r \circ T_r + \lambda_r \circ \Pi_r \circ \dot{I}_r + \lambda_r \circ \Pi_r \circ \dot{I}_r \circ \Pi_r \circ T_r.
	\label{eq:BetheSalpeter-Hedin_final}
	\end{align}
Their loop expansion reproduces \Eqs{eq:SBE-mfRG_details} and \Fig{fig:SBE_multiloop_equations}.

Finally, in step (iii), we have the relations
\begin{subequations}
\begin{alignat}{2}
\chi_r  & = \Gamma^{(3)}_r & \circ \Pi_r \circ \boldI_r & = \boldI_r \circ \Pi_r \circ \bar \Gamma^{(3)},
\label{eq:SDE_chi}
\\
P_r  & = \lambda_r & \circ \Pi_r \circ \boldI_r & = \boldI_r \circ \Pi_r \circ \bar \lambda_r  
\label{eq:SDE_P}
.
\end{alignat}
\end{subequations}
Differentiation of \Eq{eq:SDE_chi} yields the mfRG flow of $\chi_r$ as in Eq.~(44) and Fig.~8 of Ref.~\cite{Kugler2018e}.
Replacing the variables as above one more time, we differentiate \Eq{eq:SDE_P}:
\begin{align}
\label{eq:flow_Pr-start}
\dot{P}_r =  \boldI_r\circ \Pi_r \circ \dot{\bar{\lambda}}_r +  \boldI_r\circ \dot{\Pi}_r \circ \bar{\lambda}_r
.
\end{align}
After inserting \Eqs{eq:SDE_lambda} and \eqref{eq:BetheSalpeter-Hedin_final}, we eventually obtain:
\begin{align}
\label{eq:flow_Pr}
\dot{P}_r &= \lambda_r\circ\left(\dot{\Pi}_r+\Pi_r\circ\dot{I}_r\circ\Pi_r\right)\circ\bar\lambda_r
.
\end{align}

The relation between $\dot{P}_r$ and $\dot{w}_r$ follows from the Dyson equation \eqref{eq:w_r-Dyson-in-list} as
\begin{align}
\dot{w}_r &= U \fcirc \dot{P}_r \fcirc w_r + U \fcirc P_r \fcirc \dot{w}_r.
\end{align}
Solving this for $\dot{w}_r$ yields
\begin{align}
\label{eq:flow_wr}
\dot{w}_r &= (\boldI_r-U\fcirc P_r)^{-1}\fcirc U \fcirc\dot{P}_r\fcirc w_r = w_r\fcirc\dot{P}_r\fcirc w_r,
\end{align}
having inserted the inverted Dyson equation \eqref{eq:w_r-polarization}. A loop expansion of \Eq{eq:flow_Pr} yields:
\begin{align}
	\nonumber\dot{P}^{(1)}_r &= \lambda_r\circ\dot{\Pi}_r\circ\bar\lambda_r,\\
	\nonumber\dot{P}^{(2)}_r &= 0,\\
	\dot{P}^{(\ell+2)}_r &= \lambda_r\circ\Pi_r\circ\dot{\gamma}_{\bar{r}}^{(\ell)}\circ\Pi_r\circ\bar\lambda_r.
	\label{eq:P_r-mfRG}
\end{align}
Inserting the loop expansion $\dot{P}_r^{(\ell)}$ into \Eq{eq:flow_wr} for $\dot{w}_r$ yields
the same flow equation for $w_r$ as in our \Eqs{eq:SBE-mfRG_details} and \Fig{fig:SBE_multiloop_equations}.

Depending on the specific model, it can be more efficient to calculate the flow of the polarization, $\dot{P}_r$, by \Eqs{eq:P_r-mfRG} instead of the flow of the screened interaction, $\dot{w}_r$, by \Eqs{eq:SBE-mfRG_details}. The screened interaction on the contrary can be obtained by the inverted Dyson equations \eqref{eq:w_r-polarization}.

Altogether, \Eqs{eq:alt_der_dot_Mr2}, \eqref{eq:BetheSalpeter-Hedin_final}, \eqref{eq:flow_Pr} and \eqref{eq:flow_wr} (with $T_r$ given by $\Gamma-\nabla_{\bar{r}}$, \Eq{eq:define-starting-point-Gamma-1}) build a system of closed fRG equations, as full derivatives of the SBE equations~\eqref{eq:SBE-equations}.
Hence, combined with an appropriate self-energy flow (cf.~\Eq{eq:mfRG:mfRG:Selfenergy_flow} and Ref.~\cite{Kugler2018e}), they yield regulator-independent results. To integrate the flow equations in practice, one employs the mfRG loop expansions \eqref{eq:SBE-mfRG_details} and \eqref{eq:P_r-mfRG}.

\subsection{mfRG flow of the SBE approximation}
\label{sec:Simplified-SBE-flow}

To reduce numerical costs, it may sometimes be desirable to approximate the flow of the vertex treating only objects with less than all three frequency arguments. The simplest choice is to restrict the flow to functions depending on a single frequency. In the present context, this corresponds to keeping all objects except $w_r$ constant.  
With $\dot{\bar\lambda}_r = 0 = \dot\lambda_r$, the flow of the polarization \eqref{eq:flow_Pr-start} is simply
\begin{align}
	\label{eq:Poor-man's-fRG}
	\dot{P}_r = \lambda_r\circ\dot\Pi_r\circ\boldI_r = \boldI_r\circ\dot\Pi_r\circ\bar\lambda_r.
\end{align}
Hence, the flow equations of $P_r$ and $w_r$ completely decouple, and one effectively obtains a vertex consisting of three independent series of ladder diagrams.
Nevertheless, such a flow may be helpful for code-developing purposes.

An approximation of the vertex with objects of at most two frequency arguments is given by the SBE approximation \cite{Krien2019}, which sets $\varphi^{U\text{irr}} = 0$. More generally, one may also keep $\varphi^{U\text{irr}} \neq 0$ constant during the flow, e.g., as obtained from DMFT (called SBE-D$\Gamma$A in Ref.~\cite{Krien2019}). This was used in a $1\ell$ implementation of DMF$^2$RG in Ref.~\cite{Bonetti2021}. In the following, we will refer to the approximation of using a non-flowing $U$-irreducible part, $\dot{\varphi}^{U\text{irr}}=0$, as SBE approximation, regardless of whether $\varphi^{U\text{irr}}$ is set to zero or not.

We now derive mfRG flow equations for the SBE approximation, so that 
$\dot{R}=0$, as before, and furthermore $\dot M_r = 0$. 
For the most part, the SBE equations~\eqref{eq:SBE-equations} remain unchanged. Only the BSE for $M_r$ \eqref{eq:M_r-BSE-lambda-w-lambda-version} is not considered anymore, since now $\varphi^{U\mathrm{irr}}=R-U+\sum_r M_r$ is used as an input.
The corresponding flow equations can be obtained as in \Sec{Appendix:alt_deriv2}.
The flow of the polarization, the screened interaction and the Hedin vertices, prior to any transformation, is still given by \Eqs{eq:flow_Pr-start}, \eqref{eq:flow_wr} and \eqref{eq:BetheSalpeter-Hedin2} (collected here for convenience)
\begin{subequations}
		\label{eq:SBE-mfRG_SBE-approximation_lambda-r}
		\begin{align}
			\nonumber\dot{P}_r &=  \boldI_r\circ \dot{\Pi}_r \circ \bar{\lambda}_r + \boldI_r\circ \Pi_r \circ \dot{\bar{\lambda}}_r\\
			&= \dot{{\lambda}}_r \circ \Pi_r \circ \boldI_r +  {\lambda}_r\circ \dot{\Pi}_r \circ \boldI_r,
			\label{eq:flowPr2}\\
			\dot w_r &= w_r \fcirc \dot P_r \fcirc w_r,\\
			\label{eq:dot-bar-lambda_r}
			\dot{\bar{\lambda}}_r &= T_r\circ\dot{\Pi}_r\circ\boldI_r  + \dot{T}_r\circ\Pi_r\circ\boldI_r,
			\\
			\dot{\lambda}_r &= \boldI_r\circ\dot{\Pi}_r\circ T_r + \boldI_r\circ\Pi_r\circ\dot{T}_r.
		\end{align}	
However, the flow of $T_r = I_r - U + M_r$ now has no $\dot M_r$ contribution. 
It is induced solely by $\dot I_r = \dot{\nabla}_{\bar{r}}$, the flow of the $U$-reducible contributions from complementary channels,
\begin{align}
	\dot T_r &= \dot\nabla_{\bar{r}},
\end{align}
\end{subequations}
and thus is fully determined by $\dot{\bar \lambda}_{\bar{r}}$, $\dot \lambda_{\bar{r}}$ and $\dot w_{\bar{r}}$.

Equations \eqref{eq:SBE-mfRG_SBE-approximation_lambda-r} can be rewritten by inserting the flow of the higher-point objects into the lower-point objects:
\begin{subequations}
	\label{eq:SBE-mfRG_SBE-approximation_lambda-r-rewritten}
	\begin{align}
		\dot{\bar{\lambda}}_r &= T_r\circ\dot{\Pi}_r\circ\boldI_r  + \dot{\nabla}_{\bar{r}}\circ\Pi_r\circ\boldI_r,
		\\
		\dot{\lambda}_r &= \boldI_r\circ\dot{\Pi}_r\circ T_r + \boldI_r\circ\Pi_r\circ\dot{\nabla}_{\bar{r}},\\
		\nonumber\dot{P}_r &=  \boldI_r\circ \dot{\Pi}_r \circ \bar{\lambda}_r + \boldI_r\circ \Pi_r \circ T_r \circ \dot \Pi_r \circ \boldI_r\\
		\nonumber&\phantom{=} + \boldI_r\circ \Pi_r \circ \dot{\nabla}_{\bar{r}} \circ \Pi_r \circ \boldI_r\\
		\nonumber&= \boldI_r\circ \dot{\Pi}_r \circ \bar{\lambda}_r + \lambda_r\circ \dot \Pi_r \circ \boldI_r\\
		&\phantom{=}-\boldI_r\circ\dot\Pi_r\circ\boldI_r + \boldI_r\circ \Pi_r \circ \dot{\nabla}_{\bar{r}} \circ \Pi_r \circ \boldI_r.
	\end{align}
\end{subequations}
In the last line, we expressed $\boldI_r\circ\Pi_r\circ T_r$ in terms of the Hedin vertex $\lambda_r -\boldI_r$. Equations \eqref{eq:SBE-mfRG_SBE-approximation_lambda-r-rewritten} are similar
to the previous flow equations \eqref{eq:BetheSalpeter-Hedin_final} and \eqref{eq:flow_Pr} of the more general case, but some occurrences of the Hedin vertices $\bar\lambda_r, \lambda_r$ on the right there are here replaced by their zeroth-order term $\boldI_r$. Evidently, the contributions needed to upgrade
these $\boldI_r$ to  $\bar\lambda_r, \lambda_r$ are omitted when setting $\dot M_r = 0$.

A loop expansion of the above equations 
then yields
	\begin{subequations}
		\label{eq:SBE-mfRG_SBE-approximation}
		\begin{align}
			\nonumber\dot P_r^{(1)} &= \boldI_r \circ \dot{\Pi}_r \circ \bar\lambda_r + \lambda_r \circ \dot{\Pi}_r \circ \boldI_r 
			-\boldI_r\circ\dot{\Pi}_r\circ\boldI_r,\\	 
			\nonumber
			\dot{\bar{\lambda}}_r^{(1)}&= T_r\circ\dot{\Pi}_r\circ\boldI_r,
			\\ 
			\dot{\lambda}_r^{(1)} &= \boldI_r\circ\dot{\Pi}_r\circ T_r,
						\\ 
			\nonumber\dot{P}_r^{(2)}&=0,\\
			\nonumber
			\dot{\bar{\lambda}}_r^{(\ell+1)}&=\dot{\nabla}_{\bar{r}}^{(\ell)}\circ\Pi_r\circ\boldI_r,
			\\
			\dot{\lambda}_r^{(\ell+1)}&=\boldI_r\circ\Pi_r\circ\dot{\nabla}_{\bar{r}}^{(\ell)},
			\\
			\nonumber\dot{P}_r^{(\ell+2)}&= \boldI_r\circ\Pi_r\circ\dot{\nabla}_{\bar{r}}^{(\ell)}\circ\Pi_r\circ \boldI_r,\\
			\dot w_r^{(\ell)} &= w_r \fcirc \dot{P}_r^{(\ell)}\fcirc w_r.
		\end{align}
\end{subequations}
Apart from the fact that $\dot M_r$ is not needed here,
the other flow equations are also simpler than \Eqs{eq:SBE-mfRG_details} without $\dot M_r$, obtained from the full SBE equations. To be specific, \Eqs{eq:SBE-mfRG_details} contain $ \bar \lambda_r$ or $\lambda_r$ on the right of the flow equations for $\dot{\bar{\lambda}}_r^{(\ell)}$ or $\dot{{\lambda}}_r^{(\ell)}$, whereas the simplified \Eqs{eq:SBE-mfRG_SBE-approximation} contain $\boldI_r$ there, and, for $\ell \ge 2$, only one term where \Eqs{eq:SBE-mfRG_details} had two.

When using the above flow equations for the SBE approximation,
the self-energy flow \eqref{eq:mfRG:mfRG:Selfenergy_flow} should also be re-derived from either the SDE or the Hedin equation for $\Sigma$ (e.g.\ Eq.~(23) in Ref.~\cite{Krien2019b}).
Since the present paper 
focuses on vertex parametrizations, we leave a derivation 
of a suitably modified self-energy flow for future work. 
Here, it suffices to note that,
when used together with such a modified self-energy flow,
\Eqs{eq:SBE-mfRG_SBE-approximation} are again total derivatives of a closed set of equations.
So, integrating the flow until loop convergence
would yield the regulator-independent solution of the SBE approximation.

Transforming the self-consistent equations of the SBE approximation on the vertex level to an equivalent mfRG flow reveals its simplistic nature, with relations like $\dot{\lambda}^{(1)}_r = \boldI_r \fcirc \dot{\Pi}_r \fcirc T_r$, and demonstrates how fRG offers an intuitive way to go beyond that, by using, e.g., $\dot{\lambda}^{(1)}_r = \lambda_r \fcirc \dot{\Pi}_r \fcirc T_r$ (still treating only functions of at most two frequencies). 
However, the latter flow would be regulator-dependent \textit{per se}. It remains to be seen how severe the lack of regulator independence for this flow, as used, e.g., in Ref.~\cite{Bonetti2021}, is.

The simplified schemes presented in this section (i.e., \Eqs{eq:Poor-man's-fRG} and \eqref{eq:SBE-mfRG_SBE-approximation}) are closed flow equations on the vertex level and thus offer an appealing way for approaching the full SBE mfRG equations \eqref{eq:SBE-mfRG_details}. Thereby, SBE ingredients with more complicated frequency dependence can be taken into account successively during code development. To what extent they can succeed in actually capturing the essential physics of a given problem will have to be investigated on a case-by-case basis.
Generally, we showed that mfRG offers a way to make the choice of a certain approximation regulator independent, either for the simplistic flow of the SBE approximation or for the full SBE mfRG flow reproducing the PA.

\section{Asymptotic classes}
\label{sec:AsymptoticClasses}

In numerical implementations of parquet mfRG \cite{Tagliavini2019,Hille2020,Thoenniss2020,Kiese2020,Chalupa2021},
it is useful to handle the numerical complexity of the vertex by decomposing it into asymptotic classes with well-defined high-frequency behaviors. It is convenient to compute the flow of these asymptotic classes using their own flow equations; here, we recapitulate their derivation. We also elucidate the close relation between vertex parametrizations using the parquet decomposition with asymptotic classes or the SBE decomposition, deriving explicit equations relating their ingredients. These equations may facilitate the adaption of codes devised for parquet mfRG to SBE mfRG applications.

\subsection{Definition of asymptotic classes}
\label{subsec:Asymptotic-classes}

The parametrization of two-particle reducible vertices $\gamma_r$ via asymptotic classes was introduced in Ref.~\cite{Wentzell2020} to conveniently express their high-frequency asymptotics through simpler objects with fewer frequency arguments.
One makes the ansatz
\begin{align}
	\label{eq:Asymptotic_classes_introduction}
	&\gamma_r(\wr^\pprime,\vr^\pprime,\vr')  \\
	\nonumber
	&\ = \K1r(\wr^\pprime) + \K2r(\wr^\pprime,\vr^\pprime) + \Kb2r(\wr^\pprime,\vr') + \K3r(\wr^\pprime,\vr^\pprime,\vr') .
\end{align}
Here, $\K1r$ contains all diagrams having both $\nu^\pprime_r$ legs connected to the same bare vertex and both $\nu_r'$ legs connected to another bare vertex. 
(For a diagrammatic depiction, see App.~\ref{Appendix:Diagrams_for_Ki},
Fig.~\ref{fig:asymptotic-classes-general}.)
These diagrams are thus independent of $\vr$, $\vr'$ and stay finite in the limit $|\vr| \to \infty$, $|\vr'| \to \infty$,
\begin{subequations}
\label{eq:Keldysh_vertex_parametrization:Asymptotic_classes:Limit_K2}
\begin{align}
	\lim_{|\vr| \to \infty} \ \lim_{|\vr'| \to \infty} \gamma_r(\wr,\vr,\vr') = \K1r(\wr).
	\label{eq:Keldysh_vertex_parametrization:Asymptotic_classes:Limit_K1}
\end{align}
$\K2r$ (or $\Kb2r$) analogously contains the part of the vertex having 
both $\nu'_r$  (or $\nu^\pprime_r$) legs connected to the same bare vertex while the two $\nu^\pprime_r$  (or $\nu'_r$) legs are connected to different bare vertices.
Hence, it is finite for $|\vr'| \to \infty$ (or $|\vr^\pprime| \to \infty$) but vanishes for $|\vr^\pprime| \to \infty$ (or $|\vr'| \to \infty$):
\begin{align}
	\nonumber
	\lim_{|\vr'| \to \infty} \gamma_r(\wr^\pprime,\vr^\pprime,\vr') &= \K1r(\wr^\pprime) + \K2r(\wr^\pprime, \vr^\pprime) ,
	\\
	\lim_{|\vr^\pprime| \to \infty} \gamma_r(\wr^\pprime,\vr^\pprime,\vr') &= \K1r(\wr^\pprime) + \Kb2r(\wr^\pprime, \vr') .
\end{align} 	
$\K3r$ exclusively contains diagrams having both $\nu^\pprime_r$ legs connected
to different bare vertices, and likewise for both $\nu'_r$ legs. Such diagrams depend on all three frequencies and thus decay if any of them is sent to infinity. 
When taking the above limits for bubbles involving 
channels $r'$ different from $r$, we obtain zero,
\begin{align}
	\lim_{|\vr^\pprime| \to \infty} \gamma_{r'\neq r} = \lim_{|\vr'| \to \infty} \gamma_{r'\neq r} = 0 ,
	\label{eq:Keldysh_vertex_parametrization:Asymptotic_classes:Limit_gamma_r_bar}
\end{align}
\end{subequations}
as each $\Pi_{r'}$ in $\gamma_{r'}$ has a denominator containing $\omega_{r'\neq r}$, which is a linear combination of $\omega^\pprime_r$, $\nu^\pprime_r$ and $\nu'_r$.

Since $R$ explicitly depends on all frequencies, it decays to the bare vertex $U$ at high frequencies, and the asymptotic classes can be obtained by taking limits of the full vertex. 
Explicitly, $\K1r$ can be obtained from
\begin{subequations}
	\label{eq:Keldysh_vertex_parametrization:Asymptotic_classes:Limit_K1_and_K2_from_fullVertex}	
	\begin{align}
		\lim_{|\vr| \to \infty} \ \lim_{|\vr'| \to \infty} \Gamma(\wr,\vr,\vr')  = U + \K1r(\wr),
	\end{align}
	taking the double limit in such a way that 
	$\nu_r \pm \nu_r'$ is not constant, to ensure that all bosonic frequencies $|\omega_{r' \neq r}|$ go to $\infty$ \cite{Wentzell2020}. Similarly, 
	$\K2r$, $\Kb2r$ can be obtained from objects $\Gammar$, $\Gammabr$ defined via the limits
	\begin{flalign}
		\label{eq:Gammar}
		\Gammar (\wr, \nu_r) & = \lim_{|\vr'| \to \infty} \Gamma(\wr,\vr,\vr')  
		= U + \K1r + \K2r 
		, \hspace{-1cm} & 
		\\ 
		\label{eq:Gammabr}
		\Gammabr (\wr, \nu_r')& = \lim_{|\vr| \to \infty} \Gamma(\wr,\vr,\vr') 
		= U + \K1r + \Kb2r
		. \hspace{-1cm} & 
	\end{flalign} 
	For each of the latter two limits, we denote the complementary part of the vertex
	(vanishing in said limit) by 
	\begin{flalign}
		\label{eq:bGammar}
		\bGammar (\wr, \nu_r, \nu_r') & = \Gamma  \!-\!  \Gammar =  
		\Kb2r +\K3r+\gamma_{\bar{r}}+R - U , 
		\hspace{-1cm} &
		\\ 
		\label{eq:bGammabr}
		\bGammabr (\wr, \nu_r, \nu_r') & = \Gamma  \!-\!  \Gammabr = 
		\K2r+\K3r+\gamma_{\bar{r}}+R - U .
		\hspace{-1cm} &
	\end{flalign}
\end{subequations}
By taking suitable limits in the BSEs \eqref{eq:Bethe-Salpeter-equations}, the asymptotic classes can be expressed through the full vertex $\Gamma$ and the bare interaction $U$ \cite{Wentzell2020}:
\begin{subequations}
	\label{eq:Building_K12_from_Gamma}
	\begin{align}
		\label{eq:K1_compact_notation_with_Pi}
		\K1r(\wr) &= U \circ (\Pi_r+\Pi_r \circ \Gamma \circ \Pi_r) \circ  U,\\
		\label{eq:K2_compact_notation_with_Pi}
		\K2r (\wr,\vr) &= \Gamma \circ \Pi_r \circ  U-\K1r,\\
		\label{eq:K2b_compact_notation_with_Pi}
		\Kb2r (\wr,\vr') &= U \circ \Pi_r \circ \Gamma-\K1r.
	\end{align}
\end{subequations}

Hence, they are directly related to the three-point vertices $\bar\Gamma^{(3)}_r$, $\Gamma^{(3)}_r$ and susceptibilities $\chi_r$ (cf.~\Eq{eq:3point-susceptibility} and Ref.~\cite{Wentzell2020}) as
\begin{subequations}
	\begin{align}
		\chi_r(\wr) &= U^{-1}\fcirc \K1r(\wr) \fcirc U^{-1},\\
		\bar\Gamma^{(3)}_r(\wr,\vr) &= [U + \K1r + \K2r ](\wr,\vr)\fcirc U^{-1},\\
		\Gamma^{(3)}_r(\wr,\vr') &= U^{-1} \fcirc [U + \K1r + \Kb2r](\wr,\vr').
	\end{align}
\end{subequations}
$\K1r$ diagrams are therefore mediated by the bosonic fluctuations described by the susceptibility $\chi_r$, whereas $\K2r$ and $\Kb2r$ describe the coupling of fermions to these bosonic fluctuations via the three-point vertices $\bar\Gamma^{(3)}_r$ and $\Gamma^{(3)}_r$. This hints at the close relation between asymptotic classes and SBE components which is further discussed in \Sec{sec:SBE_Ki}.

\subsection{mfRG equations for asymptotic classes} 
\label{Appendix:mfRG_equations_Ki}

When the vertex is parametrized through its asymptotic classes, it is convenient to compute the latter directly during the flow, without numerically sending certain frequencies to infinity. This facilitates systematically adding or neglecting higher asymptotic classes. Therefore, we now derive explicit mfRG flow equations for the asymptotic classes, starting from the general multiloop flow equations \eqref{eq:mfRG-equations_gamma_r}, similar to the derivation of the mfRG flow equations for the SBE ingredients in \Sec{sec:mfRG_equations_SBE}. 
(For a diagrammatic derivation, see Refs.~\cite{Walter2021,Aguirre2020}.)

The parametrization \eqref{eq:Asymptotic_classes_introduction} of 
$\gamma_r$ in terms of asymptotic classes holds analogously at each loop order,
\begin{align}
	\label{eq:loop-expansion-of-asymptotic-expansion}
	\dot \gamma_{r}^{(\ell)} = 
	\dK{1}{r(\ell)} +  \dK{2}{r(\ell)} + \dK{2'}{r(\ell)} + \dK{3}{r(\ell)}  . 
\end{align}
Then, each summand can be obtained 
from \Eqs{eq:mfRG-equations_gamma_r} for $\dot \gamma_{r}^{(\ell)}$
by taking  suitable limits of the fermionic frequencies $\nu_r, \nu'_r$, 
as specified in \Eqs{eq:Keldysh_vertex_parametrization:Asymptotic_classes:Limit_K2}.
For example, consider a bubble of type $\Gamma \circ 
\dot \Pi_r \circ \tilde \Gamma$, in the $r$ representation 
of \Eq{eq:Bubble_summation_frequencies}. 
In the limit $|\nu_r|\to\infty$, the first vertex reduces to $\Gammabr$ 
(Eq.~\eqref{eq:Gammabr}), while for $|\nu'_r|\to\infty$, 
the second vertex reduces to $\tilde \Gammar$ 
(Eq.~\eqref{eq:Gammar}). Using \Eq{eq:Bubble_summation_frequencies}, we thus obtain
\begin{subequations}
	\label{eq:vertex-limits-nur-nurprime}
	\begin{align}
		\lim_{|\nu_r|\to\infty}\Gamma\circ \dot \Pi_r\circ \tilde \Gamma 
		&= \Gammabr\circ \dot \Pi_r\circ \tilde \Gamma,\\
		\lim_{|\nu'_r|\to\infty}\Gamma\circ \dot \Pi_r\circ \tilde \Gamma 
		&= \Gamma\circ \dot \Pi_r\circ \tilde \Gammar.
	\end{align}
	By contrast, when taking these limits for bubbles involving 
	channels $r'$ different from $r$, we obtain zero, 
	\begin{flalign}
		\label{eq:vertex-limits-nur-nurprime-wrong-channel}
		\lim_{|\nu_r|\to\infty} \!\!
		\Gamma\circ \dot \Pi_{r' \neq r} \circ \tilde \Gamma &= 0 , \ 
		\lim_{|\nu'_r|\to\infty} \!\!
		\Gamma\circ \dot \Pi_{r' \neq r} \circ \tilde \Gamma \, = 0 ,
		\hspace{-1cm} & 
	\end{flalign}
\end{subequations}%
by similar reasoning as that leading to \Eq{eq:Keldysh_vertex_parametrization:Asymptotic_classes:Limit_gamma_r_bar}.
In this manner, the $1\ell$ flow equation \eqref{eq:mfRG-equations_gamma_r_1l} for $\dot \gamma_r^{(1)}$  readily yields
\begin{subequations}
	\label{eq:mfRG_K1K2K3_l}
	\begin{align}
		\dK1{r\,(1)} &= \Gammabr\circ\dot{\Pi}_r\circ\Gammar 
		,
		\nonumber		\\
		\dK2{r\,(1)} &= \bGammabr\circ\dot{\Pi}_r\circ\Gammar
		,
		\nonumber \\
		\dKb2{r\,(1)} &= \Gammabr\circ\dot{\Pi}_r\circ\bGammar 
		,
		\nonumber \\
		\dK3{r\,(1)} &= \bGammabr\circ\dot{\Pi}_r\circ\bGammar
		.
		\label{eq:mfRG_K1K2K3_1l}
	\end{align}
	Similarly, 
	the two-loop contribution $\dot \gamma_r^{(2)}$, \Eq{eq:mfRG-equations_gamma_r_2l},  yields
	\begin{align}
		\dK1{r\,(2)} &= 0
		, 
		\nonumber \\
		\dK2{r\,(2)} &= \dot{\gamma}_{\bar{r}}^{(1)}\circ\Pi_r\circ\Gammar
		, 
		\nonumber \\
		\dKb2{r\,(2)} &= \Gammabr\circ\Pi_r\circ \dot{\gamma}_{\bar{r}}^{(1)}
		, 
		\nonumber \\
		\dK3{r\,(2)} &= \dot{\gamma}_{\bar{r}}^{(1)}\circ\Pi_r\circ\bGammar+\bGammabr\circ\Pi_r\circ \dot{\gamma}_{\bar{r}}^{(1)}
		.
		\label{eq:mfRG_K1K2K3_2l}
	\end{align}
	Due to \Eq{eq:Keldysh_vertex_parametrization:Asymptotic_classes:Limit_gamma_r_bar}, $	\dK1{r\,(2)}$ vanishes and $\dK2{r\,(2)}$ or 
	$\dK{2'}{r\,(2)}$ contain no terms with $\dot{\gamma}_{\bar{r}}^{(1)}$ 
	on their right or left sides, respectively.  
	Finally, \Eq{eq:mfRG-equations_gamma_r_l} for 
	$\dot \gamma_r^{(\ell+2)} $,  with $\ell \geq 1$, yields 
	\begin{align}
		\dK1{r\,(\ell + 2)} 
		&= \Gammabr\circ \Pi_r\circ\dot{\gamma}_{\bar{r}}^{(\ell)}\circ\Pi_r\circ\Gammar
		, 
		\nonumber \\
		\dK2{r\,(\ell + 2)} 
		&= \dot{\gamma}_{\bar{r}}^{(\ell+1)}\circ\Pi_r\circ\Gammar + \bGammabr\circ\Pi_r\circ\dot{\gamma}_{\bar{r}}^{(\ell)}\circ\Pi_r\circ\Gammar
		, 
		\nonumber \\
		\dKb2{r\,(\ell + 2)} 
		&= \Gammabr\circ\Pi_r\circ\dot{\gamma}_{\bar{r}}^{(\ell)}\circ\Pi_r\circ\bGammar + 
		\Gammabr\circ\Pi_r\circ\dot{\gamma}_{\bar{r}}^{(\ell+1)}, 
		\nonumber \\
		\nonumber\dK3{r\,(\ell + 2)} 
		&= \dot{\gamma}_{\bar{r}}^{(\ell+1)}\circ\Pi_r\circ\bGammar 
		+ \bGammabr\circ\Pi_r\circ\dot{\gamma}_{\bar{r}}^{(\ell)}\circ\Pi_r\circ\bGammar\\
		&\quad + \bGammabr\circ\Pi_r\circ\dot{\gamma}_{\bar{r}}^{(\ell+1)}
		. \label{eq:Kdot3rlplus2}
	\end{align}
\end{subequations}
Here, $\dK1{r(\ell + 2)} \neq 0$ since $\dot{\gamma}_{\bar{r}}^{(1)}$ appears in the middle
in the central term of Eq.~\eqref{eq:mfRG-equations_gamma_r_l}; hence, \Eq{eq:Keldysh_vertex_parametrization:Asymptotic_classes:Limit_gamma_r_bar}
does not apply.

Note that these equations can also be used in the context of DMF$^2$RG \cite{Taranto2014,Vilardi2019}. There, only the full vertex $\Gamma$ is given as an input. While $\K1r$, $\K2r$ and $\Kb2r$ can be deduced from $\Gamma$ by sending certain frequencies to infinity (cf.\ \Eqs{eq:Keldysh_vertex_parametrization:Asymptotic_classes:Limit_K1_and_K2_from_fullVertex}) {or using \Eqs{eq:Building_K12_from_Gamma}}, it is not possible to similarly extract $\K3r$ in a given channel as some frequency limit of the full vertex $\Gamma$. 
However, the classes $\K3r$ do not enter the right-hand sides of the flow equations \eqref{eq:mfRG_K1K2K3_l} individually, but only the combination $R+ \mathcal{K}_3  = R+\sum_r\K3r$. This is already clear from the general formulation of the mfRG flow equations \eqref{eq:mfRG-equations_gamma_r}. 
Consider, e.g., the $1\ell$ contribution $\dK2{r(1)}$ of 
\Eq{eq:mfRG_K1K2K3_1l}. There, $\bGammabr$ contains $R+ \K3r + \gamma_{\bar{r}}  = R+ \mathcal{K}_3  + \sum_{r'\neq r} (\K1{r'} + \K2{r'} + \Kb2{r'})$,  and hence only requires knowledge of the full 
$R + \mathcal{K}_3 $.
This holds equivalently for all insertions of the full vertex into flow equations at any loop order. Now, insertions of the \textsl{differentiated} vertex in loop order $\ell$ into the flow equations of order $\ell+1$ and $\ell+2$ \textsl{do} require a channel decomposition $\dot{\mathcal{K}}_3 = \sum_r\dK3r$. For example,
the two-loop  contribution  $\dK{2}{r\,(2)}$ of \Eq{eq:mfRG_K1K2K3_2l} contains $\dot{\gamma}_{\bar{r}}^{(1)}$, 
which, by \Eq{eq:loop-expansion-of-asymptotic-expansion}, involves differentiated vertices $\dK3{r'\neq r \, (1)}$. 
These \textsl{are} available via \Eq{eq:mfRG_K1K2K3_1l}.
Therefore, in the DMF$^2$RG context, one would start with $\K1r$, $\K2r$, $\Kb2r$ and the full $R+ \Ktot3 $ from DMFT, compute the differentiated vertices $\dK{i}{r}$ independently (including $\dK3r$), successively insert them in higher loop orders, and eventually update $\Ktot3$ using $\dKtot{3} = \sum_{\ell, r} \dK3{r \, (\ell)}$ in each step of the flow 
(recall that $R$ does not flow, $\dot{R}=0$).
The same reasoning also applies to the multi-boson terms $M_r$.

\subsection{Relating SBE ingredients and asymptotic classes}
\label{sec:SBE_Ki}

The asymptotic classes and SBE ingredients are closely related \cite{Bonetti2021}. 
This is not surprising as the properties of both follow from the assumption 
that the bare vertex contains no frequency dependence, except
for frequency conservation. For convenience, we collect these relations below.

Comparison of \Eqs{eq:definition_screened_interaction} and \eqref{eq:K1_compact_notation_with_Pi}
yields 
\begin{align}
	w_r(\wr) & = U + \K1r(\wr).
	\label{eq:Appendix:SBE:Channels:w_r}
\end{align}
Similarly, using 
\Eqs{eq:definition_Hedin-vertex}, \eqref{eq:definition_screened_interaction}, \eqref{eq:K2_compact_notation_with_Pi}, and \eqref{eq:K2b_compact_notation_with_Pi}, 
we can write the products of Hedin vertices and the screened interaction as
\begin{subequations}
	\label{eq:def_HedinVertex}
	\begin{align}
		\bar{\lambda}_r \fcirc w_r
		&= U + \Gamma \circ \Pi_r \circ U = U + \K1r + \K2r 
		,
		\\
		w_r  \fcirc \lambda_r &= U +  U \circ \Pi_r \circ \Gamma= U + \K1r + \Kb2r 
		. 
	\end{align}
\end{subequations}
We now insert \Eq{eq:Appendix:SBE:Channels:w_r} for $U \!+\! \K1r$
and solve for $\lambda_r$, $\bar \lambda_r$, formally defining $w_r^{-1}$ through $w_r \fcirc w_r^{-1} = w_r^{-1} \fcirc w_r = \boldI_r$. Thus, we obtain
\begin{align}
	\bar{\lambda}_r = \boldI_r + \K2r \fcirc w_r^{-1},\quad
	\lambda_r = \boldI_r + w_r^{-1}\fcirc\Kb2r,
	\label{eq:lambdas_K12}
\end{align}
which, when inserted into \Eq{eq:nabla_r}, yields
\begin{align}
	\nonumber\nabla_r &= 
	\left(\boldI_r+\K2r\fcirc w_r^{-1}\right)\fcirc w_r \fcirc \left(\boldI_r+w_r^{-1}\fcirc\Kb2r\right)\\
	&= U + \K1r + \K2r + \Kb2r + \K2r\fcirc w_r^{-1}\fcirc\Kb2r.
	\label{eq:Relation_nabla_r-K1-K2}
\end{align}
Depending on model details, it may happen that 
	not all components of $w_r^{-1}$ are uniquely defined.
	However, the right-hand sides of
	\Eqs{eq:lambdas_K12}--\eqref{eq:Relation_nabla_r-K1-K2} are
	unambiguous as the SBE ingredients are well-defined through \Eqs{eq:SBE-equations}.

Recalling that $\gamma_r = \nabla_r - U + M_r$, we conclude that
\begin{align}
	\label{eq:M_r=K3+K2wK2}
M_r = \K3r - \K2r\fcirc w_r^{-1}\fcirc\Kb2r  \, . 
\end{align}
Hence, $\nabla_r$
contains a part of $\K3r$,  
namely $\K2r \fcirc  w_r^{-1}  \fcirc \Kb2r$, which can be fully expressed through functions that each 
depend on at most two frequencies.
$M_r$ contains the remaining part of $\K3r$, which must be explicitly parametrized through three frequencies and thus is numerically most expensive. A recent study of the Hubbard model showed that $\sum_r M_r$ is strongly localized in frequency space, particularly in the strong-coupling regime \cite{Bonetti2021}. 
This allows for a cheaper numerical treatment of the vertex part truly depending on three frequencies
and constitutes the main computational advantage of the SBE decomposition.

\begin{figure}
	\centering
	\tikzm{overview_of_decompositions}{
		\begin{scope}[node distance=0.5cm and 0.1cm]
			\node [inner sep=0,anchor=west,text width=3.8cm, outer sep = 0.cm, minimum height = 0cm] (fullvertex_text) 										{\text{full vertex}:};
			\node [inner sep=0,anchor=west,text width=3.8cm] (parquet_text) 	[below=of fullvertex_text]		{\text{parquet decomp.}:};
			\node [inner sep=0,anchor=west,text width=3.8cm, minimum height = 0cm] (asymp_text) 	[below=of parquet_text]		{$\hookrightarrow$\text{ asymp. classes}:};
			\node [inner sep=0,anchor=west,text width=3.8cm, minimum height = 1cm] (Urred_text) 	[below=of asymp_text]		{$\hookrightarrow$\text{  \Ur-reduc.}:};
			\node [inner sep=0,anchor=west,text width=3.8cm] (SBE_text) 	[below=of Urred_text]		{\text{SBE decomp.}:};
			\node [inner sep=0,anchor=west] (parquetL) 	[right=of parquet_text]		{$\textcolor{myrot}R$};
			\node [inner sep=0,anchor=west] (parquetC) 	[right=of parquetL]		{$+$};
			\node [inner sep=0,anchor=west] (parquetR) 	[right=of parquetC]		{$ \sum_r  \textcolor{myrot}{\gamma_r}$};
			\node [inner sep=0,anchor=west, outer sep = 0.cm, minimum height = 0cm] (fullvertex) 	[above=of parquetC]		{$\textcolor{myrot}\Gamma$};
			\node [inner sep=0,anchor=west, minimum height = 0cm] (asympC) 	[below=0.55cm and 0cm of parquetC]		{$+$};
			\node [inner sep=0,anchor=west, minimum height = 0cm] (asympL) 	[left=of asympC]		{$\textcolor{myrot}R$};
			\node [inner sep=0,anchor=west, minimum height = 0cm] (asympR) 	[right=of asympC]		{$\sum_r  \big[{\textcolor{mygruen}{\K1r} + \textcolor{myblau}{\K2r} + \textcolor{myblau}{\Kb2r}} + \textcolor{myrot}{\K3r}\big]$};
			\node [inner sep=0,anchor=west, minimum height = 1cm] (UrredC) 	[below=0.55cm and 0cm of asympC]		{$+$};
			\node [inner sep=0,anchor=west, minimum height = 1cm] (UrredL) 	[left=of UrredC]		{${\textcolor{myrot}R - U} + U$};
			\node [inner sep=0,anchor=west, minimum height = 1cm] (UrredR) 	[right=of UrredC]		{$\sum_r \big[{\textcolor{myblau}{\bar{\lambda}_r}\fcirc \textcolor{mygruen}{w_r}\fcirc\textcolor{myblau}{\lambda_r} - U} + \textcolor{myrot}{M_r}\big]$};
			\node [inner sep=0,anchor=west] (SBEC) 	[below=of UrredC]		{$+$};
			\node [inner sep=0,anchor=west] (SBEL) 	[left=of SBEC]		{$\textcolor{myrot}{\varphi^{U\text{irr}}}-2U $};
			\node [inner sep=0,anchor=west] (SBER) 	[right=of SBEC]		{$ \sum_r \textcolor{myblau}{\bar{\lambda}_r}\fcirc \textcolor{mygruen}{w_r}\fcirc\textcolor{myblau}{\lambda_r}$};
			\node [inner sep=0,anchor=west] (asympR_notK3) 	[below left=0.1cm and -1.8 of asympR]		{};
			\node [inner sep=0,anchor=west] (asympR_K3) 	[below right=0.cm and -0.3 of asympR]		{};
			\node [inner sep=0,anchor=west] (UrredL_RminusU) 	[below left=-0.2cm and -0.5 cm of UrredL] {};
			\node [inner sep=0,anchor=west] (SBEL_phi) 	[above left=0.05cm and -0.3 cm of SBEL] {};
			\node [inner sep=0,anchor=west] (UrredR_SBE) 	[above left=-0.1cm and -1.85 cm of UrredR] {};
			\node [inner sep=0,anchor=west] (UrredR_Mabove) 	[above right=-0.3cm and -0.4 cm of UrredR] {};
			\node [inner sep=0,anchor=west] (UrredR_Mbelow) 	[below right=-0.3cm and -0.6 cm of UrredR] {};
			\draw[{Latex[scale=0.6]}-{Latex[scale=0.6]},gray] (asympR_K3) -- (UrredR_SBE);
			\draw[{Latex[scale=0.6]}-{Latex[scale=0.6]},gray] (asympR_K3) -- (UrredR_Mabove);
			\draw[{Latex[scale=0.6]}-{Latex[scale=0.6]},gray] (asympR_notK3) -- (UrredR_SBE);
			\draw[{Latex[scale=0.6]}-{Latex[scale=0.6]},gray] (UrredL_RminusU) -- (SBEL_phi);
			\draw[{Latex[scale=0.6]}-{Latex[scale=0.6]},gray] (UrredR_Mbelow) -- (SBEL_phi);
			\node [inner sep=0,anchor=west] (asympR_notK3_braceL) 	[above left=0.1cm and 1.cm of asympR_notK3]		{};
			\node [inner sep=0,anchor=west] (asympR_notK3_braceR) 	[above right =0.1cm and 1.cm of asympR_notK3]		{};
			\node [inner sep=0,anchor=west] (UrredL_RminusU_braceL) 	[above left=0.1cm and 0.45cm of UrredL_RminusU]		{};
			\node [inner sep=0,anchor=west] (UrredL_RminusU_braceR) 	[above right=0.1cm and 0.45cm of UrredL_RminusU]		{};
			\node [inner sep=0,anchor=west] (UrredR_SBE_braceL) 	[below left=0.1cm and 1.05cm of UrredR_SBE]		{};
			\node [inner sep=0,anchor=west] (UrredR_SBE_braceR) 	[below right=0.1cm and 1.05cm of UrredR_SBE]		{};
			\draw [decorate,decoration={brace,amplitude=3pt},xshift=0.pt,yshift=3pt,gray](asympR_notK3_braceR) -- (asympR_notK3_braceL) node[black,midway,yshift=0.2cm] {};
			\draw [decorate,decoration={brace,amplitude=3pt},xshift=0.pt,yshift=3pt,gray](UrredL_RminusU_braceR) -- (UrredL_RminusU_braceL) node[black,midway,yshift=0.2cm] {};
			\draw [decorate,decoration={brace,amplitude=3pt},xshift=0.pt,yshift=3pt,gray](UrredR_SBE_braceL) -- (UrredR_SBE_braceR) node[black,midway,yshift=0.2cm] {};
			\node [inner sep=0,anchor=west] (Legend1) 	[below right=-0.1cm and 2.35cm of fullvertex]		{\scriptsize frequencies:};
			\node [inner sep=0,anchor=west] (Legend2) 	[below=0.1cm and 0.0cm of Legend1]		{\scriptsize\textcolor{mygruen}1 \quad \textcolor{myblau}2 \quad \textcolor{myrot}3};
			\draw (7.6,-0.1) node [draw, minimum height = 0.75cm, yshift=-0.20cm, gray] { \phantom{frequencies}};
		\end{scope}
		\vspace{3mm}
	}
	\caption{Overview over vertex decompositions: The parquet decomposition (second line) can be grouped by asymptotic classes (third line) or \Ur-reducibility (fourth line), highlighting the relation between these two notions. Arrows link terms that can be identified:
$\K3r  =  M_r +  \K2r\protect\fcirc w_r^{-1}\protect\fcirc\Kb2r$ and 
$\K1r + \K2r + \Kb2r + \K2r \protect\fcirc w_r^{-1} \protect\fcirc \Kb2r =  \bar{\lambda}_r \protect\fcirc w_r\protect\fcirc\lambda_r - U $
  for the $\Pir$-reducible contributions, and
$\varphi^{U\text{irr}} = R - U + \sum_r M_r$ for the fully \Ur-irreducible contributions.
		The colors indicate whether the objects depend on \textcolor{mygruen}1, \textcolor{myblau}2, or \textcolor{myrot}3 frequency arguments.
}
\label{fig:Relation-SBE-asymptotic-classes}
\end{figure}

Equations \eqref{eq:Appendix:SBE:Channels:w_r}--\eqref{eq:Relation_nabla_r-K1-K2} fully express the SBE ingredients through asymptotic classes. Analogous results were obtained by similar arguments in App.~A of Ref.~\cite{Bonetti2021}. Figure~\ref{fig:Relation-SBE-asymptotic-classes} summarizes the relation between the two vertex decompositions and their ingredients.

Conversely, the asymptotic classes can also be expressed fully through the SBE ingredients. Using \Eqs{eq:splitgammar}, \eqref{eq:Asymptotic_classes_introduction}, \eqref{eq:Appendix:SBE:Channels:w_r}, and \eqref{eq:lambdas_K12}, one finds 
\begin{subequations}
	\label{subeq:Kir-asymptotic}
	\begin{align}
		\label{eq:Kir-asymptotic}
		\K1r &= w_r - U \, \\
		\K2r &= (\bar{\lambda}_r-\boldI_r)\fcirc w_r, \\
		\Kb2r &= w_r\fcirc(\lambda_r - \boldI_r), \\
		\label{eq:Kir-asymptotic-d}
		\K3r &= M_r + (\bar{\lambda}_r-\boldI_r)\fcirc w_r\fcirc(\lambda_r - \boldI_r).
	\end{align}
\end{subequations}
Moreover, \Eqs{eq:splitGamma-nablar-Iur}, \eqref{eq:Gammar}, \eqref{eq:Gammabr}, and \eqref{eq:def_HedinVertex} imply
\begin{subequations}
	\label{eq:Gammar2r2'-asymptotic}
	\begin{align}
		\Gammar & = \bar{\lambda}_r\fcirc w_r, \\
		\Gammabr & = w_r\fcirc\lambda_r , \\
		\bGammar &= \bar{\lambda}_r\fcirc w_r \fcirc (\lambda_r-\boldI_r)
		+ T_r \\
		\bGammabr &= (\bar{\lambda}_r-\boldI_r)\fcirc w_r\fcirc\lambda_r + T_r . 
		\label{eq:bGammabr-in-SBE}
	\end{align}
\end{subequations}
For the latter two equations, we used \Eq{eq:splitGamma-nablar-Iur}
in the form $\Gamma= \bar{\lambda}_r \fcirc w_r \fcirc \lambda_r + T_r$. 
Equivalently, using the definitions of the Hedin vertices in \Eq{eq:Hedin_definitions}, we can express $\K2r$, $\K3r$, and \Eqs{eq:Gammar2r2'-asymptotic} as
\begin{subequations}
	\begin{flalign}
		\K2r &= T_r \circ \Pi_r \circ w_r , 
		\hspace{-1cm} & 
		\\
		\Kb2r &= w_r \circ \Pi_r \circ T_r ,
		\hspace{-1cm} &
		\\
		\K3r &= M_r + T_r \circ \Pi_r \circ w_r \circ \Pi_r \circ T_r ,
		\hspace{-1cm} & 
		\\
		\Gammar &= w_r + T_r \circ \Pi_r \circ w_r , 
		\hspace{-1cm} & 
		\\
		\Gammabr &= w_r + w_r \circ \Pi_r \circ T_r , 
		\hspace{-1cm} & 
		\\
		\bGammar &= T_r + w_r \circ \Pi_r \circ T_r + T_r \circ \Pi_r \circ w_r \circ \Pi_r \circ T_r , 
		\hspace{-1cm} & 
		\\
		\bGammabr &= T_r + T_r \circ \Pi_r \circ w_r + T_r \circ \Pi_r \circ w_r \circ \Pi_r \circ T_r . 
		\hspace{-1cm} & 
	\end{flalign}
\end{subequations}

Since the asymptotic classes and SBE ingredients are closely related, 
the same is true for their mfRG flow. 
Indeed, it is straightforward to derive 
the mfRG SBE flow equations \eqref{eq:SBE-mfRG_details} 
from the flow equations \eqref{eq:mfRG_K1K2K3_l} for $\dK{i}{r\,(\ell)}$.
We briefly indicate the strategy, without presenting all details. 

We differentiate the equations \eqref{subeq:Kir-asymptotic}
expressing $\K{i}r$ through SBE ingredients, and subsequently
use \Eqs{eq:Hedin_definitions} to eliminate 
$\bar{\lambda}_r-\boldI_r$ 
and ${\lambda}_r-\boldI_r$. Thereby, we obtain
\begin{subequations}
	\begin{align}
		\label{eq:dK1r-in-SBE}
		\dK1r &= \dot{w}_r,\\
		\label{eq:dK2r-in-SBE}
		\dK2r &= \dot{\bar{\lambda}}_r\fcirc w_r + T_r \circ \Pi_r\circ\dot{w}_r,\\
		\dKb2r &= \dot{w}_r \circ\Pi_r \circ T_r + w_r\fcirc\dot{\lambda}_r,\\
		\nonumber \dK3r &= \dot{\bar{\lambda}}_r\fcirc w_r\circ\Pi_r \circ T_r + T_r \circ \Pi_r \circ \dot{w}_r \circ \Pi_r \circ T_r\\
		&\quad + T_r \circ \Pi_r \circ w_r \fcirc \dot{\lambda}_r + 
		\dot{M}_r .
	\end{align}
\end{subequations}
Now, we use \Eqs{eq:mfRG_K1K2K3_l} to express the $\dK{i}{r \, ({\ell})}$ on the left through 
$\Gammar$, $\Gammabr$, $\bGammar$, $\bGammabr$, and
\Eqs{eq:Gammar2r2'-asymptotic} to express the latter through SBE ingredients. 
By matching terms on the left and right in each loop order, we obtain flow equations for $\dot w^{(\ell)}$, $\dot{\bar \lambda}_r^{(\ell)}$, $\dot \lambda_r^{(\ell)}$ and $\dot M_r^{(\ell)}$.
For example, at $1\ell$ order, \Eqs{eq:mfRG_K1K2K3_1l} and 
\eqref{eq:dK1r-in-SBE} for $\dK1{r \, (1)}$ yield
\begin{align}
	\label{eq:dotwr1-check}
	\dot{w}_r^{(1)} & = \Gammabr\circ\dot{\Pi}_r\circ\Gammar =
	w_r\fcirc\lambda_r\circ\dot{\Pi}_r\circ\bar{\lambda}_r\fcirc w_r
	,
\end{align} 
consistent with \Eq{eq:SBE-mfRG_1l}. Similarly, for
$\dK2{r \, (1)}$, we obtain 
\begin{align}
	&\dot{\bar{\lambda}}_r^{(1)}\fcirc w_r + T_r \circ \Pi_r \circ \dot{w}^{(1)}_r = \bGammabr\circ\dot{\Pi}_r\circ\Gammar \\
	\nonumber&\quad = T_r \circ \dot{\Pi}_r \circ \bar{\lambda}_r\fcirc w_r + T_r \circ \Pi_r \circ w_r \fcirc \lambda_r \circ \dot{\Pi}_r \circ \bar{\lambda}_r \fcirc w_r.
\end{align}	
The second terms on the left and right cancel due to \Eq{eq:dotwr1-check}. The remaining terms, right-multiplied by $w_r^{-1}$, yield  $\dot{\bar{\lambda}}_r^{(1)}= T_r \circ\dot{\Pi}_r\circ\bar{\lambda}_r$, consistent with \Eq{eq:SBE-mfRG_1l}.
All of the equations \eqref{eq:SBE-mfRG_details} can be derived in this manner.

\section{Conclusions and outlook}
\label{sec:Conclusion}

The SBE decomposition of the four-point vertex was originally introduced in Hubbard-like models respecting SU$(2)$ spin symmetry and was written in terms of physical (e.g., spin and charge) channels \cite{Krien2019}. Inspired by 
Refs.~\cite{Krien2019a,Krien2019,Krien2019b,Krien2020a,Krien2020b,Krien2021}, we here formulated the SBE decomposition without specifying the structure of non-frequency arguments (such as position or momentum, spin, etc.) starting from the parquet equations for general fermionic models. The only restriction on the structure of the bare vertex $U$ is that,
apart from being frequency-conserving, it is otherwise constant in frequency. Our formulation can thus be used as a starting point for a rather general class of models. It can also be easily extended to the Keldysh formalism or to other types of particles such as bosons or real fermions.

In this generalized framework, we re-derived self-consistent equations for the ingredients of the SBE decomposition $\nabla_r = \bar\lambda_r\fcirc w_r\fcirc\lambda_r$, the so-called SBE equations, by separating the BSEs for the two-particle reducible vertices regarding their $U$-reducibility. The $U$-reducible $\nabla_r$ have a transparent interpretation through bosonic exchange fluctuations and Hedin vertices, describing the coupling of these bosonic fluctuations to fermions. 
As our main result, we derived multiloop flow equations for the SBE ingredients in two different ways: first by inserting the SBE decomposition into parquet mfRG and second by differentiating the SBE equations. Thereby, we presented the multiloop generalization of the $1\ell$ SBE flow of Ref.~\cite{Bonetti2021}.
In addition, we gave a detailed discussion of the relation between the SBE ingredients, $M_r$ and $\nabla_r = \bar{\lambda}_r \fcirc w_r \fcirc \lambda_r$, and the asymptotic classes 
$\K{i}r$ of the two-particle reducible vertices. 
Finally, we also presented multiloop flow equations for the $\K{i}r$ and thus provided a unified formulation for the mfRG treatment of the parquet and the SBE vertex decompositions. 

A numerical study of the SBE mfRG flow for relevant model systems, such as the single-impurity Anderson model or the Hubbard model, is left for future work. Below, we outline some open questions to be addressed.

The numerically most expensive SBE ingredient is the fully $U$-irreducible vertex $\varphi^{U\text{irr}}$, involving the multi-boson exchange terms $M_r$, because these all depend on three frequency arguments. One may hope that, 
for certain applications, it might suffice to neglect $\varphi^{U\text{irr}}$ (as done in Ref.~\cite{Harkov2021} for a DMFT treatment of the Hubbard model), or to treat it in a cheap fashion, e.g., by not keeping track of its full frequency dependence or by not letting it flow (cf.\ Ref.~\cite{Bonetti2021}). 
This spoils the parquet two-particle self-consistency while retaining SBE self-consistency. It is an interesting open question which of the main qualitative features of the parquet solution, such as fulfillment of the Mermin--Wagner theorem \cite{Bickers1992}, remain intact this way. 

One formal feature, namely regulator independence, is maintained if multiloop flow equations in the SBE approximation are used.
These equations are derived by setting $\varphi^{U\text{irr}}=0$ and $\dot M_r=0$ from the beginning (Sec.~\ref{sec:Simplified-SBE-flow}) and are actually simpler than those obtained by setting $\dot M_r=0$ in the full SBE mfRG flow. We left the derivation of a self-energy flow directly within the SBE approximation for future work. The combination of such a self-energy flow with the vertex flow of Sec.~\ref{sec:Simplified-SBE-flow} would constitute the total derivative of the SBE approximation.
Therefore, if loop convergence can be achieved when integrating these simplified flow equations, the results will be regulator independent, just as for the full SBE mfRG flow with $\varphi^{U\text{irr}}=\sum_r M_r$ and $\dot M_r \neq 0$, reproducing the PA.

Even if it turns out that a full treatment of $\varphi^{U\text{irr}}$ is required for capturing essential qualitative features of the vertex, this might still be numerically cheaper than a full treatment of $\K3{}$. The reason is that each $\K3r$ contains a contribution,
the $ \K2r\fcirc w_r^{-1}\fcirc\Kb2r$ term in
\Eq{eq:Relation_nabla_r-K1-K2}, which is included not in $M_r$
but in $\nabla_r$, and parametrized through
the numerically cheaper Hedin vertices and screened interactions, see \Fig{fig:Relation-SBE-asymptotic-classes}. If these terms decay comparatively slowly with frequency, their treatment via the $\K{i}r$ decomposition would be numerically expensive, and 
the SBE decomposition could offer a numerically cheaper alternative. A systematic comparison of the numerical costs required to compute the multiloop flow of the two decompositions should thus be a main goal of future work.
\\

\noindent
\textbf{Acknowledgments}\
We thank F.~Krien, J.~Halbinger, and N.~Ritz for critical reading of the manuscript. 
This research is part of the Munich Quantum Valley, which is supported by the Bavarian state government with funds from the Hightech Agenda Bayern Plus. 
We acknowledge funding for M.G.\ from the International Max Planck Research School for Quantum Science and Technology (IMPRS-QST),
for A.G.\ and J.v.D.\ from the Deutsche Forschungsgemeinschaft under 
Germany's Excellence Strategy EXC-2111 (Project No.\ 390814868),
and for F.B.K.\ from the Alexander von Humboldt Foundation through the Feodor Lynen Fellowship.

\vspace{-5mm}
\section*{Author contributions}

M.G., E.W., A.G., and F.B.K.\ contributed to the derivation of the presented equations. All authors jointly prepared the manuscript.

\appendix

\begin{figure*}[tb]
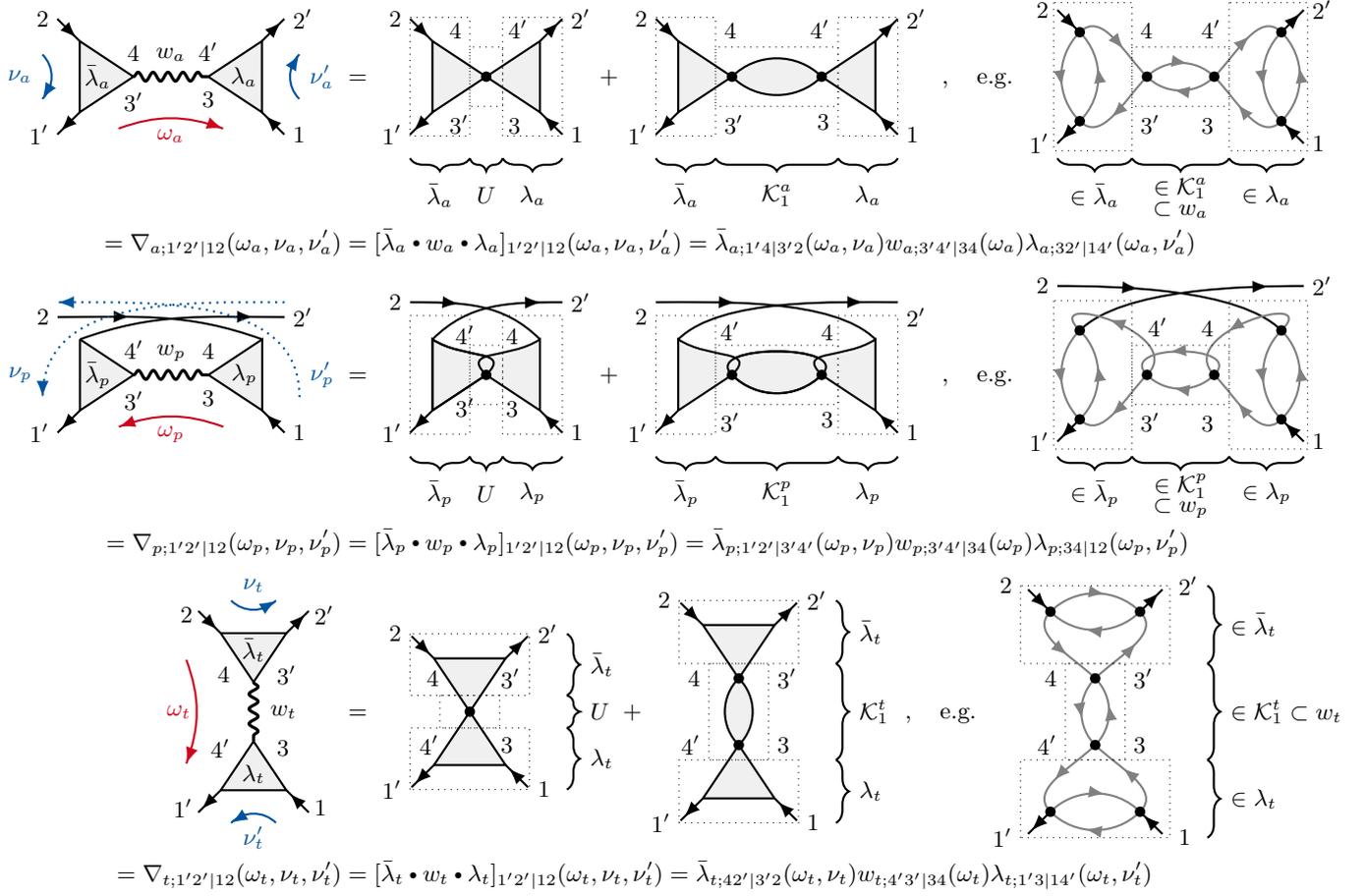

	\begin{align*}
		\tikzm{lambda_w_lambda_a_diagram}{
			\threepointvertexleftarrows{$\bar\lambda_a$}{0}{0}{1.2}
			\bosonfull{0.72}{0}{1.72}{0}
			\threepointvertexrightarrows{$\lambda_a$}{1.72}{0}{1.2}
			\node at (1.22,0.3) {$w_a$};
			\node[left] at (-0.3,-0.78) {$1'$};
			\node[left] at (-0.3,0.78) {$2$};
			\node[right] at (2.74,-0.78) {$1$};
			\node[right] at (2.74,0.78) {$2'$};
			\node[above] at (0.72,0.1) {$4$};
			\node[below] at (0.72,-0.1) {$3'$};
			\node[above] at (1.72,0.1) {$4'$};
			\node[below] at (1.72,-0.1) {$3$};
			\draw[lineWithArrowEnd,blau] (-0.5,0.3) to [out=315, in=45] (-0.5,-0.3) -- (-0.5006,-0.301);
			\node at (-0.8,0) {\bl{\small$\va$}};
			\draw[lineWithArrowEnd,blau] (2.94,-0.3) to [out=135, in=225] (2.94,0.3) -- (2.9406,0.301);
			\node at (3.24,0) {\bl{\small$\va'$}};
			\draw[lineWithArrowEnd,rot] (0.52,-0.7) to [out=20, in=160] (1.92,-0.7);
			\node at (1.22,-0.8) {\rt{\small$\wa$}};
		}
		&=
		\tikzm{lambda_w_lambda_a_bare_schematic}{
			\threepointvertexleftarrows{}{0}{0}{1.2}
			\threepointvertexrightarrows{}{0.72}{0}{1.2}
			\barevertex{0.72}{0}
			\draw[dotted] (-0.3,-0.8) rectangle (0.5,0.8);
			\draw[dotted] (0.5,-0.4) rectangle (0.94,0.4);
			\draw[dotted] (0.94,-0.8) rectangle (1.74,0.8);
			\node[left] at (-0.3,-0.78) {$1'$};
			\node[left] at (-0.3,0.78) {$2$};
			\node[above left] at (0.6,0.4) {$4\phantom{'}$};
			\node[below left] at (0.6,-0.4) {$3'$};
			\node[above right] at (0.9,0.4) {$4'$};
			\node[below right] at (0.9,-0.4) {$3\phantom{'}$};
			\node[right] at (1.74,-0.78) {$1$};
			\node[right] at (1.74,0.78) {$2'$};
			\tikzunderbrace{$\bar\lambda_a$}{-0.3}{0.5}{-1.1}{-0.5}
			\tikzunderbrace{$U$}{0.5}{0.94}{-1.1}{-0.5}
			\tikzunderbrace{$\lambda_a$}{0.94}{1.74}{-1.1}{-0.5}
		}
		\! + \!
		\tikzm{lambda_w_lambda_a_schematic}{
			\threepointvertexleftarrows{}{0}{0}{1.2}
			\Konea{}{0.72}{0}{1.2/1.4}
			\threepointvertexrightarrows{}{1.92}{0}{1.2}
			\barevertex{0.72}{0}
			\barevertex{1.92}{0}
			\draw[dotted] (-0.3,-0.8) rectangle (0.5,0.8);
			\draw[dotted] (0.5,-0.4) rectangle (2.14,0.4);
			\draw[dotted] (2.14,-0.8) rectangle (2.94,0.8);
			\node[left] at (-0.3,-0.78) {$1'$};
			\node[left] at (-0.3,0.78) {$2$};
			\node[above right] at (0.5,0.4) {$4$};
			\node[below right] at (0.5,-0.4) {$3'$};
			\node[above left] at (2.14,0.4) {$4'$};
			\node[below left] at (2.14,-0.4) {$3$};
			\node[right] at (2.94,-0.78) {$1$};
			\node[right] at (2.94,0.78) {$2'$};
			\tikzunderbrace{$\bar\lambda_a$}{-0.3}{0.5}{-1.1}{-0.5}
			\tikzunderbrace{$\K1a$}{0.5}{2.14}{-1.1}{-0.5}
			\tikzunderbrace{$\lambda_a$}{2.14}{2.94}{-1.1}{-0.5}
		}
		\ , \quad
		\text{e.g.}
		\
		\tikzm{lambda_w_lambda_a_PT}{
			\arrowslefthalffull{0.6}{0}{2}
			\tbubblebarebarebare{0}{0.6}{1}
			\draw[lineBareWithArrowCenter] (0,0.6) to [out=45, in=135] (0.9,0);
			\draw[lineBareWithArrowCenter] (0.9,0) to [out=225, in=315] (0,-0.6);
			\abubblebarebarebare{0.9}{0}{0.75}
			\draw[lineBareWithArrowCenter] (1.8,0) to [out=45, in=135] (2.7,0.6);
			\draw[lineBareWithArrowCenter] (2.7,-0.6) to [out=225, in=315] (1.8,0);
			\tbubblebarebarebare{2.7}{0.6}{1}
			\arrowsrighthalffull{2.1}{0}{2}
			\barevertex{0}{0.6}
			\barevertex{0}{-0.6}
			\barevertex{0.9}{0}
			\barevertex{1.8}{0}
			\barevertex{2.7}{0.6}
			\barevertex{2.7}{-0.6}
			\draw[dotted] (-0.35,1) rectangle (0.7,-1);
			\draw[dotted] (0.7,0.4) rectangle (2,-0.4);
			\draw[dotted] (2,1) rectangle (3.05,-1);
			\tikzunderbrace{$\in\bar\lambda_a$}{-0.3}{0.7}{-1.1}{-0.5}
			\tikzunderbrace{$\in \K1a$}{0.7}{2}{-1.1}{-0.4}
			\node at (1.35,-1.8) {$\subset w_a$};
			\tikzunderbrace{$\in\lambda_a$}{2}{3}{-1.1}{-0.5}
			\node[left] at (-0.3,-0.9) {$1'$};
			\node[left] at (-0.3,0.9) {$2$};
			\node[above right] at (0.7,0.4) {$4\phantom{'}$};
			\node[below right] at (0.7,-0.4) {$3'$};
			\node[above left] at (2,0.4) {$4'$};
			\node[below left] at (2,-0.4) {$3\phantom{'}$};
			\node[right] at (3,-0.9) {$1$};
			\node[right] at (3,0.9) {$2'$};
		}
		\\[-1mm]
		=\nabla_{a;1'2'|12}(\wa,\va,\va') &= [\bar{\lambda}_a\fcirc w_a\fcirc\lambda_a]_{1'2'|12}(\omega_a,\nu_a,\nu'_a)
		=
		\bar\lambda_{a;1'4|3'2}(\wa,\va)
		w_{a;3'4'|34}(\wa)
		\lambda_{a;32'|14'}(\wa,\va')
		\\
		\tikzm{lambda_w_lambda_p_diagram}{
			\threepointvertexleft{$\bar\lambda_p$}{0}{0}{1.2}
			\bosonfull{0.72}{0}{1.72}{0}
			\threepointvertexright{$\lambda_p$}{1.72}{0}{1.2}
			\node at (1.22,0.3) {$w_p$};
			\node at (1.22,0.74) {};
			\draw[lineWithArrowCenterEnd] (0,-0.48) -- (-0.3,-0.78);
			\draw[lineWithArrowCenterEnd] (2.74,-0.78) -- (2.44,-0.48);
			\draw[lineWithArrowCenterEnd] (0,0.48) to [out=20, in=180] (2.74,0.78);
			\draw[lineWithArrowCenterStart] (-0.3,0.78) to [out=0, in=160] (2.44,0.48);
			\node[left] at (-0.3,-0.78) {$1'$};
			\node[left] at (-0.3,0.78) {$2$};
			\node[right] at (2.74,-0.78) {$1$};
			\node[right] at (2.74,0.78) {$2'$};
			\node[above] at (0.72,0.1) {$4'$};
			\node[below] at (0.72,-0.1) {$3'$};
			\node[above] at (1.72,0.1) {$4$};
			\node[below] at (1.72,-0.1) {$3$};
			\draw[lineWithArrowEnd,blau,dotted] (2.74,0.98) to [out=180, in=90] (-0.5,-0.3) -- (-0.5,-0.3001);
			\node at (-0.8,0) {\bl{\small$\vp$}};
			\draw[lineWithArrowEnd,blau,dotted] (2.94,-0.3) to [out=90, in=0] (-0.3,0.98) -- (-0.3001,0.98);
			\node at (3.24,0) {\bl{\small$\vp'$}};
			\draw[lineWithArrowEnd,rot] (1.92,-0.7) to [out=160, in=20] (0.52,-0.7);
			\node at (1.22,-0.8) {\rt{\small$\wp$}};
		}
		&=
		\tikzm{lambda_w_lambda_p_bare_schematic}{
			\draw[lineWithArrowCenterEnd] (0,-0.48) -- (-0.3,-0.78);
			\draw[lineWithArrowCenterEnd] (0,0.48) to [out=45, in=180] (1.74,0.98);
			\draw[lineWithArrowCenterEnd] (1.74,-0.78) -- (1.44,-0.48);
			\draw[lineWithArrowCenterStart] (-0.3,0.98) to [out=0, in=135] (1.44,0.48);
			\draw[linePlain, fill=verylightgray] (0,-0.48) -- (0.72,0) .. controls ++(45:0.5) and ++(-33.7:0.5) .. (0,0.48) -- (0,-0.48);
			\draw[linePlain, fill=verylightgray] (1.44,-0.48) -- (0.72,0) .. controls ++(135:0.5) and ++(213.7:0.5) .. (1.44,0.48) -- (1.44,-0.48);
			\draw[linePlain] (0,-0.48) -- (0.72,0) .. controls ++(45:0.5) and ++(-33.7:0.5) .. (0,0.48) -- (0,-0.48);
			\barevertex{0.72}{0}
			\draw[dotted] (-0.3,-0.8) rectangle (0.5,0.8);
			\draw[dotted] (0.5,-0.4) rectangle (0.94,0.4);
			\draw[dotted] (0.94,-0.8) rectangle (1.74,0.8);
			\node[left] at (-0.3,-0.78) {$1'$};
			\node[left] at (-0.3,0.98) {$2$};
			\node[above left] at (0.68,0.3) {$4'$};
			\node[below left] at (0.68,-0.2) {$3'$};
			\node[above right] at (0.87,0.3) {$4\phantom{'}$};
			\node[below right] at (0.87,-0.2) {$3\phantom{'}$};
			\node[right] at (1.74,-0.78) {$1$};
			\node[right] at (1.74,0.98) {$2'$};
			\tikzunderbrace{$\bar\lambda_p$}{-0.3}{0.5}{-1.1}{-0.5}
			\tikzunderbrace{$U$}{0.5}{0.94}{-1.1}{-0.5}
			\tikzunderbrace{$\lambda_p$}{0.94}{1.74}{-1.1}{-0.5}
		}
		\! + \!
		\tikzm{lambda_w_lambda_p_schematic}{
			\draw[lineWithArrowCenterEnd] (0,-0.48) -- (-0.3,-0.78);
			\draw[lineWithArrowCenterEnd] (0,0.48) .. controls ++(45:0.8) and ++(180:0.8) .. (2.94,0.98);
			\draw[linePlain, fill=verylightgray] (0.72,0) to [out=315, in=225] (1.92,0) .. controls ++(45:0.6) and ++(135:0.6) .. (0.72,0);
			\draw[linePlain, fill=verylightgray] (0,-0.48) -- (0.72,0) .. controls ++(45:0.5) and ++(-33.7:0.5) .. (0,0.48) -- (0,-0.48);
			\draw[linePlain, fill=verylightgray] (2.64,-0.48) -- (1.92,0) .. controls ++(135:0.5) and ++(213.7:0.5) .. (2.64,0.48) -- (2.64,-0.48);
			\draw[linePlain] (0.72,0) to [out=315, in=225] (1.92,0) .. controls ++(45:0.6) and ++(135:0.6) .. (0.72,0);
			\draw[lineWithArrowCenterEnd] (2.94,-0.78) -- (2.64,-0.48);
			\draw[lineWithArrowCenterStart] (-0.3,0.98) .. controls ++(0:0.8) and ++(135:0.8) .. (2.64,0.48);
			\barevertex{0.72}{0}
			\barevertex{1.92}{0}
			\draw[dotted] (-0.3,-0.8) rectangle (0.5,0.8);
			\draw[dotted] (0.5,-0.4) rectangle (2.14,0.4);
			\draw[dotted] (2.14,-0.8) rectangle (2.94,0.8);
			\node[left] at (-0.3,-0.78) {$1'$};
			\node[left] at (-0.3,0.78) {$2$};
			\node[above right] at (0.45,0.35) {$4'$};
			\node[below right] at (0.45,-0.4) {$3'$};
			\node[above left] at (2.19,0.35) {$4$};
			\node[below left] at (2.19,-0.4) {$3$};
			\node[right] at (2.94,-0.78) {$1$};
			\node[right] at (2.94,0.78) {$2'$};
			\tikzunderbrace{$\bar\lambda_p$}{-0.3}{0.5}{-1.1}{-0.5}
			\tikzunderbrace{$\K1p$}{0.5}{2.14}{-1.1}{-0.5}
			\tikzunderbrace{$\lambda_p$}{2.14}{2.94}{-1.1}{-0.5}
		}
		\ , \quad
		\text{e.g.}
		\
		\tikzm{lambda_w_lambda_p_PT}{
			\draw[lineWithArrowCenterEnd] (0,-0.6) -- (-0.3,-0.9);
			\draw[lineWithArrowCenterEnd] (0,0.6) .. controls ++(45:0.8) and ++(180:0.8) .. (3,1.2);
			\draw[lineWithArrowCenterEnd] (3,-0.9) -- (2.7,-0.6);
			\draw[lineWithArrowCenterStart] (-0.3,1.2) .. controls ++(0:0.8) and ++(135:0.8) .. (2.7,0.6);
			\tbubblebarebarebare{0}{0.6}{1}
			\draw[lineBareWithArrowCenter] (0.9,0) .. controls ++(45:0.8) and ++(135:0.8) .. (0,0.6);
			\draw[lineBareWithArrowCenter] (0.9,0) to [out=225, in=315] (0,-0.6);
			\draw[lineBareWithArrowCenter] (1.8,0) to [out=225, in=315] (0.9,0);
			\draw[lineBareWithArrowCenter] (1.8,0) .. controls ++(45:0.6) and ++(135:0.6) .. (0.9,0);
			\draw[lineBareWithArrowCenter] (2.7,0.6) .. controls ++(45:0.8) and ++(135:0.8) .. (1.8,0);
			\draw[lineBareWithArrowCenter] (2.7,-0.6) to [out=225, in=315] (1.8,0);
			\tbubblebarebarebare{2.7}{0.6}{1}
			\barevertex{0}{0.6}
			\barevertex{0}{-0.6}
			\barevertex{0.9}{0}
			\barevertex{1.8}{0}
			\barevertex{2.7}{0.6}
			\barevertex{2.7}{-0.6}
			\draw[dotted] (-0.35,1) rectangle (0.7,-1);
			\draw[dotted] (0.7,0.4) rectangle (2,-0.4);
			\draw[dotted] (2,1) rectangle (3.05,-1);
			\tikzunderbrace{$\in\bar\lambda_p$}{-0.3}{0.7}{-1.1}{-0.5}
			\tikzunderbrace{$\in \K1p$}{0.7}{2}{-1.1}{-0.4}
			\node at (1.35,-1.8) {$\subset w_p$};
			\tikzunderbrace{$\in\lambda_p$}{2}{3}{-1.1}{-0.5}
			\node[left] at (-0.3,-0.9) {$1'$};
			\node[left] at (-0.3,1.2) {$2$};
			\node[above right] at (0.8,0.4) {$4'$};
			\node[below right] at (0.7,-0.4) {$3'$};
			\node[above left] at (2,0.4) {$4\phantom{'}$};
			\node[below left] at (2,-0.4) {$3\phantom{'}$};
			\node[right] at (3,-0.9) {$1$};
			\node[right] at (3,1.2) {$2'$};
		}
		\\[-1mm]
		=\nabla_{p;1'2'|12}(\wp,\vp,\vp')
		&= [\bar{\lambda}_p\fcirc w_p\fcirc\lambda_p]_{1'2'|12}(\omega_p,\nu_p,\nu'_p) =
		\bar{\lambda}_{p;1'2'|3'4'}(\omega_p,\nu_p) w_{p;3'4'|34}(\omega_p) \lambda_{p;34|12}(\omega_p,\nu'_p)
		\\
		\tikzm{lambda_w_lambda_t_diagram}{
			\threepointvertexupperarrows{$\bar\lambda_t$}{0}{0.73}{1.1}
			\bosonfull{0}{0.4}{0}{-0.4}
			\node at (0.4,0) {$w_t$};
			\threepointvertexlowerarrows{$\lambda_t$}{0}{-0.73}{1.1}
			\node[left] at (-0.7,1.3) {$2$};
			\node[right] at (0.7,1.3) {$2'$};
			\node[left] at (-0.2,0.5) {$4$};
			\node[right] at (0.2,0.5) {$3'$};
			\node[left] at (-0.2,-0.5) {$4'$};
			\node[right] at (0.2,-0.5) {$3$};
			\node[left] at (-0.7,-1.3) {$1'$};
			\node[right] at (0.7,-1.3) {$1$};
			\draw[lineBareWithArrowEnd,blau] (-0.3,1.5) to [out=315, in=225] (0.3,1.5) -- (0.301,1.5006);
			\node at (0,1.7) {{\color{blau} \small$\vt$}};
			\draw[lineBareWithArrowEnd,blau] (0.3,-1.5) to [out=135, in=45] (-0.3,-1.5) -- (-0.301,-1.5006);
			\node at (0,-1.7) {{\color{blau} \small$\vt'$}};
			\draw[lineBareWithArrowEnd,rot] (-0.9,0.7) to [out=290, in=70] (-0.9,-0.7);
			\node at (-1,0) {{\color{rot} \small$\wt$}};
		}
		&=
		\tikzm{lambda_w_lambda_t_bare_schematic}{
			\threepointvertexupperarrows{}{0}{0.36}{1.2}
			\threepointvertexlowerarrows{}{0}{-0.36}{1.2}
			\barevertex{0}{0}
			\draw[dotted] (-0.8,1.05) rectangle (0.8,0.22);
			\draw[dotted] (-0.4,0.22) rectangle (0.4,-0.22);
			\draw[dotted] (-0.8,-1.05) rectangle (0.8,-0.22);
			\draw[decoration={brace,mirror,amplitude=5},decorate,thick] (1.3,0.22) -- (1.3,1.05);
			\node[right] at (1.5,0.635) {$\bar\lambda_t$};
			\draw[decoration={brace,mirror,amplitude=5},decorate,thick] (1.3,-0.22) -- (1.3,0.22);
			\node[right] at (1.5,0) {$U$};
			\draw[decoration={brace,mirror,amplitude=5},decorate,thick] (1.3,-1.05) -- (1.3,-0.22);
			\node[right] at (1.5,-0.635) {$\lambda_t$};
			\node[left] at (-0.8,1.05) {$2$};
			\node[left] at (-0.3,0.45) {$4$};
			\node[left] at (-0.3,-0.45) {$4'$};
			\node[left] at (-0.8,-1.05) {$1'$};
			\node[right] at (0.8,1.05) {$2'$};
			\node[right] at (0.3,0.45) {$3'$};
			\node[right] at (0.3,-0.45) {$3$};
			\node[right] at (0.8,-1.05) {$1$};
		}
		\! + \!
		\tikzm{lambda_w_lambda_t_schematic}{
			\threepointvertexupperarrows{}{0}{0.81}{1.2}
			\Konet{}{0}{0.45}{0.9/1.4}
			\threepointvertexlowerarrows{}{0}{-0.81}{1.2}
			\barevertex{0}{0.45}
			\barevertex{0}{-0.45}
			\draw[dotted] (-0.8,1.5) rectangle (0.8,0.65);
			\draw[dotted] (-0.4,0.65) rectangle (0.4,-0.65);
			\draw[dotted] (-0.8,-1.5) rectangle (0.8,-0.65);
			\draw[decoration={brace,mirror,amplitude=5},decorate,thick] (1.3,0.65) -- (1.3,1.5);
			\node[right] at (1.5,1.075) {$\bar\lambda_t$};
			\draw[decoration={brace,mirror,amplitude=5},decorate,thick] (1.3,-0.65) -- (1.3,0.65);
			\node[right] at (1.5,0) {$\K1t$};
			\draw[decoration={brace,mirror,amplitude=5},decorate,thick] (1.3,-1.5) -- (1.3,-0.65);
			\node[right] at (1.5,-1.075) {$\lambda_t$};
			\node[left] at (-0.8,1.5) {$2$};
			\node[left] at (-0.4,0.45) {$4$};
			\node[left] at (-0.4,-0.45) {$4'$};
			\node[left] at (-0.8,-1.5) {$1'$};
			\node[right] at (0.8,1.5) {$2'$};
			\node[right] at (0.4,0.45) {$3'$};
			\node[right] at (0.4,-0.45) {$3$};
			\node[right] at (0.8,-1.5) {$1$};
		}
		\ , \quad
		\text{e.g.}
		\
		\tikzm{lambda_w_lambda_t_PT}{
			\arrowsupperhalffull{0}{0.75}{2}
			\abubblebarebarebare{-0.6}{1.35}{1}
			\draw[lineBareWithArrowCenter] (-0.6,1.35) to [out=225, in=135] (0,0.45);
			\draw[lineBareWithArrowCenter] (0,0.45) to [out=45, in=315] (0.6,1.35);
			\tbubblebarebarebare{0}{0.45}{0.9/1.2}
			\draw[lineBareWithArrowCenter] (0,-0.45) to [out=225, in=135] (-0.6,-1.35);
			\draw[lineBareWithArrowCenter] (0.6,-1.35) to [out=45, in=315] (0,-0.45);
			\abubblebarebarebare{-0.6}{-1.35}{1}
			\arrowslowerhalffull{0}{-0.75}{2}
			\barevertex{-0.6}{1.35}
			\barevertex{0.6}{1.35}
			\barevertex{0}{0.45}
			\barevertex{0}{-0.45}
			\barevertex{-0.6}{-1.35}
			\barevertex{0.6}{-1.35}
			\draw[dotted] (-1,1.7) rectangle (1,0.65);
			\draw[dotted] (-0.4,0.65) rectangle (0.4,-0.65);
			\draw[dotted] (-1,-1.7) rectangle (1,-0.65);
			\draw[decoration={brace,mirror,amplitude=5},decorate,thick] (1.5,0.65) -- (1.5,1.7);
			\node[right] at (1.7,1.175) {$\in\bar\lambda_t$};
			\draw[decoration={brace,mirror,amplitude=5},decorate,thick] (1.5,-0.65) -- (1.5,0.65);
			\node[right] at (1.7,0) {$\in\K1t \subset w_t$};
			\draw[decoration={brace,mirror,amplitude=5},decorate,thick] (1.5,-1.7) -- (1.5,-0.65);
			\node[right] at (1.7,-1.175) {$\in\lambda_t$};
			\node[left] at (-1,1.7) {$2$};
			\node[left] at (-0.4,0.45) {$4$};
			\node[left] at (-0.4,-0.45) {$4'$};
			\node[left] at (-1,-1.7) {$1'$};
			\node[right] at (1,1.7) {$2'$};
			\node[right] at (0.4,0.45) {$3'$};
			\node[right] at (0.4,-0.45) {$3$};
			\node[right] at (1,-1.7) {$1$};
		}
		\\[-1mm]
		=\nabla_{t;1'2'|12}(\wt,\vt,\vt')
		&= [\bar{\lambda}_t\fcirc w_t\fcirc\lambda_t]_{1'2'|12}(\omega_t,\nu_t,\nu'_t) = 
		\bar{\lambda}_{t;42'|3'2}(\omega_t,\nu_t)w_{t;4'3'|34}(\omega_t)\lambda_{t;1'3|14'}(\omega_t,\nu'_t)
	\end{align*} \vspace{-5mm}
	\caption{Illustration of the structure of $\nabla_r$ using $w_r = U + \K1r$ (\Eq{eq:Appendix:SBE:Channels:w_r}), including an exemplary sixth-order diagram. While $\bar\lambda_r$, $w_r$, $\lambda_r$ factorize \wrt their frequency dependence (since they are connected by bare vertices in $\nabla_r$), they are viewed as four-point objects \wrt the other quantum numbers (the internal indices $3,3',4,4'$ have to be summed over, cf.~\Eq{eq:Bubble_summation}).}
	\label{fig:lambda_w_lambda}
\end{figure*}

\section{Diagrams of SBE ingredients}
\label{Appendix:Diagrams_for_SBE}

Figure~\ref{fig:lambda_w_lambda} illustrates which parts of the $\Ur$-reducible diagrams $\nabla_r$ belong to the Hedin vertices $\bar{\lambda}_r, \lambda_r$ and which parts belong to the screened interactions $w_r$ (for exemplary low-order diagrams, see Figure~\ref{fig:SBE_constituents_PT}). 

\section{Diagrams of asymptotic classes}
\label{Appendix:Diagrams_for_Ki}

\begin{figure*}
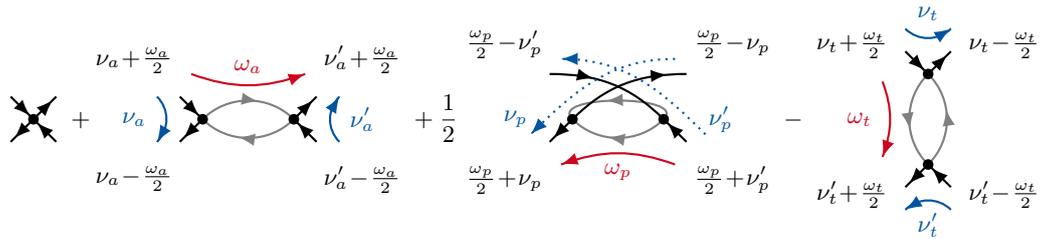

	\begin{align*}
		\tikzm{Vertex_parametrization-channeldec-PT1}{
			\barevertexwithlegs{0}{0}
		} \
		+
		\!\!
		\tikzm{Vertex_parametrization-channeldec-PT2a}{
			\arrowslefthalf{0}{0}
			\abubblebarebarebare{0}{0}{1}
			\arrowsrighthalf{1.2}{0}
			\barevertex{0}{0}
			\barevertex{1.2}{0}
			\draw[lineWithArrowEnd,blau] (-0.6,0.3) to [out=315, in=45] (-0.6,-0.3) -- (-0.6006,-0.301);
			\node at (-0.9,0) {\bl{\small$\va$}};
			\draw[lineWithArrowEnd,blau] (1.8,-0.3) to [out=135, in=225] (1.8,0.3) -- (1.8006,0.301);
			\node at (2.1,0) {\bl{\small$\va'$}};
			\draw[lineWithArrowEnd,rot] (-0.15,0.6) to [out=340, in=200] (1.35,0.6);
			\node at (0.6,0.7) {\rt{\small$\wa$}};
			\node[above] at (-0.9,0.5) {\small$\va \! + \! \tfrac{\wa}{2}$};
			\node[above] at (2.1,0.5) {\small$\va' \! + \! \tfrac{\wa}{2}$};
			\node[below] at (-0.9,-0.5) {\small$\va \! - \! \tfrac{\wa}{2}$};
			\node[below] at (2.1,-0.5) {\small$\va' \! - \! \tfrac{\wa}{2}$};
		} 
		+
		\frac{1}{2}
		\tikzm{Vertex_parametrization-channeldec-PT2p}{
			\pbubblebarebarebare{0}{0}{1}
			\arrowslefthalfp{0}{0}{1}
			\arrowsrighthalfp{1.2}{0}{1}
			\barevertex{0}{0}
			\barevertex{1.2}{0}
			\draw[lineBareWithArrowEnd,dotted,blau] (1.35,0.8) to [out=180, in=45] (-0.55,-0.2);
			\node at (-0.75,0) {{\color{blau} \small$\vp$}};
			\draw[lineBareWithArrowEnd,dotted,blau] (1.75,-0.2) to [out=135, in=0] (-0.15,0.8);
			\node at (1.95,0) {{\color{blau} \small$\vp'$}};
			\draw[lineWithArrowEnd,rot] (1.35,-0.6) to [out=160, in=20] (-0.15,-0.6);
			\node at (0.6,-0.7) {\rt{\small$\wp$}};
			\node[above] at (-0.9,0.7) {\small$\tfrac{\wp}{2} \! - \! \vp'$};
			\node[above] at (2.1,0.7) {\small$\tfrac{\wp}{2} \! - \! \vp$};
			\node[below] at (-0.9,-0.5) {\small$\tfrac{\wp}{2} \! + \! \vp$};
			\node[below] at (2.1,-0.5) {\small$\tfrac{\wp}{2} \! + \! \vp'$};
		}
		-
		\tikzm{Vertex_parametrization-channeldec-PT2t}{
			\arrowsupperhalf{0}{0.6}
			\tbubblebarebarebare{0}{0.6}{1}
			\arrowslowerhalf{0}{-0.6}
			\barevertex{0}{0.6}
			\barevertex{0}{-0.6}
			\draw[lineWithArrowEnd,blau] (-0.3,1.2) to [out=315, in=225] (0.3,1.2) -- (0.301,1.2006);
			\node at (0,1.4) {\bl{\small$\vt$}};
			\draw[lineWithArrowEnd,blau] (0.3,-1.2) to [out=135, in=45] (-0.3,-1.2) -- (-0.301,-1.2006);
			\node at (0,-1.4) {\bl{\small$\vt'$}};
			\draw[lineWithArrowEnd,rot] (-0.6,0.5) to [out=290, in=70] (-0.6,-0.5);
			\node at (-0.9,0) {\rt{\small$\wt$}};
			\node[above] at (-1,0.7) {\small$\vt \! + \! \tfrac{\wt}{2}$};
			\node[above] at (1,0.7) {\small$\vt \! - \! \tfrac{\wt}{2}$};
			\node[below] at (-1,-0.7) {\small$\vt' \! + \! \tfrac{\wt}{2}$};
			\node[below] at (1,-0.7) {\small$\vt' \! - \! \tfrac{\wt}{2}$};
		} 
	\end{align*} \vspace{-5mm}
\caption{Diagrams in second-order perturbation theory including the channel-specific frequency parametrization.}
\label{fig:2nd-order-PT}
\end{figure*}

We illustrate the channel-specific frequency para\-metrizations of the vertex (\Fig{fig:Vertex_frequency_parametrization}) in second-order perturbation theory in \Fig{fig:2nd-order-PT}.
\begin{figure*}[tb]
	\begingroup
	\allowdisplaybreaks  
		\label{eq:Keldysh_vertex_parametrization:Asymptotic_classes:Diagrammatic_decomposition}
		\begin{alignat*}{5} \vspace{-5mm}
			\tikzm{Vertex_parametrization-diagclass-gamma_a}{
				\fullvertexwithlegs{$\gamma_a$}{0}{0}{1}
				\node[left] at (-0.6,0.7) {\small$\va \! + \! \tfrac{\wa}{2}$};
				\node[right] at (0.6,0.7) {\small$\va' \! + \! \tfrac{\wa}{2}$};
				\node[left] at (-0.6,-0.7) {\small$\va \! - \! \tfrac{\wa}{2}$};
				\node[right] at (0.6,-0.7) {\small$\va' \! - \! \tfrac{\wa}{2}$};
			} 
			&=
			\tikzm{Vertex_parametrization-diagclass-K1a}{
				\arrowslefthalf{0}{0}
				\Konea{$\K1a$}{0}{0}{1}
				\arrowsrighthalf{1.4}{0}
				\draw[lineBareWithArrowEnd,blau] (-0.6,0.3) to [out=315, in=45] (-0.6,-0.3) -- (-0.6006,-0.301);
				\node at (-0.8,0) {{\color{blau} \small$\va$}};
				\draw[lineBareWithArrowEnd,blau] (2,-0.3) to [out=135, in=225] (2,0.3) -- (2.0006,0.301);
				\node at (2.2,0) {{\color{blau} \small$\va'$}};
				\draw[lineBareWithArrowEnd,rot] (-0.15,0.6) to [out=340, in=200] (1.55,0.6);
				\node at (0.7,0.7) {{\color{rot} \small$\wa$}};
			}
			&&+
			\tikzm{Vertex_parametrization-diagclass-K2a}{
				\arrowslefthalffull{0.3}{0}{1}
				\Ktwoa{$\K2a$}{0}{0}{1}
				\arrowsrighthalf{1.2}{0}
				\draw[lineBareWithArrowEnd,rot] (-0.45,0.4) to [out=315, in=45] (-0.45,-0.4) -- (-0.4507,-0.401);
				\node at (-0.65,0) {{\color{rot} \small$\va$}};
				\draw[lineBareWithArrowEnd,blau] (1.8,-0.3) to [out=135, in=225] (1.8,0.3) -- (1.8006,0.301);
				\node at (2,0) {{\color{blau} \small$\va'$}};
				\draw[lineBareWithArrowEnd,rot] (-0.15,0.7) to [out=330, in=190] (1.35,0.5);
				\node at (0.6,0.7) {{\color{rot} \small$\wa$}};
			}
			&&+
			\tikzm{Vertex_parametrization-diagclass-Kb2a}{
				\arrowslefthalf{0}{0}
				\Ktwoab{$\Kb2a$}{0}{0}{1}
				\arrowsrighthalffull{0.9}{0}{1}
				\draw[lineBareWithArrowEnd,blau] (-0.6,0.3) to [out=315, in=45] (-0.6,-0.3) -- (-0.6006,-0.301);
				\node at (-0.8,0) {{\color{blau} \small$\va$}};
				\draw[lineBareWithArrowEnd,rot] (1.65,-0.4) to [out=135, in=225] (1.65,0.4) -- (1.6507,0.401);
				\node at (1.85,0) {{\color{rot} \small$\va'$}};
				\draw[lineBareWithArrowEnd,rot] (-0.15,0.5) to [out=350, in=210] (1.35,0.7);
				\node at (0.6,0.7) {{\color{rot} \small$\wa$}};
			}
			&&+
			\tikzm{Vertex_parametrization-diagclass-K3a}{
				\fullvertexwithlegs{$\K3a$}{0}{0}{1}
				\draw[lineBareWithArrowEnd,rot] (-0.75,0.4) to [out=315, in=45] (-0.75,-0.4) -- (-0.7507,-0.401);
				\node at (-0.95,0) {{\color{rot} \small$\va$}};
				\draw[lineBareWithArrowEnd,rot] (0.75,-0.4) to [out=135, in=225] (0.75,0.4) -- (0.7507,0.401);
				\node at (0.95,0) {{\color{rot} \small$\va'$}};
				\draw[lineBareWithArrowEnd,rot] (-0.4,0.7) to [out=315, in=225] (0.4,0.7) -- (0.401,0.7007);
				\node at (0,0.9) {{\color{rot} \small$\wa$}};
			}
			\\[-1mm]
			\gamma_a (\wa, \va, \va') \quad \qquad
			&= \qquad \qquad \K1a (\rt{\wa}) &&+ 
			\quad \qquad \K2a (\rt{\wa, \va}) &&+ 
			\quad \qquad \Kb2a (\rt{\wa, \va'}) 
			&&+ \quad \K3a (\rt{\wa, \va, \va'}) ,
			\\[3mm]
			\tikzm{Vertex_parametrization-diagclass-gamma_p}{
				\fullvertexwithlegs{$\gamma_p$}{0}{0}{1}
				\node[left] at (-0.6,0.7) {\small$\tfrac{\wp}{2} \! - \! \vp'$};
				\node[right] at (0.6,0.7) {\small$\tfrac{\wp}{2} \! - \! \vp$};
				\node[left] at (-0.6,-0.7) {\small$\tfrac{\wp}{2} \! + \! \vp$};
				\node[right] at (0.6,-0.7) {\small$\tfrac{\wp}{2} \! + \! \vp'$};
			}
			&=
			\tikzm{Vertex_parametrization-diagclass-K1p}{
				\Konea{$\K1p$}{0}{0}{1}
				\arrowslefthalfp{0}{0}{1.4/1.2}
				\arrowsrighthalfp{1.4}{0}{1.4/1.2}
				\draw[lineBareWithArrowEnd,dotted,blau] (1.55,0.8) to [out=180, in=45] (-0.45,-0.1);
				\node at (-0.7,0) {{\color{blau} \small$\vp$}};
				\draw[lineBareWithArrowEnd,dotted,blau] (1.85,-0.1) to [out=135, in=0] (-0.15,0.8);
				\node at (2.1,0) {{\color{blau} \small$\vp'$}};
				\draw[lineBareWithArrowEnd,rot] (1.55,-0.6) to [out=160, in=20] (-0.15,-0.6);
				\node at (0.7,-0.7) {{\color{rot} \small$\wp$}};
				\node at (0,0.85) {};
			}
			&&+
			\tikzm{Vertex_parametrization-diagclass-K2p}{
				\Ktwoa{$\K2p$}{0}{0}{1}
				\draw[lineWithArrowCenterEnd] (0,-0.3) -- (-0.3,-0.6);
				\draw[lineWithArrowCenterEnd] (0,0.3) to [out=30, in=180] (1.5,0.6);
				\arrowsrighthalfp{1.2}{0}{1}
				\draw[lineBareWithArrowEnd,dotted,rot] (1.35,0.8) to [out=180, in=70] (-0.45,-0.3);
				\node at (-0.7,0) {{\color{rot} \small$\vp$}};
				\draw[lineBareWithArrowEnd,dotted,blau] (1.65,-0.1) to [out=135, in=0] (-0.15,0.8);
				\node at (1.9,0) {{\color{blau} \small$\vp'$}};
				\draw[lineBareWithArrowEnd,rot] (1.35,-0.5) to [out=170, in=30] (-0.15,-0.7);
				\node at (0.6,-0.7) {{\color{rot} \small$\wp$}};
				\node at (0,0.85) {};
			}
			&&+
			\tikzm{Vertex_parametrization-diagclass-Kb2p}{
				\Ktwoab{$\Kb2p$}{0}{0}{1}
				\arrowslefthalfp{0}{0}{1}
				\draw[lineWithArrowCenterEnd] (1.5,-0.6) -- (1.2,-0.3);
				\draw[lineWithArrowCenterStart] (-0.3,0.6) to [out=0, in=150] (1.2,0.3);
				\draw[lineBareWithArrowEnd,dotted,blau] (1.35,0.8) to [out=180, in=45] (-0.45,-0.1);
				\node at (-0.7,0) {{\color{blau} \small$\vp$}};
				\draw[lineBareWithArrowEnd,dotted,rot] (1.65,-0.3) to [out=110, in=0] (-0.15,0.8);
				\node at (1.9,0) {{\color{rot} \small$\vp'$}};
				\draw[lineBareWithArrowEnd,rot] (1.35,-0.7) to [out=150, in=10] (-0.15,-0.5);
				\node at (0.6,-0.7) {{\color{rot} \small$\wp$}};
				\node at (0,0.85) {};
			}
			&&+
			\tikzm{Vertex_parametrization-diagclass-K3p}{
				\fullvertexwithlegs{$\K3p$}{0}{0}{1}
				\draw[lineBareWithArrowEnd,rot] (0.4,-0.75) to [out=135, in=45] (-0.4,-0.75) -- (-0.401,-0.7507);
				\node at (0,-0.95) {{\color{rot} \small$\wp$}};
				\draw[lineBareWithArrowEnd,dotted,rot] (0.4,0.75) to [out=200, in=70] (-0.75,-0.4);
				\node at (-0.95,0) {{\color{rot} \small$\vp$}};
				\draw[lineBareWithArrowEnd,dotted,rot] (0.75,-0.4) to [out=110, in=340] (-0.4,0.75);
				\node at (0.95,0) {{\color{rot} \small$\vp'$}};
			}
			\\[-1mm]
			\gamma_p (\wp, \vp, \vp') \quad \qquad 
			&=
			\qquad \qquad \K1p (\rt{\wp}) &&+ 
			\quad \qquad \K2p (\rt{\wp, \vp}) &&+ 
			\quad \qquad \Kb2p (\rt{\wp, \vp'}) 
			&&+ 
			\quad \K3p (\rt{\wp, \vp, \vp'}) ,
			\\[1mm]
			\tikzm{Vertex_parametrization-diagclass-gamma_t}{
				\fullvertexwithlegs{$\gamma_t$}{0}{0}{1}
				\node[left] at (-0.6,0.7) {\small$\vt \! + \! \tfrac{\wt}{2}$};
				\node[right] at (0.6,0.7) {\small$\vt \! - \! \tfrac{\wt}{2}$};
				\node[left] at (-0.6,-0.7) {\small$\vt' \! + \! \tfrac{\wt}{2}$};
				\node[right] at (0.6,-0.7) {\small$\vt' \! - \! \tfrac{\wt}{2}$};
			}
			&=
			\qquad \quad \tikzm{Vertex_parametrization-diagclass-K1t}{
				\arrowsupperhalf{0}{0.7}
				\Konet{$\K1t$}{0}{0.7}{1}
				\arrowslowerhalf{0}{-0.7}
				\draw[lineBareWithArrowEnd,blau] (-0.3,1.3) to [out=315, in=225] (0.3,1.3) -- (0.301,1.3006);
				\node at (0,1.5) {{\color{blau} \small$\vt$}};
				\draw[lineBareWithArrowEnd,blau] (0.3,-1.3) to [out=135, in=45] (-0.3,-1.3) -- (-0.301,-1.3006);
				\node at (0,-1.5) {{\color{blau} \small$\vt'$}};
				\draw[lineBareWithArrowEnd,rot] (-0.6,0.85) to [out=290, in=70] (-0.6,-0.85);
				\node at (-0.7,0) {{\color{rot} \small$\wt$}};
			}
			&&+
			\qquad \;\; 
			\tikzm{Vertex_parametrization-diagclass-K2t}{
				\Ktwot{$\K2t$}{0}{0.6}{1}
				\arrowsupperhalffull{0}{0.3}{1}
				\arrowslowerhalf{0}{-0.6}
				\draw[lineBareWithArrowEnd,rot] (-0.4,1.05) to [out=315, in=225] (0.4,1.05) -- (0.401,1.0507);
				\node at (0,1.25) {{\color{rot} \small$\vt$}};
				\draw[lineBareWithArrowEnd,blau] (0.3,-1.2) to [out=135, in=45] (-0.3,-1.2) -- (-0.301,-1.2006);
				\node at (0,-1.4) {{\color{blau} \small$\vt'$}};
				\draw[lineBareWithArrowEnd,rot] (-0.7,0.75) to [out=300, in=80] (-0.5,-0.75);
				\node at (-0.7,0) {{\color{rot} \small$\wt$}};
			}
			&&+\qquad \;\;
			\tikzm{Vertex_parametrization-diagclass-Kb2t}{
				\Ktwotb{$\Kb2t$}{0}{0.6}{1}
				\arrowsupperhalf{0}{0.6}
				\arrowslowerhalffull{0}{-0.3}{1}
				\draw[lineBareWithArrowEnd,blau] (-0.3,1.2) to [out=315, in=225] (0.3,1.2) -- (0.301,1.2006);
				\node at (0,1.4) {{\color{blau} \small$\vt$}};
				\draw[lineBareWithArrowEnd,rot] (0.4,-1.05) to [out=135, in=45] (-0.4,-1.05) -- (-0.401,-1.0507);
				\node at (0,-1.25) {{\color{rot} \small$\vt'$}};
				\draw[lineBareWithArrowEnd,rot] (-0.5,0.75) to [out=280, in=60] (-0.7,-0.75);
				\node at (-0.7,0) {{\color{rot} \small$\wt$}};
			}
			&&+
			\tikzm{Vertex_parametrization-diagclass-K3t}{
				\fullvertexwithlegs{$\K3t$}{0}{0}{1}
				\draw[lineBareWithArrowEnd,rot] (-0.4,0.75) to [out=315, in=225] (0.4,0.75) -- (0.401,0.7507);
				\node at (0,0.95) {{\color{rot} \small$\vt$}};
				\draw[lineBareWithArrowEnd,rot] (0.4,-0.75) to [out=135, in=45] (-0.4,-0.75) -- (-0.401,-0.7507);
				\node at (0,-0.95) {{\color{rot} \small$\vt'$}};
				\draw[lineBareWithArrowEnd,rot] (-0.75,0.4) to [out=315, in=45] (-0.75,-0.4) -- (-0.7507,-0.401);
				\node at (-0.95,0) {{\color{rot} \small$\wt$}};
			}
			\\[-1mm]
			\gamma_t (\wt, \vt, \vt') \quad \qquad 
			&= \qquad \qquad \K1t (\rt{\wt}) &&+ 
			\quad \qquad \K2t (\rt{\wt, \vt}) &&+ 
			\quad \qquad \Kb2t (\rt{\wt, \vt'}) &&+ 
			\quad  \K3t (\rt{\wt, \vt, \vt'}) .
		\end{alignat*}
	\vspace{-5mm}
	\endgroup 
	\caption{
		Illustration of the decomposition of the two-particle
		reducible vertices $\gamma_r$ into asymptotic classes,
		$\K1r + \K2r + \K{2'}r + \K3r$.}
	\label{fig:asymptotic-classes-general}
\end{figure*}
The bosonic frequency $\wr$ is transferred through the bubble in which each diagram is reducible, while the fermionic frequencies $\vr,\vr'$ parametrize the frequency dependence on each side of the bubble.
Evidently, the internal propagator lines only depend on the bosonic transfer frequency of the corresponding channel (and the internal integration frequency). The external fermionic frequency
$\nu_r$ flows in and out at the same bare vertex, 
and so does $\nu'_r$ at another bare vertex, such that the value of each diagram is independent of $\vr,\vr'$.
This notion can be generalized \cite{Wentzell2020}, leading to the decomposition of each $\Pir$-reducible vertex $\gamma_r$ into four different asymptotic classes, 
$\K1r + \K2r + \K{2'}r + \K3r$,
depicted diagrammatically in Fig.~\ref{fig:asymptotic-classes-general}. A formal definition is given by \Eqs{eq:Keldysh_vertex_parametrization:Asymptotic_classes:Limit_K2} in the main text.

\section{Relation to SBE in physical channels}
\label{Appendix:physical_diagrammatic}

The SBE decomposition was originally defined in terms of the \underline{ch}arge, \underline{sp}in, and \underline{si}nglet pairing channels \cite{Krien2019}. These involve specific linear combinations of the spin components, chosen to diagonalize the spin structure in the BSEs for SU$(2)$-symmetric systems \cite{Bickers2004}.
Assuming SU(2) spin symmetry, we show below how these ``physical'' SBE channels
are related to the ``diagrammatic'' SBE channels used in the main text.

By spin conservation, each incoming spin $\sigma \in \{ \uparrow, \downarrow \}$ must also come out of a vertex.
The nonzero components thus are
\begin{align}
	\label{eq:vertex_spinComponents}
	\Gamma^\ssb = \Gamma^\sig11   , \quad 
	\hat\Gamma^{\ssb} = \Gamma^\sig12  , \quad
	\Gamma^\sss = \Gamma^\sig00 .
\end{align}
Furthermore, crossing symmetry relates 
$\Gamma\updown$ and $\hat\Gamma^{{\uparrow\downarrow}}$,
and SU(2) spin symmetry yields
$\Gamma^\sss = \Gamma^{\ssb} + \hat\Gamma^{\ssb}$ \cite{Rohringer2013}.

On the level of the full vertex, one defines the charge, spin, and singlet or triplet pairing channels as \cite{Bickers2004,Rohringer2012}
\begin{align}
	\Gamma^\text{ch/sp} = \Gamma\upup   \pm     \Gamma\updown ,
	\quad
	\Gamma^\text{tr/si} = \Gamma\updown \pm \hat\Gamma\updown .
	\label{eq:Appendix:SBE:Channels:Gamma_chsp}
\end{align}
This notation carries over to all {vertex objects like $\nabla_r^\alpha$, $\lambda_r^\alpha$ and $w_r^\alpha$}, with $\alpha$ denoting ch, sp, si, or tr.

The bare vertex has $U\upup = 0$ and 
$U\updown = - \hat U\updown$, so that%
\begin{subequations}
	\label{eq:Appendix:SBE:Channels:Gamma0}
	\begin{align}
		U^\text{ch/sp} &= U\upup \pm U\updown = \pm U\updown ,
		\\
		U^\text{si} &= U\updown - \hat{U}\updown = 2 U\updown . 
		\label{eq:Appendix:SBE:Channels:Gamma0_s}
	\end{align}
\end{subequations}
The bare interaction $U^{\text{tr}}$ in the triplet pairing channel vanishes and does not give a $U$-reducible contribution \cite{Krien2019}.

We now show that, if the ingredients of the SBE decomposition \Eq{eq:FinalSBE-decomposition} are expressed through the physical charge and spin components (ch, sp) rather than the diagrammatic components ($\uparrow\uparrow$, $\uparrow\downarrow$) used here, one indeed obtains the original form of the SBE decomposition depicted in Fig.~1 of Ref.~\cite{Krien2019}.

This is trivial to see for the fully $U$-irreducible part $\varphi^{U\text{irr}}$ (analogous to \Eq{eq:Appendix:SBE:Channels:Gamma_chsp}) and the bare vertex $U$ (\Eq{eq:Appendix:SBE:Channels:Gamma0}). It remains to show that for the \Ur-reducible terms $\nabla_r = \bar\lambda_r \fcirc w_r \fcirc \lambda_r$, the components $\nabla_r^\alpha$ have the form given in Fig.~1 of Ref.~\cite{Krien2019}, with $\alpha=\text{ch}$ or $\text{sp}$.

We start with the $t$ channel. 
Defining sign factors for charge and spin 
channels, $s^\text{ch} = 1$ and  $s^\text{sp} = -1$, we have
\begin{align}
	\nabla_t^\alpha 
	&= \nabla_t\upup + s^\alpha \nabla_t\updown
	\nonumber \\
	&= \bar\lambda_t^{\sigma\uparrow|\sigma\uparrow} w_t^{\sigma'\sigma|\sigma'\sigma} \lambda_t^{\uparrow\sigma'|\uparrow\sigma'}
	+ s^\alpha
	\bar\lambda_t^{\sigma\downarrow|\sigma\downarrow} w_t^{\sigma'\sigma|\sigma'\sigma} \lambda_t^{\uparrow\sigma'|\uparrow\sigma'}
	.
\end{align}
Here, we sum as usual over spin indices $\sigma$, $\sigma'$. Making use of 
$w_t\upup = w_t\downdown$, $w_t\downup = w_t\updown$, and similarly for $\bar\lambda_t$, $\lambda_t$, we can collect the summands as
\begin{align}
	\nabla_t^\alpha 
	&= (\bar{\lambda}_t\upup + s^\alpha \bar{\lambda}_t\updown)
	(w_t\upup + s^\alpha w_t\updown)
	(\lambda_t\upup + s^\alpha \lambda_t\updown)
	\nonumber \\
	&= \bar\lambda_t^\alpha w_t^\alpha \lambda_t^\alpha ,
\end{align}
which is equivalent to $\nabla^\text{ph}$ in Ref.~\cite{Krien2019}. (Note that in our convention of depicting diagrams, all diagrams are mirrored along the diagonal from the top left to bottom right (i.e., the bottom left and top right legs are exchanged) compared to the convention used in Ref.~\cite{Krien2019}: The ph ($\overline{\text{ph}}$) channel corresponds to the $t$ ($a$) channel.)

We continue with the $a$ channel, which is related to the $t$ channel by crossing symmetry,
\begin{align}
	\label{eq:crossingsymmetry-omega-t-omega-a}
	\hat\Gamma\updown (\wa, \va, \va') 
	= -\Gamma\updown (\wt=\wa, \vt=\va, \vt'=\va') .
\end{align}
The frequency arguments on the right are defined according
to the $t$-channel conventions $(\omega_t, \nu_t, \nu'_t)$, 
and then evaluated at the $a$-channel frequencies occuring on the left.
In particular, we have (cf.~Eq.~(11) of Ref.~\cite{Krien2019})
\begin{align}
	&\Gamma^\alpha (\wa, \va, \va')
	\nonumber \\
	&= - \tfrac12 \left[
	\Gamma^\text{ch} + (1+2s^\alpha) \Gamma^\text{sp}
	\right]
	(\wt=\wa, \vt=\va, \vt'=\va') .
\end{align}
The \Ua-reducible diagrams $\nabla_a$ can therefore be expressed through the \Ut-reducible diagrams $\nabla_t$:
\begin{align}
	&\nabla_a^\alpha (\wa, \va, \va')
	\nonumber \\
	&= -\tfrac12 \left[
	\bar\lambda_t^\text{ch} w_t^\text{ch} \lambda_t^\text{ch} + (1+2s^\alpha) \bar\lambda_t^\text{sp} w_t^\text{sp} \lambda_t^\text{sp}
	\right]
	(\wa, \va, \va') ,
\end{align}
reproducing $\nabla^{\overline{\text{ph}}}$ in Ref.~\cite{Krien2019}. The frequency arguments on the right have the same meaning as in  \Eq{eq:crossingsymmetry-omega-t-omega-a}.

Last, we consider the $p$ channel. With \sutwo\ symmetry, $\nabla_p\upup = \nabla_p\updown + \hat\nabla_p\updown$, we have
\begin{align}
	\nabla_p^\alpha
	&= \nabla_p\upup + s^\alpha \nabla_p\updown
	= \hat\nabla_p\updown + (1+s^\alpha) \nabla_p\updown
	\nonumber \\
	&=
	\bar\lambda_p^{\uparrow\downarrow|\sigma\bar\sigma}
	w_p^{\sigma\bar\sigma|\sigma'\bar\sigma'}
	\lambda_p^{\sigma'\bar\sigma'|\uparrow\downarrow}
	\nonumber \\
	&\phantom{=} + (1+s^\alpha)
	\bar\lambda_p^{\uparrow\downarrow|\sigma\bar\sigma}
	w_p^{\sigma\bar\sigma|\sigma'\bar\sigma'}
	\lambda_p^{\sigma'\bar\sigma'|\downarrow\uparrow}
	.
\end{align}
Note that the spins in the first and second pair of spin indices of $w_p$ have to be opposite, $\sigma\bar\sigma$ and $\sigma'\bar\sigma'$, since they connect to the same bare vertex (cf.~\Fig{fig:lambda_w_lambda}), and $U^{\sigma\sigma}=0$.
Furthermore, the crossing relation $U\updown = -\hat{U}\updown$ implies $w_p\updown = -\hat{w}_p\updown$. By use of this, we can combine the terms in the spin sums as
\begin{align}
	\nabla_p^\alpha
	&= \tfrac{s^\alpha}{2}
	(\bar\lambda_p\updown - \hat{\bar{\lambda}}_p\updown)
	(w_p\updown - \hat{w}_p\updown)
	(\lambda_p\updown - \hat\lambda_p\updown)
	\nonumber \\
	&= \tfrac{s^\alpha}{2} \bar\lambda_p^\text{si} w_p^\text{si} \lambda_p^\text{si}
	,
\end{align}
which gives $\nabla^\text{pp}$ in Ref.~\cite{Krien2019}.

In summary, we thus reproduce the decomposition of Ref.~\cite{Krien2019}:
\begin{subequations}
	\label{eq:SBE_decomposition_algebraic}
	\begin{align}
		\Gamma^\alpha &=
		\varphi^{U\text{irr},\alpha} + \nabla_{a}^\alpha + \nabla_{p}^\alpha + \nabla_{t}^\alpha - 2 U^\alpha ,	
		\intertext{where the the $\Ur$-reducible parts are defined
			as}	
		\label{eq:SBE_decomposition_algebraic_Nabla_a}
		\nabla_{a}^\alpha(\omega_a, \nu_a, \nu_a') &=
		-\tfrac{1}{2} \nabla_{t}^\text{ch}(\omega_a, \nu_a, \nu_a')
		\nonumber \\
		&		
		\quad- (\tfrac{3}{2} - 2 \delta_{\alpha, \text{sp}})	\nabla_{t}^\text{sp}(\omega_a, \nu_a, \nu_a')
		,
		\\
		\nabla_{p}^\alpha (\omega_p, \nu_p, \nu_p') &=
		(\tfrac{1}{2} - \delta_{\alpha, \text{sp}}) [\bar{\lambda}^{\text{si}}_p w^{\text{si}}_p \lambda^{\text{si}}_p](\omega_p, \nu_p, \nu_p'),
		\\
		\nabla_{t}^\alpha(\omega_t, \nu_t, \nu_t') &=
		[\bar{\lambda}^{\alpha}_t w^{\alpha}_t \lambda^{\alpha}_t](\omega_t, \nu_t, \nu_t').
	\end{align}
\end{subequations}

\section{Correlators and susceptibilities}
\label{app:3-point-correlators}

Reference~\cite{Krien2019} established that the 
SBE ingredients $\bar\lambda_r$, $w_r$, $\lambda_r$ are related to
three-point correlators and generalized susceptibilities. 
For completeness, we illustrate here how these relations arise within the present framework. 
The starting point is the general relation between the four-point correlator $G^{(4)}$ and the four-point vertex $\Gamma$,
\begin{align}
G^{(4)}_{12|1'2'} = \langle c_1c_2\bar{c}_{2'}\bar{c}_{1'}\rangle &= G_{1|1'}G_{2|2'}- G_{1|2'}G_{2|1'} \nonumber \\
&\phantom{=}+G_{1|5'}G_{2|6'}\Gamma_{5'6'|56}G_{5|1'}G_{6|2'}.
\end{align}
By combining two fermionic fields, one obtains the bosonic exchange field $\psi$, the pairing field $\phi$, and its conjugate $\bar \phi$,
\begin{subequations}
	\label{eq:bosons}
	\begin{align}
		\psi_{12'}(\omega) &= \sum_\nu {c}_{1}(\nu-\tfrac{\omega}{2})\bar{c}_{2'}(\nu+\tfrac{\omega}{2}) =\bar{\psi}_{2'1}(-\omega),\\
		\phi_{12}(\omega) &= \sum_\nu c_1(\tfrac{\omega}{2}+\nu)c_2(\tfrac{\omega}{2}-\nu),\\
		\bar{\phi}_{1'2'}(\omega) &= \sum_{\nu'}\bar{c}_{2'}(\tfrac{\omega}{2}-\nu')\bar{c}_{1'}(\tfrac{\omega}{2} +\nu').
	\end{align}
\end{subequations}
Three-point correlators and bosonic two-point correlators involving these fields can be obtained from $G^{(4)}$ by summing over the frequency $\nu^{(\prime)}_r$ in the channel-specific parametrization (cf. \Eq{eq:r-representation-Gamma} and \Fig{fig:Vertex_frequency_parametrization}):
\begin{subequations}
	\label{eq:bosonic_correlators}
	\begin{align}
		\bar{G}_{r;12|1'2'}^{(3)}(\omega_r,\nu_r) &= \sum_{\nu'_r} G_{12|1'2'}^{(4)}(\omega_r,\nu_r,\nu'_r),\\
		{G}_{r;12|1'2'}^{(3)}(\omega_r,\nu'_r) &= \sum_{\nu_r} 	G_{12|1'2'}^{(4)}(\omega_r,\nu_r,\nu'_r),\\
		D_{r;12|1'2'}(\omega_r) &= \sum_{\nu_r,\nu'_r} 	G_{12|1'2'}^{(4)}(\omega_r,\nu_r,\nu'_r).
	\end{align}
\end{subequations}
For example, in the $p$ channel, we have,
\begin{align}
	\label{eq:example_G^(3)_p}
	\bar{G}^{(3)}_{p;12|1'2'} = \bigl\langle c_{1} c_{2}\bar \phi_{1'2'}\bigr\rangle, \quad D_{p;12|1'2'} = \bigl\langle\phi_{12}\bar\phi_{1'2'}\bigr\rangle.
\end{align}

The four-point correlator $G^{(4)}$ is closely related to the \textsl{generalized} susceptibilities $\chi_r^{(4)}$ \cite{Rohringer2012}:
\begin{subequations}
	\label{eq:chi(4)}
	\begin{align}
	&\chi_{a;12|1'2'}^{(4)}(\omega_a,\nu_a,\nu'_a) \nonumber \\
	\nonumber&= 
	G^{(4)}_{12|1'2'}(\omega_a,\nu_a,\nu'_a) + \delta_{\omega_a,0}G_{1|2'}(\nu_a)G_{2|1'}(\nu'_a)\\
	&=
	\delta_{\nu_a\nu'_a}\Pi_{a;12|1'2'}(\omega_a,\nu_a) + [\Pi_a\!\circ\! \Gamma\!\circ\! \Pi_a]_{12|1'2'}(\omega_a,\nu_a,\nu'_a),
	\\
	\label{eq:chi(4)_p}
	 &\chi_{p;12|1'2'}^{(4)}(\omega_p,\nu_p,\nu'_p) 
	= \tfrac{1}{4}G^{(4)}_{12|1'2'}(\omega_p,\nu_p,\nu'_p)
	\nonumber \\
	\nonumber&= \delta_{\nu_p\nu'_p}\tfrac{1}{2}\Pi_{p;12|1'2'}(\omega_p,\nu_p) -\delta_{\nu_p,-\nu'_p}\tfrac{1}{2}\Pi_{p;12|2'1'}(\omega_p,\nu_p)
	\\
	&\phantom{=} +[\Pi_p\!\circ\! \Gamma\!\circ\!  \Pi_p]_{12|1'2'}(\omega_p,\nu_p,\nu'_p),
	\\
 &\chi_{t;12|1'2'}^{(4)}(\omega_t,\nu_t,\nu'_t) \nonumber \\
	\nonumber &=G_{12|1'2'}^{(4)}(\omega_t,\nu_t,\nu'_t) - \delta_{\omega_t,0}G_{1|1'}(\nu'_t)G_{2|2'}(\nu_t) \\
	&=  \delta_{\nu_t\nu'_t}\Pi_{t;12|1'2'}(\omega_t,\nu_t)
	+ [\Pi_t \!\circ\! \Gamma\!\circ\! \Pi_t]_{12|1'2'}(\omega_t,\nu_t,\nu'_t).
	\end{align}
\end{subequations}
In analogy to \Eqs{eq:bosonic_correlators}, we then obtain three-point functions $\bar\chi_r^{(3)}, \chi_r^{(3)}$ and physical susceptibilities $\chi_r$ by summing over frequencies:
\begin{subequations}
	\label{eq:chi(3)=sum_chi(4)}
	\begin{align}
		\bar{\chi}_{r;12|1'2'}^{(3)}(\omega_r,\nu_r) &= \sum_{\nu'_r} \chi_{r;12|1'2'}^{(4)}(\omega_r,\nu_r,\nu'_r),\\
		{\chi}_{r;12|1'2'}^{(3)}(\omega_r,\nu'_r) &= \sum_{\nu_r} 	\chi_{r;12|1'2'}^{(4)}(\omega_r,\nu_r,\nu'_r),\\
		\chi_{r;12|1'2'}(\omega_r) &= \sum_{\nu_r,\nu'_r} 	\chi_{r;12|1'2'}^{(4)}(\omega_r,\nu_r,\nu'_r).
		\label{eq:chi_r-appendix}
	\end{align}
\end{subequations}
The prefactor $\tfrac{1}{4}$ in \Eq{eq:chi(4)_p} ensures that the susceptibility $\chi_r$ in \Eq{eq:chi_r-appendix} is consistent with its counterpart in the main text (cf.~\Eq{eq:susceptibility}).

To make a connection between $\bar{\chi}^{(3)}_r, \chi^{(3)}_r$, $\chi_r$ and SBE objects, we use \Eqs{eq:chi(4)}, multiply by the bare interaction $U$, and express the result in terms of the four-point vertex:
\begin{subequations}
\begin{align}
\bar{\chi}_{r}^{(3)}\fcirc U &= 
\Pi_r\circ(U+\Gamma\circ\Pi_r\circ U),\\
U\fcirc \chi_r^{(3)} &=(U + U\circ\Pi_r\circ\Gamma)\circ\Pi_r,\\
U\fcirc \chi_r \fcirc U &=  U\circ\Pi_r\circ U + U\circ \Pi_r \circ \Gamma \circ \Pi_r \circ U.
\end{align}
\end{subequations}
Finally, comparing
these expressions to \Eqs{eq:definition_Hedin-vertex}--\eqref{eq:lambda_r-w_r-susceptibility} shows
their relation to the SBE ingredients:
\begin{subequations}
	\label{eq:chi(SBE)}
	\begin{align}
		\bar{\chi}_r^{(3)} &= \Pi_r \circ \bar\lambda_r\fcirc w_r \fcirc U^{-1} = \Pi_r \circ \bar\Gamma^{(3)}_r,\\
		\chi_r^{(3)} &= U^{-1}\fcirc w_r\fcirc\lambda_r \circ\Pi_r = \Gamma^{(3)}_r\circ\Pi_r ,\\
		\chi_r &=  U^{-1} \fcirc ( w_r - U ) \fcirc U^{-1} .
	\end{align}
\end{subequations}
These relations are analogous to those given in Eqs.~(6), (8) and (15) in Ref.~\cite{Krien2019}. Relations between the bosonic correlators $\bar G_r^{(3)}$, $G_r^{(3)}$, $D_r$ from \Eqs{eq:bosonic_correlators} and the SBE ingredients $\bar\lambda_r$, $w_r$, $\lambda_r$ are analogous up to disconnected terms and can be readily constructed from \Eqs{eq:chi(3)=sum_chi(4)}, \eqref{eq:chi(4)},  and \eqref{eq:chi(SBE)}. For example, in the $a$ channel, we have
	\begin{align}
		\nonumber\bar{G}^{(3)}_{a;12|1'2'}(\omega_a,\nu_a) &= [\Pi_a\circ\bar\lambda_a\fcirc w_a\fcirc U^{-1}]_{12|1'2'}(\omega_a,\nu_a)\\
		\nonumber&\phantom{=}-\delta_{\omega_a,0}G_{1|2'}(\nu_a)\sum_{\nu'_a}G_{2|1'}(\nu'_a),\\
		\nonumber D_{a;12|1'2'}(\omega_a) &= [U^{-1}\fcirc(w_a-U)\fcirc U^{-1}]_{12|12'}(\omega_a)\\
		&\phantom{=} -\delta_{\omega_a,0}\sum_{\nu_a}G_{1|2'}(\nu_a)\sum_{\nu'_a}G_{2|1'}(\nu'_a).
	\end{align}

\section{Susceptibilities for Hubbard interaction}
\label{Appendix:susceptibilities}

The susceptibilities defined in \Eq{eq:susceptibility} and in App.~\ref{app:3-point-correlators} exhibit general dependencies \wrt their non-frequency indices $12|1'2'$.
In the following, we show how they are related to physical charge, spin, and pairing susceptibilities. To this end, we focus on models with a local (momentum-independent) bare interaction, which has only spin degrees of freedom subject to the Pauli principle. In the $a$ and $t$ channel, \Eq{eq:w_r_chi} with $\K1r=w_r-U$ then reads
\begin{subequations}
	\label{eq:K1_chi_at}
	\begin{align}
		\K1{a;\sigma\sigma'|\sigma\sigma'}
		&= 
		U^{\sigma\bar\sigma|\bar\sigma'\sigma'} \,
		\chi_a^{\bar\sigma'\bar\sigma|\bar\sigma'\bar\sigma} \,
		U^{\bar\sigma'\sigma'|\sigma\bar\sigma}
		,
		\\
		\K1{t;\sigma\sigma'|\sigma\sigma'}
		&=
		U^{\bar\sigma'\sigma'|\bar\sigma'\sigma'} \,
		\chi_t^{\bar\sigma\bar\sigma'|\bar\sigma\bar\sigma'} \,
		U^{\sigma\bar\sigma|\sigma\bar\sigma}
		.
	\end{align}
\end{subequations}
We further specify $U^{\sigma\bar\sigma|\bar\sigma'\sigma'} = u(\delta_{\sigma\sigma'} - \delta_{\sigma\bar\sigma'})$, with the (scalar) bare interaction strength $u$.
With \sutwo\ symmetry, $\chi_r^{\sigma_1\sigma_1'|\sigma_2\sigma_2'} = \chi_r^{\bar\sigma_1\bar\sigma_1'|\bar\sigma_2\bar\sigma_2'}$, \Eq{eq:K1_chi_at} thus simplifies to 
\begin{align}
	\chi_{a/t}^{\sigma\sigma'|\sigma\sigma'} = \K1{a/t;\sigma\sigma'|\sigma\sigma'}/u^2 .
	\label{eq:chi_K1_at}
\end{align}
In the $p$ channel, we have
\begin{subequations}
	\begin{align}
		\K1{p;\sigma\sigma'|\sigma\sigma'}
		&= \sum_{\sigma_1\sigma_2} 
		U^{\sigma\sigma'|\sigma_1\bar\sigma_1} \,
		\chi_p^{\sigma_1\bar\sigma_1|\sigma_2\bar\sigma_2} \,
		U^{\sigma_2\bar\sigma_2|\sigma\sigma'}
		\\
		&=
		U^{\sigma\sigma'|\sigma\sigma'} \,
		\tilde\chi_p^{\sigma\sigma'|\sigma\sigma'} \,
		U^{\sigma\sigma'|\sigma\sigma'}
		.
		\label{eq:K1_chi_p_SU2}
	\end{align}
\end{subequations}
Here, the second line \eqref{eq:K1_chi_p_SU2} follows from \sutwo\ and crossing symmetry. It employs
\begin{align}
	\label{eq:susceptibility_tilde}
	\tilde\chi_p(\wp) &= 
	\nonumber[\boldI_p\circ\tilde\Pi_p\circ\boldI_p](\omega_p)\\
	&\phantom{=} +[\boldI_p\circ\tilde\Pi_p\circ\Gamma\circ\tilde\Pi_p\circ\boldI_p](\omega_p),
\end{align}
where $\tilde\Pi_{p;34|3'4'} = G_{3|3'}G_{4|4'} = 2\Pi_{p;34|3'4'}$ does \textsl{not} include a prefactor 1/2 (introduced in \Eq{eq:Definition_Pi_p} in order to avoid double counting within internal spin sums), since there are no spin sums in \Eq{eq:K1_chi_p_SU2}. 
(This definition of the $p$ susceptibility agrees with the related literature, e.g., Ref.~\cite{Rohringer2012}.)
With $U^{\sigma\sigma'|\sigma\sigma'} = -u \delta_{\sigma\bar\sigma'}$, we can write
\begin{align}
	\tilde\chi_p^{\sigma\sigma'|\sigma\sigma'} = \delta_{\sigma\bar\sigma'}\, 
	\K1{p;\sigma\sigma'|\sigma\sigma'} / u^2  ,
\end{align}
in analogy to \Eq{eq:chi_K1_at}.

The relation between these ``diagrammatic'' susceptibilities $\chi_r$ and their ``physical'' counterparts can be made explicit by means of the bilinears
\begin{subequations}
	\begin{alignat}{2}
		\rho_{\sigma\sigma'} &= \bar{c}_\sigma c_{\sigma'}
		, & \quad
		\delta \rho_{\sigma\sigma'} 
		&= \rho_{\sigma\sigma'} 
		- \langle\rho_{\sigma\sigma}\rangle \delta_{\sigma\sigma'}
		\\
		\rho^-_{\sigma\sigma'} &= c_\sigma c_{\sigma'}
		,
		& 
		\rho^+_{\sigma\sigma'} & = \bar{c}_{\sigma'} \bar{c}_\sigma .
	\end{alignat}
\end{subequations}
Then, we have in the imaginary-time domain
\begin{subequations}
	\label{eq:physical_susceptibilities}
	\begin{align}
		\chi_a^{\sigma\sigma'|\sigma\sigma'} (\tau) 
		&= -	\langle \delta \rho_{\sigma'\sigma}(\tau) \delta \rho_{\sigma\sigma'}(0) \rangle 
		,
		\\
		\tilde\chi_p^{\sigma\sigma'|\sigma\sigma'} (\tau) 
		&= \langle \rho^-_{\sigma\sigma'}(\tau) \rho^+_{\sigma\sigma'}(0) \rangle 
		,
		\\
		\chi_t^{\sigma\sigma'|\sigma\sigma'} (\tau) 
		&= \langle \delta n_\sigma(\tau) \delta n_{\sigma'}(0) \rangle
		. 
	\end{align}
\end{subequations}
with $n_\sigma = \rho_{\sigma\sigma}$.
Choosing the spin arguments as $\chi_r^{\uparrow\downarrow} = \chi_r^{\uparrow\downarrow|\uparrow\downarrow}$,
we furthermore get
\begin{subequations}
	\begin{align}
		\chi_a^{\uparrow\downarrow} (\tau) 
		&= -	\langle S_-(\tau) S_+ \rangle 
		,
		\\
		\tilde\chi_p^{\uparrow\downarrow} (\tau) 
		&= \langle \Delta_{\mathrm{si}} (\tau) \Delta_{\mathrm{si}}^\dag(0) \rangle 
		,
		\\
		\chi_t^{\uparrow\downarrow} (\tau) 
		&= 	\langle \delta n_\uparrow(\tau) \delta n_\downarrow(0) \rangle
		.
	\end{align}
\end{subequations}
Hence, $\chi_a^{\uparrow\downarrow}$ describes spin fluctuations 
($S_-=\bar{c}_\downarrow c_\uparrow$, $S_+=\bar{c}_\uparrow c_\downarrow$)
and $\tilde\chi_p^{\uparrow\downarrow}$ singlet pairing fluctuations
($\Delta_{\mathrm{si}} = c_\uparrow c_\downarrow$).
By SU(2) spin symmetry, 
$\tfrac12 \chi_a^{\uparrow\downarrow} (\tau) 
= -\langle S_z(\tau) S_z \rangle$,
with $S_z = \tfrac12 (n_\uparrow-n_\downarrow) = \tfrac12 (\delta n_\uparrow - \delta n_\downarrow)$.
It then follows that
\begin{align}
	\nonumber
	\chi_t^{\uparrow\downarrow} (\tau) - \tfrac12 \chi_a^{\uparrow\downarrow} (\tau)
	& =
	\tfrac12 (\langle \delta n_\uparrow(\tau) \delta n_\uparrow \rangle + \langle \delta n_\uparrow(\tau) \delta n_\downarrow \rangle)
	\\
	& =
	\tfrac14 \langle \delta n(\tau) \delta n \rangle
\end{align}
describes charge fluctuations
with $n = n_\uparrow + n_\downarrow$.

\bibliographystyle{epj}

\end{document}